\documentclass{emulateapj}
\usepackage{graphics,graphicx,rotating}
\usepackage[colorlinks,allcolors=blue]{hyperref}
\usepackage[all]{hypcap} 
\usepackage{xspace}
\usepackage{amssymb}
\usepackage{amsmath}  
\usepackage[percent]{overpic}    
\usepackage{epstopdf}
\usepackage[titletoc]{appendix}
\usepackage{subfigure}
\usepackage{xcolor}
\usepackage{tabularx,ragged2e}
\usepackage{natbib}
\begin{document}

\title{The (black hole mass)\sbond  (color) relations for
  early- and late-type galaxies: red and blue sequences}
\shorttitle{Black hole mass\sbond   color correlation}
\shortauthors{Dullo et al.} 
\author{{\color{blue} Bililign T.\ Dullo}$^{1}$, {\color{blue} Alexandre Y. K. Bouquin}$^{1,2,3}$, {\color{blue} Armando Gil de
  Paz}$^{1}$, {\color{blue} Johan H. Knapen}$^{2,3}$, and {\color{blue} Javier Gorgas}$^{1}$}
\affil{\altaffilmark{1} Departamento de F\'isica de la Tierra y
  Astrof\'isica, Instituto de F\'isica de Part\'iculas y del Cosmos IPARCOS,
  Universidad Complutense de Madrid, E-28040 Madrid, Spain}
\affil{\altaffilmark{2} Instituto de Astrof\'{i}sica de Canarias, V\'{i}a L\'{a}ctea S/N, 
E-38205 La Laguna, Spain}
\affil{\altaffilmark{3} Departamento de Astrof\'{i}sica, Universidad de La Laguna, E-38206 La 
Laguna, Spain}

\begin{abstract}
 
Tight correlations between supermassive black hole (SMBH) mass
  ($M_{\rm BH}$) and the properties of the host galaxy have useful
  implications for our understanding of the growth of SMBHs and
  evolution of galaxies. Here, we present newly observed correlations
  between $M_{\rm BH}$ and the
  host galaxy total \mbox{UV$-$ [3.6]} color ($\mathcal{C_{\rm UV,tot}}$,
  Pearson's r = $0.6-0.7$) for a sample of 67 galaxies (20 early-type
  galaxies and 47 late-type galaxies) with directly measured
  $M_{\rm BH}$ in the \textit{GALEX}/S$^{4}$G survey.  The colors are
  carefully measured in a homogeneous manner using the galaxies' FUV,
  NUV and 3.6 $\micron$ magnitudes and their multi-component
  structural decompositions in the literature.  We find that more
  massive SMBHs are hosted by (early- and late-type) galaxies with
  redder colors, but the $M_{\rm BH}- \mathcal{C_{\rm UV,tot}}$
  relations for the two morphological types have slopes that differ at
  $\sim 2 \sigma$ level. Early-type galaxies define a red sequence in
  the $M_{\rm BH}- \mathcal{C_{\rm UV,tot}}$ diagrams, while late-type
  galaxies trace a blue sequence. Within the assumption that the
  specific star formation rate of a galaxy (sSFR) is well traced by
  $L_{\rm UV}/L_{\rm 3.6}$, it follows that the SMBH masses for
  late-type galaxies exhibit a steeper dependence on sSFR than those
  for early-type galaxies. The
  \mbox{$M_{\rm BH}- \mathcal{C_{\rm UV,tot}}$} and
  \mbox{$M_{\rm BH}-L_{\rm 3.6,tot}$} relations for the sample
  galaxies reveal a comparable level of vertical scatter in the log
  $M_{\rm BH}$ direction, roughly $5\%-27\%$ more than the vertical
  scatter of the $M_{\rm BH}-\sigma$ relation. Our
  $M_{\rm BH}- \mathcal{C_{\rm UV,tot}}$ relations suggest different
  channels of SMBH growth for early- and late-type galaxies,
  consistent with their distinct formation and evolution
  scenarios. These new relations offer the prospect of estimating SMBH
  masses reliably using only the galaxy color. Furthermore, we show
  that they are capable of estimating intermediate black hole masses
  in low-mass, early- and late-type galaxies.

\end{abstract}


\section{Introduction}

Almost all local galaxies are believed to harbor a supermassive
black hole (SMBH, $M_{\rm BH} \sim 10^{5}-10^{9} M_{\sun}$) at their center
(\citealt{1998AJ....115.2285M,1998Natur.395A..14R,2005SSRv..116..523F}). The
connection between SMBHs and their host galaxies has been a subject of
 interest, since \citet[see also
\citealt{1989IAUS..134..217D}]{1995ARA&A..33..581K} first reported a
linear correlation between SMBH mass ($M_{\rm BH}$) and the luminosity
of the host bulge (i.e., the entire galaxy in case of elliptical
galaxies), see \citet{2013ARA&A..51..511K,2016ASSL..418..263G} for
recent reviews.  SMBH masses scale with a number of host galaxy
properties including stellar velocity dispersion ($\sigma$,
\citealt{2000ApJ...539L...9F,2000ApJ...539L..13G}), bulge luminosity
($L_{\rm bulge}$) and bulge mass ($M_{\rm bulge}$,
\citealt{1995ARA&A..33..581K, 1998AJ....115.2285M}), depleted stellar
core \citep[e.g., ][]{2019ApJ...886...80D} and stellar concentration
\citep{2001ApJ...563L..11G}. Not only did these scaling relations
allow us to predict $M_{\rm BH}$ in galaxies, but they also led to a
suggestion that the fueling and growth rate of the central SMBH are
intimately coupled to the star formation rate and stellar mass
build-up of the host galaxy. However, the exact nature of the physical
mechanism driving this connection remains unclear.

Accretion of gas onto SMBHs triggers active galactic nucleus (AGN)
feedback, critical for the regulation of the star formation and growth
of the host galaxy (e.g., \citealt{1998A&A...331L...1S,
  1999MNRAS.308L..39F,
  2005Natur.435..629S,2005Natur.433..604D,2006MNRAS.365...11C,
  2006ApJS..163....1H}).  The distinct formation histories, colors and
structural properties of early- and late-type galaxies may therefore
reflect two different channels of gas accretion and SMBH growth for
the two morphological types (see
\citealt{2009ApJ...694..599H,2009ApJ...690...20S,2010ApJ...711..284S,2014MNRAS.440..889S,2018MNRAS.473.5237K}). Observations
show that early-type (i.e., lenticular and elliptical) galaxies do not
correlate with SMBHs in the same manner as late-type (i.e., spiral)
galaxies do (e.g.,
\citealt{2013ApJ...764..184M,2016ApJ...818...47S,2018MNRAS.473.5237K,2019arXiv190304738S}). Early-type
galaxies are typically red with low level star-formation rate, while
late-type galaxies have blue colors and are actively star
forming. This dichotomy gives rise to a bimodal distribution of
galaxies in the color-magnitude diagram---early-type galaxies define a
red-sequence separate from late-type galaxies which reside in a blue
cloud (e.g.,
\citealt{1964AJ.....69..635C,1977A&A....59..317V,2004ApJ...600..681B,2009ApJ...706L.173B}).
 
In the hierarchical galaxy formation scenario, early-type galaxies are
built through mergers of smaller systems and accretion events (e.g.,
\citealt{1978MNRAS.183..341W,2006ApJ...636L..81N,2009ApJS..181..135H,2009ApJS..181..486H,2016MNRAS.458.2371R,2017MNRAS.470.3507M}). Major
merger driven inflow of gas into the nuclear regions of the newly
formed merger remnant can produce rapid bursts of star formation and
fuel the SMBH
(\citealt{1991ApJ...370L..65B,1996ApJ...471..115B,2006MNRAS.372..839N,2010MNRAS.407.1529H}). For
late-type (i.e., spiral) galaxies, one of the most advocated formation
scenarios is secular evolution involving non-axisymmetric stellar
structures, such as bars and spiral arms which can drive an inflow of
gas from the disk into the nuclear region and onto the central SMBH
\citep{1982ApJ...257...75K,1996ApJ...457L..73C,1997AJ....114.2366C,2004ARA&A..42..603K,2005MNRAS.358.1477A,2007MNRAS.381..401L,2008AJ....136..773F,2008MNRAS.388.1708G,2009MNRAS.393.1531G,2016MNRAS.459.4109T,2016MNRAS.462.3800D,2019ApJ...871....9D}.

\begin{figure*}
\hspace{1.80cm}
\includegraphics[angle=270,scale=0.745]{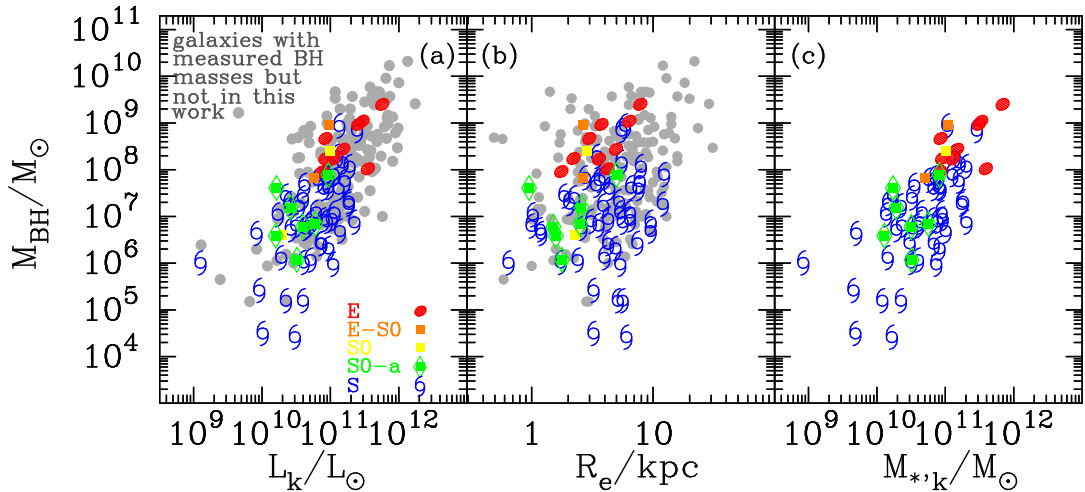}
\caption{Galaxies with measured SMBH masses ($M_{\rm BH}$). 
  $M_{\rm BH}$ plotted  as a function of (a)  total $K_{s}$-band
  luminosity and (b) the half-light radius
  ($R_{\rm e}$) for a sample
  of 245
  galaxies with measured $M_{\rm BH}$ \citep[his
  Table~2]{2016ApJ...831..134V}.  Colored symbols denote our sample of
  67 galaxies, whereas filled
  gray circles show the remaining 178 galaxies with measured
  $M_{\rm BH}$ in \citet[his
  Table~2]{2016ApJ...831..134V}. Panel (c): $M_{\rm BH}$ versus total  $K_{s}$-band
  stellar mass ($M_{*,k}$) for our sample derived from the  total $K_{s}$-band
  luminosities assuming the $K_{s}$-band mass-to-light ratio
  $M_{*}/L_{k}= 0.10\sigma^{0.45}$ \citep{2016ApJ...831..134V}. Morphological classifications are
  from HyperLeda. }

\label{Fig1} 
 \end{figure*}

A related issue is the observed departures from single
power-law relations (e.g., bends and offsets) in SMBH
scaling diagrams at the low-mass and high-mass ends or when the galaxy
sample contains S\'ersic and core-S\'ersic galaxies.  In particular,
\citet[see also \citealt{2013ApJ...764..151G}]{2012ApJ...746..113G}
reported two separate \mbox{$M_{\rm BH}-L_{\rm bulge}$} relations
with distinct slopes for S\'ersic and core-S\'ersic galaxies. Core-S\'ersic galaxies are luminous
($M_{B} \la -20.5 \pm 0.5$ mag) galaxies which exhibit a flattening in
their inner stellar light distribution due to a central deficit of
light relative to the inward extrapolation of their outer
\citet{1968adga.book.....S} light profile
(\citealt{1997AJ....114.1771F,2003AJ....125.2951G,2009ApJS..181..486H,2012ApJ...755..163D,2013ApJ...768...36D,2014MNRAS.444.2700D,2015ApJ...798...55D,2017MNRAS.471.2321D,2018MNRAS.475.4670D, 2019ApJ...886...80D}). They
are thought to have formed from a few number of gas-poor major merger
events\footnote{ The
depleted cores of core-S\'ersic galaxies are thought to be scoured by coalescing
binary SMBHs formed in the gas-poor merger events
\citep{1980Natur.287..307B,1991Natur.354..212E,2006ApJ...648..976M}.}, but a small fraction of them can host molecular gas reservoirs that feeds an ongoing, low level star formation \citep{2019arXiv190308884D}. In contrast, the low- and intermediate luminosity ($M_{B} \ga -20.5 $
mag) S\'ersic galaxies, with no depleted cores, are product of
gas-rich mergers
\citep{2009ApJS..181..135H,2012ApJ...755..163D,2014MNRAS.444.2700D,2016MNRAS.462.3800D,2019ApJ...871....9D}.
Departures from the best-fitting single power-law relation in a BH
scaling diagram may hold implications for SMBH and galaxy
co-evolution, however, as the majority of SMBH scaling relations to
date are based on host galaxy properties that trace solely the old
stellar populations, disregarding the young and intermediate-aged
stars, the exact mechanism establishing the supposed coupling between
the growth of the SMBHs and the build-up of their host galaxies is
unclear.   

Here, we explore a scenario in which the complex interplay between the
details of the SMBH growth, efficiency of AGN feedback, regulated star
formation histories and major merger histories of the host galaxy
establish a relation between the SMBH and color of the host galaxy
(see also \citealt{2010ApJ...711..284S,2014MNRAS.440..889S}). There are
emerging evidences showing a  link between $M_{\rm BH}$ and the
star formation rate in nearby galaxies. \citet{2017ApJ...844..170T} reported
an inverse correlation between specific star formation rate (sSFR) and
specific supermassive black hole mass. \citet{2018Natur.553..307M}
observed a trend between SMBH mass and host galaxy star formation
histories, where the star formation in galaxies
hosting more massive SMBHs was quenched early and more efficiently
than in those with less massive SMBHs (see also
\citealt{2019MNRAS.tmp..406V}). Given the UV$-$ [3.6] color
($\mathcal{C_{\rm UV}}$) is a good proxy for sSFR
\citep[e.g.,][]{2018ApJS..234...18B}, a correlation between
$M_{\rm BH}$ and UV$-$ [3.6] color is expected. The UV flux is a proxy
for the current star formation rate since it traces massive, young
stars.  The emission at 3.6 $\micron$, which is less affected by dust
extinction, is a good proxy for stellar mass since it traces primarily
older stellar populations. However, a small fraction (5\%$-$15\%) of
the 3.6 $\micron$ flux can be due to contributions from
intermediate-age stars, polycyclic aromatic hydrocarbons and hot dust
\citep{2012ApJ...744...17M}.  Hitherto, the
\mbox{$M_{\rm BH} - {\rm color}$ } relation was overlooked, in part,
owing to the narrow wavelength baselines commonly used to determine
colors.      

In this paper, we present (for the first time) correlations between
$M_{\rm BH}$ and $\mathcal{C_{\rm UV}}$ colors for 67
\textit{GALEX}/S$^{4}$G galaxies with directly measured SMBH masses
\citep{2016ApJ...831..134V} and homogeneously determined
\textit{GALEX} FUV and NUV, and \textit{Spitzer} 3.6 $\micron$
magnitudes \citep{2018ApJS..234...18B}. Dividing the sample galaxies
by morphology, we fit two different
\mbox{$M_{\rm BH} - \mathcal{C_{\rm UV}}$} relations with distinct
slopes for early- and late-type galaxies. 

The \mbox{$M_{\rm BH} - \mathcal{C_{\rm UV}}$} relation has multiple,
important applications. It allows the SMBHs in early- and late-type
galaxies to be predicted free from uncertainties due to distances and
mass-to-light ratios, although there are uncertainties due to K-corrections
 for more distant galaxies. Furthermore, using
\mbox{$M_{\rm BH} - \mathcal{C_{\rm UV}}$} relation we hope to
understand properly the poorly constrained low-mass end of the SMBH
scaling relations. In so doing, we can predict SMBHs or
intermediate-mass black holes (IMBHs) with masses
$\sim 100 - 10^{5} M_{\sun}$ in low-mass systems and bulgeless spiral galaxies. 
In addition, since colors are easy to measure even for high-redshift
galaxies, studying the galaxy color and BH mass evolution at different
epochs may provide further clues on the different channels of BH growth. 

The paper is organized as follows. Section~\ref{Sec2.0} describes the
sample selection.  Sections~\ref{Sec2.1} and \ref{Sec2.2} describe the
SMBH data, and the UV and 3.6 $\micron$ apparent, asymptotic magnitudes. The
derivation of bulge, disk and total magnitudes for the sample galaxies
along with the corresponding dust corrections and error measurements
are discussed in Sections \ref{Sec2.3} and ~\ref{Sec2.4},
respectively. We go on to discuss the regression techniques employed for fitting
the BH scaling relations in Section \ref{Sec3} and present the results
from our regression analyses in Sections \ref{Sec3.1} and
\ref{Sec3.2}.  Section~\ref{Sec4} provides a discussion of our
results, including the origin and implications of the
\mbox{$M_{\rm BH} - \mathcal{C_{\rm UV}}$}
relations. Section~\ref{Sec5} summarizes our main conclusions.   

There are four appendices at the end of this paper (Appendices
\ref{AppA}, \ref{ApB},  \ref{AppD} and \ref{AppE}). Appendix~\ref{AppA}
discuses our implementation of the MCMC Bayesian statistical
method. Notes on five notable outliers in the
\mbox{$M_{\rm BH} - \mathcal{C_{\rm UV}}$} relations are given in
Appendix~\ref{ApB}. Appendix~\ref{AppD} includes a table, listing apparent total magnitudes, flux
ratios, dust corrections and direct SMBH masses for our sample galaxies. We tentatively predict BH masses in a sample of 1382 
\textit{GALEX}/S$^{4}$G galaxies with no measured  BH masses  using our \mbox{$M_{\rm BH} - \mathcal{C_{\rm UV}}$} relations and tabulated them in  Appendix~\ref{AppE}. 

\section{Sample and data}\label{Sec2}
\subsection{Sample Selection}\label{Sec2.0}

Our investigation of correlations between the SMBH mass ($M_{\rm BH}$)
and (\mbox{UV $-$ [3.6]}) galaxy color uses UV and \mbox{3.6 $\micron$} magnitudes,
determined in a homogeneous manner, for a large sample of galaxies
with measured SMBH masses. Henceforth, we designate the (\mbox{UV $-$
  [3.6]}) color as
$\mathcal{C}$. \citet{2015ApJ...800L..19B,2018ApJS..234...18B}
provided far-UV (FUV, $\lambda_{\rm eff} \sim 1526$ \AA), near-UV
(NUV, $\lambda_{\rm eff} \sim 2267$ \AA) and \mbox{3.6 $\micron$} asymptotic
magnitudes for a sample of 1931 nearby galaxies taken from the
\textit{Spitzer} Survey of Stellar Structure in Galaxies (S$^{4}$G)
sample \citep{2010PASP..122.1397S}.  
They used ultraviolet (UV) and near-infrared (3.6 $\micron$) imaging
data obtained with the \textit{Galaxy Evolution Explorer, GALEX},
\citep{2005ApJ...619L...1M,2007ApJS..173..185G} and the Infrared Array
Camera (IRAC) on the \textit{Spitzer Space Telescope}, respectively.
We use the SMBH sample presented in \citet[his
Table~2]{2016ApJ...831..134V}.  They publish a compilation of directly
measured SMBH masses ($M_{\rm BH}$), half-light radii ($R_{\rm e}$),
and total $K_{s}$-band luminosities ($L_{k}$) for a large sample of
245 galaxies.  We selected all galaxies that were in common to
\citet{2018ApJS..234...18B} and \citet{2016ApJ...831..134V}, resulting
in a sample of 67 galaxies studied in this paper. Homogenized mean
central velocity dispersions ($\sigma$) and morphological
classifications (elliptical E, elliptical-lenticular E-S0, lenticular
S0, lenticular-spiral S0-a and spiral S) the sample galaxies were
obtained from Hyperleda\footnote{http://leda.univ-lyon1.fr}
(\citealt{2003A&A...412...45P,2014A&A...570A..13M}). In the analysis of the scaling
relations (Tables~\ref{Table1}, \ref{Table4}, \ref{Table5} and
\ref{TableBH}), the galaxies are divided into two, broad morphological
classes, early-type galaxies (9 Es, 2 E-S0s, 2 S0s and 7 S0-a) and
late-type galaxies (47 Ss). Most of our late-type galaxies are disk
dominated. The full list of galaxies and their properties are
presented in Appendix~\ref{AppD}.  

In Fig.~\ref{Fig1}, we check on potential sample selection biases by
comparing $M_{\rm BH}$, $R_{\rm e}$, and $L_{k}$ of our sample and
those of other (178) known galaxies with measured black hole masses
using data from \citet[his Table~2]{2016ApJ...831..134V}.  Both the
early- and late-type galaxies in our sample probe large range in
$M_{\rm BH}$, $R_{\rm e}$, and $L_{k}$ to allow a robust investigation
of the $ M_{\rm BH}-\mathcal{C}$ relations. We also find that our
early- and late-type galaxies span a wide range in stellar mass
($M_{*,k}$), Fig.~\ref{Fig1} c. The galaxy stellar masses we compute
using $L_{k}$ and assuming the $K_{s}$-band mass-to-light ratio
$M_{*}/L_{k}= 0.10\sigma^{0.45}$ \citep{2016ApJ...831..134V}.
Late-type galaxies in our sample have $M_{*,k}$ that range from
$10^{9} M_{\sun}$ to $ 2\times10^{11} M_{\sun}$, and for ~75\% of them
$M_{*,k}/M_{\sun} \ga 2\times10^{10} $. For the early-type galaxies,
$10^{10} \la M_{*,k}/M_{\sun} \la 10^{12} $, and ~90\% of these
galaxies have $M_{*,k}/M_{\sun} \ga 2\times10^{10} $. It worth noting
that the \textit{GALEX}/S$^{4}$G sample galaxies
(\citealt{2018ApJS..234...18B}) were chosen to have radio-derived
radial velocities of $V_{\rm radio}<3000 $ Km s$^{-1}$ in
HyperLEDA. As such, the sample lacks {\sc hi}-poor, extremely
massive galaxies including brightest cluster galaxies (BCGs).

\subsection{Black hole masses}\label{Sec2.1}

Among our selected sample of 67 galaxies, 64 have SMBH masses
($M_{\rm BH}$) determined from stellar, gas or maser kinematic
measurements. For the remaining 3/67 galaxies (NGC 4051, NGC 4593 and
NGC 5273), $M_{\rm BH}$ were based on reverberation mapping
\citep{1972ApJ...171..467B,1993PASP..105..247P}. Of the 67 galaxies,
62 harbor black holes that are supermassive (i.e.,
$M_{\rm BH} \ga 10^{6} M_{\sun}$); see Appendix~\ref{AppD}. While the
remaining 5 sample galaxies have
$2\times 10^{4}M_{\sun} \la (M_{\rm BH}) \la 3\times 10^{5} M_{\sun}$,
we also refer to these black holes as SMBHs. The sample includes 25
galaxies with $M_{\rm BH}$ upper limits, comprised of 23 late-type
galaxies and 2 early-type galaxies. We discuss how the inclusion of
upper limits affects the BH scaling relations in Section~\ref{SecT3}.
The uncertainties on $M_{\rm BH}$ were taken from
\citet{2016ApJ...831..134V}.

\begin{center}
\begin{table*} 
\setlength{\tabcolsep}{0.08415812628109948in}
\begin{sideways}
\begin {minipage}{210mm}
  \caption{$M_{\rm BH} - \mathcal C_{\rm FUV,tot}$ scaling relations
    for early- and late-type galaxies.}
\label{Table1}
\begin{tabular}{@{}lclclclclclc|c|c@{}}
  \hline
  \hline
&&& \multicolumn{4}{c}{$Y$=$\beta$$X$+$\alpha$, $X$=$\mathcal C_{\rm
  FUV,tot}$$-$6.5, $Y$=log M$_{\rm
                                                       BH}$}\\
&&&\multicolumn{4}{c}{\bf early-type galaxies, Fig.~\ref{Fig2}, left}\\
Regression method&~~~~~~~~~~~~~~~~~~~~$Y|X$&~~~~~~~&~~~~~~~~~~~$X|Y$ &~~&~~~~
                                                Bisector&&$r$&$\Delta
                                                               ~\rm
                                                               [dex]$&$\epsilon$
                                                                        [dex]&$N$\\
  &~~~~~~~~~$\alpha$&$\beta$ &$\alpha$&$\beta$ &$\alpha$&$\beta$&&\\
  (1)&(2)&(3)&(4)&(5)&(6)&(7)&(8)&(9)&(10)&(11)\\
  \multicolumn{1}{c}{} \\              
  \hline      
 BCES & 8.12 $\pm$ 0.18  & 1.28 $\pm$ 0.40 & 8.30  $\pm$ 0.31 & 2.65
                                                                $\pm$
                                                                0.82 &
                                                                       {\bfseries   8.18}  {\bf$\pm$ 0.20}  &   {\bf1.75 $\pm$   0.41} & 0.61 & 0.86 & --- &  18\\  
 {\sc linmix\_err} &8.15 $\pm$ 0.22  & 1.25 $\pm$  0.44 & 8.34  $\pm$
                                                          0.23  & 2.47
                                                                  $\pm$
                                                                  0.84
& --- & 1.71 $\pm$ 0.51 & ---& ---& $0.69\pm0.20 $ &  18\\
OLS & 8.10 $\pm$ 0.17  & 1.04 $\pm$  0.30 & 8.32  $\pm$ 0.27  & 2.59
                                                                $\pm$
                                                                0.77 &
                                                                       8.17
                                                                       $\pm$
                                                                       0.18
                                           &  1.57  $\pm$  0.27 & ---
                                                                     &
                                                                       0.87 & ---&  18\\
&&&&& Symmetric \\
MCMC& ---&--- & --- & ---&  8.27 $\pm$  0.25    &  2.06  $\pm$  0.63 & ---& 0.85 & ---&  18\\

 \hline    
&&&\multicolumn{4}{c}{$Y$=$\beta$$X$+$\alpha$, $X$=$\mathcal C_{\rm
   FUV,tot}-$3.3, $Y$=log M$_{\rm
                                                       BH}$}\\
&&&\multicolumn{4}{c}{{\bf late-type galaxies,  Fig.~\ref{Fig2}, left}}\\

 &~~~~~~~~~~~~~~~~~~~~$Y|X$&~~~~~~~~&~~~~~~~~~~~$X|Y$ &~~&~~~~
                                                Bisector&&$r$&$\Delta
                                                               ~\rm
                                                               [dex]$&$\epsilon$
                                                                        [dex]&$N$\\
  
 &~~~~~~~~~$\alpha$&$\beta$ &$\alpha$&$\beta$ &$\alpha$&$\beta$&&\\         
  \hline      
 BCES & 7.01 $\pm$  0.13  &  0.82 $\pm$ 0.20 & 7.06  $\pm$ 0.19 &
                                                                  1.35
                                                                  $\pm$
                                                                  0.36
  &  {\bf7.03  $\pm$  0.14 }&   {\bf1.03 $\pm$  0.13} & 0.60 & 0.87 &
                                                                      --- &  45\\ 
 {\sc linmix\_err} &7.07 $\pm$ 0.13  & 0.87 $\pm$  0.26 & 7.09  $\pm$
                                                          0.13 & 1.59
                                                                 $\pm$
                                                                 0.38
&  --- & 1.17 $\pm$ 0.27 & --- & ---& $0.69\pm0.22$&  45\\
OLS &  6.97 $\pm$  0.12  &  0.62 $\pm$  0.16 &  7.10   $\pm$ 0.22 &  1.82 $\pm$ 0.36 & 7.02  $\pm$ 0.14 &  1.05  $\pm$ 0.12 & ---& 0.87 & --- & 45\\
&&&&& Symmetric\\
MCMC& ---&--- & --- & ---&  6.95  $\pm$ 0.11    &    0.87  $\pm$  0.17 & --- & 0.81 & ---&  45\\
 \hline
&&& \multicolumn{4}{c}{$Y$=$\beta$$X$+$\alpha$, $X$=$\mathcal C_{\rm
    NUV,tot}-$5.0, $Y$=log M$_{\rm
                                                       BH}$}\\
&&&\multicolumn{4}{c}{\bf early-type galaxies, Fig.~\ref{Fig2}, right}\\
&~~~~~~~~~~~~~~~~~~~~$Y|X$&~~~~~~~&~~~~~~~~~~~$X|Y$ &~~&~~~~
                                                Bisector&&$r$&$\Delta
                                                               ~\rm
                                                                [dex]$&$\epsilon$
                                                                        [dex]&$N$\\
 &~~~~~~~~~$\alpha$&$\beta$ &$\alpha$&$\beta$ &$\alpha$&$\beta$&&\\
  \hline      
  BCES & 8.07 $\pm$ 0.15  & 1.57 $\pm$ 0.36 &   8.21 $\pm$ 0.29 &  2.74 $\pm$  0.99 &{\bf 8.12  $\pm$  0.17}  & {\bf 1.95 $\pm$  0.28} & 0.70 & 0.72 & --- &  19\\           
 {\sc linmix\_err} &8.10 $\pm$ 0.20  & 1.66 $\pm$  0.55 & 8.19  $\pm$
                                                          0.20  & 2.63
                                                                  $\pm$
                                                                  0.79
&  --- & 2.04 $\pm$ 0.60& --- & --- &  $0.66 \pm0.18$ &  19\\
OLS &  8.06 $\pm$ 0.15  & 1.27 $\pm$  0.28 & 8.18  $\pm$ 0.24  & 2.65 $\pm$ 0.68 &   8.11  $\pm$ 0.16  &   1.76  $\pm$ 0.20 & --- & 0.68 &---&  19\\

&&&&& Symmetric\\
MCMC& ---&--- & --- & ---&  8.06   $\pm$  0.26  &    2.33  $\pm$  0.74   
                                                   & --- & 0.82 &
                                                                   ---&  19\\
\hline
&&& \multicolumn{4}{c}{$Y$=$\beta$$X$+$\alpha$, $X$=$\mathcal C_{\rm
    NUV,tot}-$2.7, $Y$=log M$_{\rm
                                                       BH}$}\\
&&&\multicolumn{4}{c}{{\bf late-type galaxies,  Fig.~\ref{Fig2}, right}}\\
&~~~~~~~~~~~~~~~~~~~~$Y|X$&~~~~~~~&~~~~~~~~~~~$X|Y$ &~~&~~~~
                                                Bisector&&$r$&$\Delta
                                                               ~\rm
                                                               [dex]$&$\epsilon
                                                                     $
                                                                     [dex]&$N$\\
 &~~~~~~~~~$\alpha$&$\beta$ &$\alpha$&$\beta$ &$\alpha$&$\beta$&&\\   
  \hline      
 BCES & 6.96 $\pm$ 0.15  & 1.24 $\pm$ 0.35& 6.96  $\pm$ 0.17 &  1.61 $\pm$  0.33 &{\bf 6.96  $\pm$  0.15}  & {\bf  1.38 $\pm$  0.23} & 0.65 & 0.86 & --- &  45\\                
 {\sc linmix\_err} &7.02 $\pm$ 0.12  & 1.08 $\pm$  0.30 & 8.32  $\pm$ 0.27  & 1.86 $\pm$ 0.41 &  --- & ---  & --- & ---& $0.66\pm0.18 $&  45\\
OLS &  6.94 $\pm$ 0.12  &  0.82 $\pm$  0.17 & 6.94 $\pm$ 0.18  & 1.94
                                                                 $\pm$
                                                                 0.35
                 &   6.94  $\pm$ 0.13  &    1.24  $\pm$ 0.13 & --- &
                                                                      0.83
                                                                        &---&
                                                                              45\\
&&&&& Symmetric \\
MCMC& ---&--- & --- & ---&  6.90  $\pm$ 0.13  &  1.35  $\pm$   0.26 & --- & 0.85 & ---&  45\\
\hline
\hline    
\end{tabular} 
Notes.--- Correlation between SMBH mass ($M_{\rm BH}$) and total UV
$-$ [3.6] color ($\mathcal C_{\rm UV,tot})$ for our early-
and late-type galaxies with directly measured $M_{\rm BH}$.  $\mathcal C_{\rm FUV,tot}=$ $m_{\rm FUV}$ $-$
$m_{\rm 3.6~ \mu \rm m}$, $\mathcal C_{\rm NUV,tot}=$ $m_{\rm NUV}$
$-$ $m_{\rm 3.6~ \mu \rm m}$, where $m_{\rm FUV}$, $m_{\rm NUV}$ and
$m_{3.6~ {\micron}}$ are the FUV, NUV and $m_{3.6~ {\micron}}$ total
apparent magnitudes of the galaxies (Table~\ref{Table6}). Col. (1)
regression method. Cols. (2) and (3) are the intercepts ($\alpha$) and
slopes ($\beta$) from the ($Y|X$) regressions. Cols. (4) and (5) are
$\alpha$ and $\beta$ obtained from the ($Y|X$) regression fits, while
cols. (6) and (7) show $\alpha$ and $\beta$ from the symmetrical
bisector regressions. The preferred slopes and intercepts are
highlighted in bold. Cols. (8) Pearson correlation coefficient
($r$). Cols. (9) the root-mean-square (rms) scatter around the fitted
{\sc bces} bisector relation in the \mbox {log $M_{\rm BH}$} direction
($\Delta$).  Col. (10) intrinsic scatter ($\epsilon$), see the text
for details. Col. (11) number of data points contributing to the
regression fits.
\end {minipage}
\end{sideways}
\end{table*}
\end{center}

\subsection{ UV and 3.6 $\micron$ apparent
  asymptotic magnitudes}\label{Sec2.2}

\citet{2015ApJ...800L..19B,2018ApJS..234...18B} used the 3.6 $\micron$
surface brightness profiles from \citet{2015ApJS..219....3M} to
determine the 3.6 $\micron$ magnitudes for their galaxies. They
extracted the UV surface brightness profiles of the galaxies following
the prescription of \citet{2015ApJS..219....3M}. Briefly, the sky
levels in galaxy images were first determined before the images were
masked to avoid bright foreground and background objects \citep[their
section 3.1]{2018ApJS..234...18B}. The FUV, NUV and 3.6 $\micron$
radial surface brightness profiles were then extracted using a series
of elliptical annuli with fixed ellipticity and position angle, after
excluding the innermost regions, $ R < 3\arcsec$,
\citep{2015ApJS..219....3M,2015ApJ...800L..19B,2018ApJS..234...18B}.
Each annulus has a width of 6$\arcsec$ and measurements were taken up
to 3 $\times$ the major axis of the galaxy $D25$ elliptical isophote.
To derive the FUV, NUV and 3.6 $\micron$ asymptotic magnitudes from
the corresponding surface brightness profiles,
\citet{2018ApJS..234...18B} extrapolated the growth curves. The UV,
and 3.6 $\micron$ data reach depths of $\sim$27 mag arcsec$^{-2}$ and
$\sim$26.5 mag arcsec$^{-2}$, respectively, allowing the growth curves
to flatten out for most galaxies.  The method yielded robust
asymptotic magnitudes with errors that are within the FUV and NUV zero-point
uncertainties of 0.05 and 0.03 AB mag \citep{2007ApJS..173..682M}.   

\section{Results}\label{SecT3}

\subsection{Calculating dust-corrected  total magnitudes}\label{Sec2.3}

The FUV, NUV and 3.6 $\micron$  (total) asymptotic magnitudes of our sample
were corrected for Galactic extinction by
\citet{2015ApJ...800L..19B,2018ApJS..234...18B} using the $E$($B-V$)
reddening values taken from \citet{1998ApJ...500..525S}. 
For the analyses in the paper, we define the total galaxy
luminosity as the sum of the luminosities of the bulge and disk,
excluding additional galaxy structural components such as bars and
rings. The bulk of the galaxies in our sample are multi-component
systems. Fractional luminosities of the individual structural
components are needed to obtain accurate bulge and disk magnitudes
from the total asymptotic magnitudes. We note that, from here on, the term
`bulge' is used when referring to both the spheroids of elliptical
galaxies and the bulges of disk galaxies. Fortunately,
\citet{2015ApJS..219....4S} performed detailed 2D multi-component
decompositions of \mbox{3.6 $\micron$} images of all our galaxies into
nuclear sources, bulges, disks and bars, and they presented fractional
bulge and disk luminosities which we use to calculate the 3.6
$\micron$ bulge and disk magnitudes. For two galaxies (NGC 3368 and NGC 4258) with
poor fits in \citet{2015ApJS..219....4S}, the 3.6 $\micron$ fractional
luminosities were taken from \citet{2016ApJS..222...10S}.  The
bulge-to-disk flux ratio ($B/D$) of a galaxy depends on both the
observed wavelength and galaxy morphological type (e.g.,
\citealt{2004A&A...415...63M, 2006A&A...456..941M,
  2008MNRAS.388.1708G, 2016MNRAS.460.3458K}). The UV $B/T$ and $D/T$ ratios
tabulated in Appendix~\ref{AppD} were derived using
the 3.6 $\micron$ $B/D$ ration together with the (extrapolation of the)
\mbox{$B/D$-passband-morphological type} diagram from \citet[his
Fig.~5]{2004A&A...415...63M}.  

For the disk galaxies, we additionally applied the inclination ($i$) dependent,
internal dust attenuation corrections from \citet[their equations 1
and 2 and Table 1]{2008ApJ...678L.101D} to determine dust-corrected
bulge and disk magnitudes.  Since \citet{2008ApJ...678L.101D} did not
provide 3.6 $\micron$ dust corrections, we rely on the prescription
for their reddest bandpass (i.e., $K$)  to correct the 3.6 $\micron$
magnitudes. These corrections are given by:

 \begin{equation}
m^{\rm corr}_{\rm bulge,UV} =m^{\rm obs}_{\rm bulge,UV}
-1.10-0.95[1-{\rm cos} (i)]^{2.18}, 
\label{Eq1}
 \end{equation}
\begin{equation}
m^{\rm corr}_{\rm disk,UV} =m^{\rm obs}_{\rm disk,UV}
-0.45-2.31[1-{\rm cos} (i)]^{3.42}, 
\label{Eq2}
 \end{equation}

 \begin{equation}
m^{\rm corr}_{\rm bulge,3.6} =m^{\rm obs}_{\rm bulge,3.6}
-0.11-0.79[1-{\rm cos} (i)]^{2.77}, 
\label{Eq3}
 \end{equation}
\begin{equation}
m^{\rm corr}_{\rm disk,3.6} =m^{\rm obs}_{\rm disk,3.6}
-0.04-0.46[1-{\rm cos} (i)]^{4.23},
\label{Eq4}
 \end{equation}
 where cos($i$) = $b/a$, i.e., the minor-to-major axis ratios, which
 were computed
 for our galaxies from the minor and major galaxy diameters given in NED.

 For each sample galaxy, we derived the dust-corrected total galaxy
 magnitude as

\begin{equation}
m^{\rm corr}_{\rm total} =-2.5 {\rm log} (10^{-0.4m^{\rm corr}_{\rm disk} }+10^{-0.4m^{\rm corr}_{\rm bulge}})
\label{Eq5}
 \end{equation}

\begin{figure*}
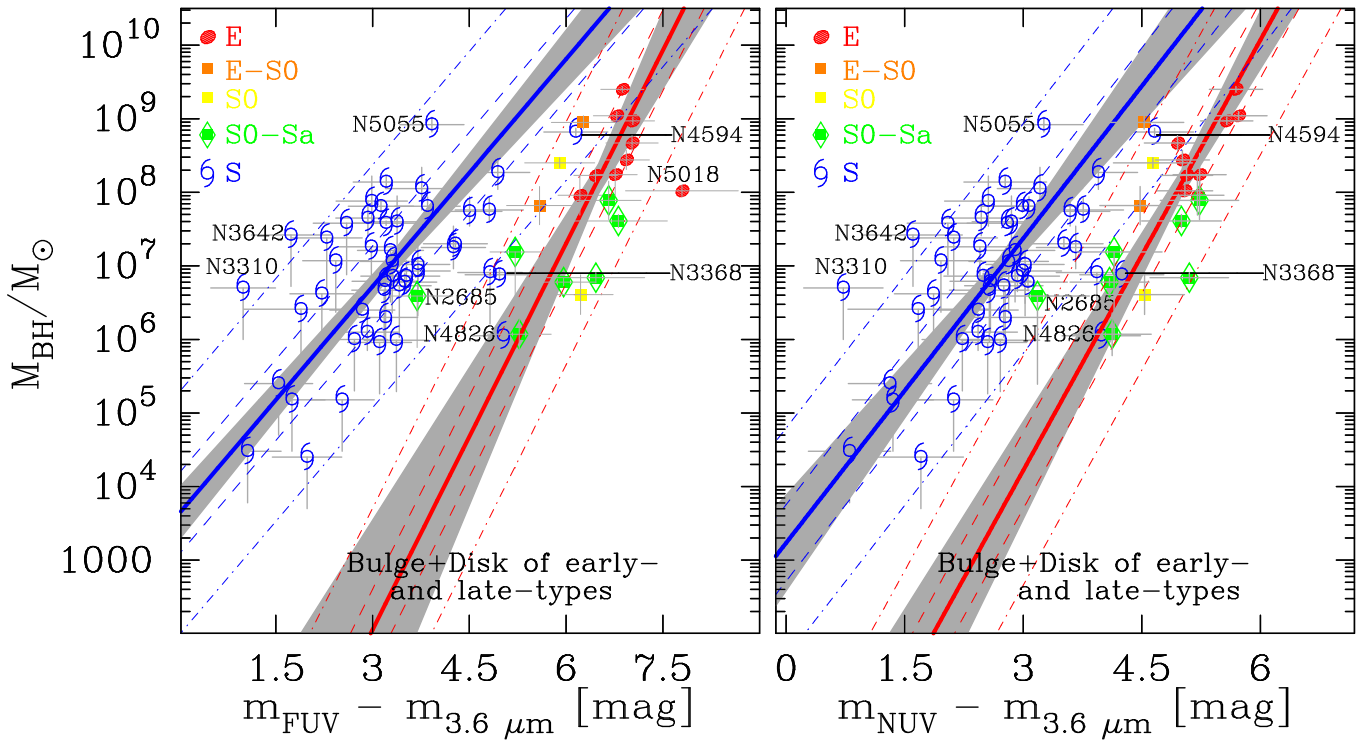

\hspace{-0.1541cm}
\includegraphics[angle=270,scale=0.728]{Total_mag_fuvNF.ps}
\includegraphics[angle=270,scale=0.728]{Total_mag_nuvNF.ps}
\label{Fig2} 
\caption{Correlations of directly measured SMBH masses ($M_{\rm BH}$)
  with the total (i.e., bulge+disk) UV $-$ 3.6 $\micron$ colors
  ($\mathcal{C_{\rm UV,tot}}$) of their host galaxies.
  $M_{\rm BH}- \mathcal{C_{\rm FUV, tot}}$ relations (left) and
  $M_{\rm BH}- \mathcal{C_{\rm NUV, tot}}$ relations (right) for
  early- and late-type galaxies. Our early-type morphological bin
  comprises E, E-S0, S0, and S0-a. Late-type galaxies (i.e., Sa, Sb,
  Sc, Sd, Sm and Irr) are plotted in blue.  Early- and late-type
  galaxies, which are fit separately, define two distinct red and blue
  sequences with significantly different slopes (see
  Table~\ref{Table1}).  The solid red and solid blue lines are the
  symmetric {\sc bces} bisector fits to our early- and late-type data,
  respectively, the shaded regions cover the associated $1\sigma$
  uncertainties on these fits (Table~\ref{Table1}). The dashed and
  dash-dotted lines delineate the one and three times the intrinsic
  scatter, respectively. Errors on $M_{\rm BH}$ and
  $\mathcal{C_{\rm UV,tot}}$ are shown, and for galaxies with black hole upper
  limits we  only show the lower uncertainty on $M_{\rm BH}$ (see the text
  for further detail).  }
 \end{figure*}

\subsection{Uncertainties on magnitudes}\label{Sec2.4}

Analyses of the (black hole)-(color) relations can be affected by
uncertainties on the bulge, disk and total magnitudes
(Section~\ref{Sec2.3} and Appendix~\ref{AppD}) that are dominated by
systematic errors. We account for four potential sources of
systematic uncertainty. There are uncertainties on the UV and 3.6
$\micron$ asymptotic magnitudes introduced by dust contamination,
imperfect sky background determination and poor masking of bright
sources \citep{2018ApJS..234...18B}. For the disk galaxies, there are 
uncertainties due to our dust correction. There is also a need to account
for uncertainties on the 3.6 $\micron$ $B/T$ and $D/T$ ratios 
\citep{2015ApJS..219....4S} used to covert the asymptotic magnitudes
into disk, bulge and total magnitudes. For the UV magnitudes, an
additional source of systematic uncertainty is the derivation of the
$B/T$ and $D/T$ ratios (Section~\ref{Sec2.3}). The total uncertainties
associated with the UV and 3.6 $\micron$ magnitudes were calculated by
adding the individual contributions in the error budget in
quadrature. Appendix~\ref{AppD} lists total magnitudes and associated
errors for our sample galaxies. The quoted magnitudes are in the AB
system, unless noted otherwise.

\subsection{ Regression Analysis}\label{Sec3}

Inherent differences in the employed statistical methods may
systematically affect the derived BH scaling relations and it is
therefore vital to explore this issue. We performed linear regression
fits to
\mbox{$M_{\rm BH}-\mathcal{C}(=m^{\rm corr}_{\rm UV} - m^{\rm
    corr}_{\rm 3.6} $)}, $M_{\rm BH}-L_{\rm 3.6}$ and
$M_{\rm BH}-\sigma$ data using three traditional regression
techniques: the Bivariate Correlated Errors and intrinsic Scatter
({\sc bces}) code (\citealt{1996ApJ...470..706A}), the Bayesian linear
regression routine ({\sc linmix\_err}, \citealt{2007ApJ...665.1489K})
and Ordinary Least-Squares ({\sc ols}) code
(\citealt{1992ApJ...397...55F}), Figs.~\ref{Fig2} and \ref{Fig5}.  The
{\sc bces} routine \citep{1996ApJ...470..706A} was implemented in our
work using the python module by \citet{2012Sci...338.1445N}.  While
both the {\sc bces} and \mbox{\sc linmix\_err} methods take into
account the intrinsic scatter and uncertainties in both $M_{\rm BH}$
and the host galaxy properties, only the latter can deal with
`censored' data, e.g., black hole mass upper limits. The {\sc ols}
routine does not account for errors, we use this method to provide
$M_{\rm BH}-\mathcal{C}$ relations independent of measurement errors.

In an effort to assess the robustness of the above linear regression fits,
we have additionally performed symmetric, MCMC Bayesian linear
regression fits for our (\mbox{$M_{\rm BH}, \mathcal{C}$}) data, which
account for black hole mass upper limits (see Appendix
\ref{AppA}).

Fitting the $Y=\beta X + \alpha$ line with the {\sc bces}, {\sc
  linmix\_err} and {\sc ols} codes, we present the results from the
($Y|X$), ($X|Y$) and bisector regression analyses
(Tables~\ref{Table1}, \ref{Table4} and \ref{Table5}). We also present
the results from our symmetric,  MCMC Bayesian regressions (Table~\ref{Table1}). The
($Y|X$), ($X|Y$) regressions minimize the residuals around the fitted
regression lines in the $Y$ and $X$ directions, respectively. The
symmetrical bisector line bisects the ($Y|X$) and ($X|Y$) lines. The
{\sc linmix\_err} code does not return bisector regressions, thus we computed the
the slope of line that bisects the ($Y|X$) and ($X|Y$)
lines. Throughout this work, we focus on the relations from the
symmetrical {\sc bces} bisector regressions (Figs.~\ref{Fig2}
and \ref{Fig5}).


\begin{center}
\begin{table*}  
\setlength{\tabcolsep}{0.032109948in}
\begin {minipage}{180mm}
\caption{$M_{\rm BH}-\sigma$  relation}
\label{Table4}
\begin{tabular}{@{}lclclclclclc|c|c@{}}
  \hline
  \hline
  &\multicolumn{8}{c}{{$Y$=$\beta$$X$+$\alpha$, $X$=log
      $\sigma$$-$2.2, $Y$=log M$_{\rm
                                                       BH}$ ({\bf
      early-type, Fig.~\ref{Fig5}, left})}}\\ 
Regression method&~~~~~~~~~~~~~~~~~~~~$Y|X$&~~~~~~~&$X|Y$ &~~&~~~~
                                                Bisector&&$r$&$\Delta
                                                               ~\rm
                                                               [dex]$&$\epsilon$
                                                                     [dex]&$N$\\
  &~~~~~~~~~$\alpha$&$\beta$ &$\alpha$&$\beta$ &$\alpha$&$\beta$&&\\
  (1)&(2)&(3)&(4)&(5)&(6)&(7)&(8)&(9)&(10)&(11)\\
  \multicolumn{1}{c}{} \\              
  \hline      
  \hline      
  {\sc bces}&  7.84 $\pm$ 0.14  & 4.33 $\pm$ 0.80 & 7.81 $\pm$ 0.23 &  7.10$\pm$ 1.58 &{\bf  7.84  $\pm$  0.16}  &{\bf   5.42 $\pm$   0.90}& 0.72 & 0.67 & $0.70\pm0.16$ &  20\\ 
   \hline   
  &\multicolumn{8}{c}{{$Y$=$\beta$$X$+$\alpha$, $X$=log
    $\sigma$$-$2.0, $Y$=log M$_{\rm
                                                       BH}$
({\bf late-type, Fig.~\ref{Fig5}, left})}}\\ &~~~~~~~~~~~~~~~~~~~~$Y|X$&~~~~~~~&$X|Y$ &~~&~~~~
                                                Bisector&&$r$&$\Delta
                                                               ~\rm
                                                               [dex]$&$\epsilon$
                                                                                 [dex]&$N$\\
&~~~~~~~~~$\alpha$&$\beta$ &$\alpha$&$\beta$ &$\alpha$&$\beta$&&\\
        
  {\sc bces} & 6.98 $\pm$ 0.10  & 4.00 $\pm$  0.53 & 7.00  $\pm$ 0.11  & 5.00 $\pm$ 0.99 &{\bf  6.98  $\pm$  0.10} &{\bf 4.49 $\pm$   0.48 }& 0.75 & 0.70 &  $0.63\pm0.10$ &  45\\
   \hline   
  &\multicolumn{8}{c}{{$Y$=$\beta$$X$+$\alpha$, $X$=log $\sigma$$-$2.0,
                                                $Y$=log M$_{\rm
                                                       BH}$
  ({\bf all galaxies, Fig.~\ref{Fig5}, left})}}\\ &~~~~~~~~~~~~~~~~~~~~$Y|X$&~~~~~~~&$X|Y$ &~~&~~~~
                                                Bisector&&$r$&$\Delta
                                                               ~\rm
                                                               [dex]$&$\epsilon$
                                                                                 [dex]&$N$\\
  &~~~~~~~~~$\alpha$&$\beta$ &$\alpha$&$\beta$ &$\alpha$&$\beta$&&\\
  {\sc bces} & 6.99 $\pm$ 0.09 & 4.23 $\pm$  0.40 & 6.92 $\pm$ 0.10  &  5.39 $\pm$ 0.72 &{\bf  6.96  $\pm$ 0.09}  &{\bf  4.65  $\pm$ 0.35 }& 0.78 & 0.68 &$0.62\pm0.07$&  65\\
  \hline

\end{tabular} 
Notes.---  Similar to Table~\ref{Table1}, but here showing linear
regression analyses of the correlation between SMBH mass ($M_{\rm
                                                       BH}$) and
                                                     velocity
                                                     dispersion ($\sigma$). 
\end {minipage}
\end{table*}
\end{center}

\begin{center}
\begin{table*} 
\setlength{\tabcolsep}{0.02358109948in}
\begin {minipage}{180mm}
\caption{$M_{\rm BH}-L_{\rm 3.6,tot}$  relation}
\label{Table5}
\begin{tabular}{@{}lclclclclclc|c|c@{}}
  \hline
  \hline
  &\multicolumn{8}{c}{{$Y$=$\beta$$X$+$\alpha$, $X$=$M_{\rm 3.6,
                                                 tot}$$+$18.5, $Y$=log M$_{\rm
                                                       BH}$} ({\bf
                                                               all
                                                               galaxies,
                                                               Fig.~\ref{Fig5},
                                                               right})}\\
 Regression method&~~~~~~~~~~~~~~~~~~~~$Y|X$&~~~~~~~&$X|Y$ &~~&~~~~
                                                Bisector&&$r$&$\Delta
                                                               ~\rm
                                                               [dex]$&$\epsilon$
                                                             [dex]&$N$\\
   &~~~~~~~~~$\alpha$&$\beta$ &$\alpha$&$\beta$ &$\alpha$&$\beta$&&\\
  (1)&(2)&(3)&(4)&(5)&(6)&(7)&(8)&(9)&(10)&(11)\\
  \multicolumn{1}{c}{} \\              
  \hline      
  \hline      
  
  {\sc bces} & 6.96 $\pm$ 0.11 & -0.37 $\pm$  0.06 & 6.80 $\pm$0.17  & -0.65 $\pm$ 0.12 &{\bf   6.88  $\pm$ 0.12 } &{\bf -0.49 $\pm$ 0.06}& -0.66 & 0.84 &$0.69\pm0.09$&  67\\
  \hline

\end{tabular} 
Notes.--- Similar to Table~\ref{Table1}, but here showing linear
regression analyses of the correlation between the SMBH masses ($M_{\rm
                                                       BH}$) and 3.6 {\micron}
                                                     total absolute
                                                     magnitude of the
                                                     galaxies ($M_{\rm
                                                       3.6, tot}$),
                                                     see the text for details. 
\end {minipage}
\end{table*}
\end{center}

\subsection{The $M_{\rm BH}$ $-$  $\mathcal{C_{\rm UV,tot}}$ relations}\label{Sec3.1}

In this section, we investigate the correlations between $M_{\rm BH}$
and total  (i.e., bulge+disk) colors $\mathcal{C_{\rm FUV,tot}}$
(=$m^{\rm corr}_{\rm FUV,total} - m^{\rm corr}_{\rm 3.6,total} $) and
$\mathcal{C_{\rm NUV,tot}}$ 
(=$m^{\rm corr}_{\rm NUV,total} - m^{\rm corr}_{\rm 3.6,total} $) for
our sample of 67 galaxies comprised of 20 early-type galaxies and 47
late-type galaxies (Appendix~\ref{AppD}).  In Fig.~\ref{Fig2}, we plot
these correlations (Table~\ref{Table6}), with data points color-coded
based on morphological type.  The regression analyses reveal that
early- and late-type galaxies define two distinct red and blue
sequences with markedly different slopes in the
$M_{\rm BH}- \mathcal{C}_{\rm UV,tot}$ diagrams, regardless of the
applied regression methods (Table~\ref{Table1}).  We find that the
slopes for early- and late-type galaxies are different at
$\sim 2 \sigma$ level (Appendix \ref{AppA}) and the significance levels
for rejecting the null hypothesis of  these two morphological types having
the same slope are 1.7\%$-$6.7\%. We note in passing that this trend of different slopes for early- 
and late-type galaxies holds for the correlations between $M_{\rm BH}$
and the bulge colors of the two Hubble types 
($\mathcal{C_{\rm UV,bulge}}$, Dullo et al. in prep). For both early- and late-type
galaxies, the $\mathcal{C}_{\rm FUV,tot}$ and $M_{\rm BH}$ data
correlate with a Pearson correlation coefficient $r\sim 0.60$
(Table~\ref{Table1}). The \mbox{$\mathcal{C}_{\rm NUV,tot}$,
  $M_{\rm BH}$} data have Pearson correlation coefficients
$r \sim 0.70 $ and 0.65 for early- and late-type galaxies,
respectively. In the $M_{\rm BH}- \mathcal{C}$ regression
analyses, we have excluded 1 early-type galaxy (\mbox{NGC 2685}) and 2
late-type galaxies (\mbox{NGC 3310} and \mbox{NGC 4826}) which offset from the
relations by more than three times the intrinsic scatter
(Fig.~\ref{Fig2}). The early-type galaxy \mbox{NGC 5018} was also excluded
from the $M_{\rm BH}- \mathcal{C}_{\rm FUV}$ relations, resulting in
18 early-type galaxies. Interestingly, all these four outliers are have peculiar
 characteristics  that are discussed in Appendix
\ref{outlier}.

Symmetrical {\sc bces} bisector fits to the
(\mbox{$M_{\rm BH}, \mathcal{C_{\rm FUV,tot}}$}) and
(\mbox{$M_{\rm BH},\mathcal{C_{\rm NUV,tot}}$}) data yield relations
for early-type galaxies with slopes of 1.75 $\pm $ 0.41 and 1.95
$\pm $ 0.28, respectively (Table~\ref{Table1}), such that
$M_{\rm BH} \propto (L_{\rm FUV,tot}/L_{\rm 3.6,tot})^{-4.38 \pm
  1.03}$ and
$M_{\rm BH} \propto (L_{\rm NUV,tot}/L_{\rm 3.6,tot})^{-4.88 \pm
  0.70}$.  This is to be compared with the derived {\sc bces} bisector
$M_{\rm BH}- \mathcal{C_{\rm FUV,tot}}$ and
$M_{\rm BH}- \mathcal{C_{\rm NUV,tot}}$ relations for the late-type
galaxies having shallower slopes of 1.03 $\pm $ 0.13 and 1.38 $\pm $
0.23, such that
$M_{\rm BH} \propto (L_{\rm FUV,tot}/L_{\rm 3.6,tot})^{-2.58 \pm
  0.33}$ and
$M_{\rm BH} \propto (L_{\rm NUV,tot}/L_{\rm 3.6,tot})^{-3.45 \pm
  0.58}$. 
  
With the assumption that  the UV-to-3.6 {\micron} luminosity ratio
($L_{\rm UV}/L_{3.6}$) is a proxy for specific star formation rate,
sSFR, (e.g., \citealt{2018ApJS..234...18B}), it implies that the
growth of black holes in late-type galaxies have a steeper dependence
on sSFR (i.e., \mbox{$M_{\rm BH} \propto \rm sSFR_{\rm FUV}^{-2.58}$})
than early-type galaxies
(\mbox{$M_{\rm BH} \propto \rm sSFR_{\rm FUV}^{-4.38}$}). 
 That is, at
  a given value of sSFR, late-type galaxies tend to have more massive
  BHs than early-type galaxies. The caveat of using FUV magnitudes as a proxy for the current star
formation rate is that a significant fraction of the FUV light in
$\sim$ 5\% of massive early-type galaxies may come from extreme
horizontal branch stars instead of young upper main sequence stars, a
phenomenon dubbed `UV upturn' (e.g.,
\citealt{1979ApJ...228...95C,1999ARA&A..37..603O,2011ApJS..195...22Y}). 
This is likely due to the rarity of very massive early-type galaxies in our sample.
 Nonetheless, 
we found that none of our early-type galaxies are UV upturns, when
using the criteria given by \citet[][their Table
1]{2011ApJS..195...22Y}.  Owing to a stronger sensitivity of the
FUV-band to the galaxy star formation rate (SFR) than the NUV-band,
the $M_{\rm BH}- \mathcal{C_{\rm FUV,tot}}$ relations are
systematically shallower than the corresponding
$M_{\rm BH}- \mathcal{C_{\rm NUV,tot}}$ relations, although they are
consistent with overlapping $1\sigma$ uncertainties
(Table~\ref{Table1}).  

The root-mean-square (rms) scatter ($\Delta $)
around the fitted {\sc bces} bisector relations in the \mbox {log
  $M_{\rm BH}$} direction are $\Delta_{\rm FUV,early} \sim 0.86$ dex,
$\Delta_{\rm FUV,late} \sim 0.87$ dex,
$\Delta_{\rm NUV,early} \sim 0.72$ dex and
$\Delta_{\rm NUV,late} \sim 0.86$ dex. We report intrinsic scatters
($\epsilon$) for our $M_{\rm BH}- \mathcal{C}_{\rm UV,tot}$ relations
as derived by \mbox{\sc linmix\_err} to be
$\epsilon_{\rm FUV,early} \sim 0.69 \pm 0.20$,
$\epsilon_{\rm FUV,late} \sim 0.69 \pm 0.22$,
$\epsilon_{\rm NUV,early} \sim 0.66 \pm 018$ and
$\epsilon_{\rm NUV,late} \sim 0.66 \pm 0.18$.

\begin{figure*}
\hspace{2.720cm}
\includegraphics[angle=270,scale=0.5]{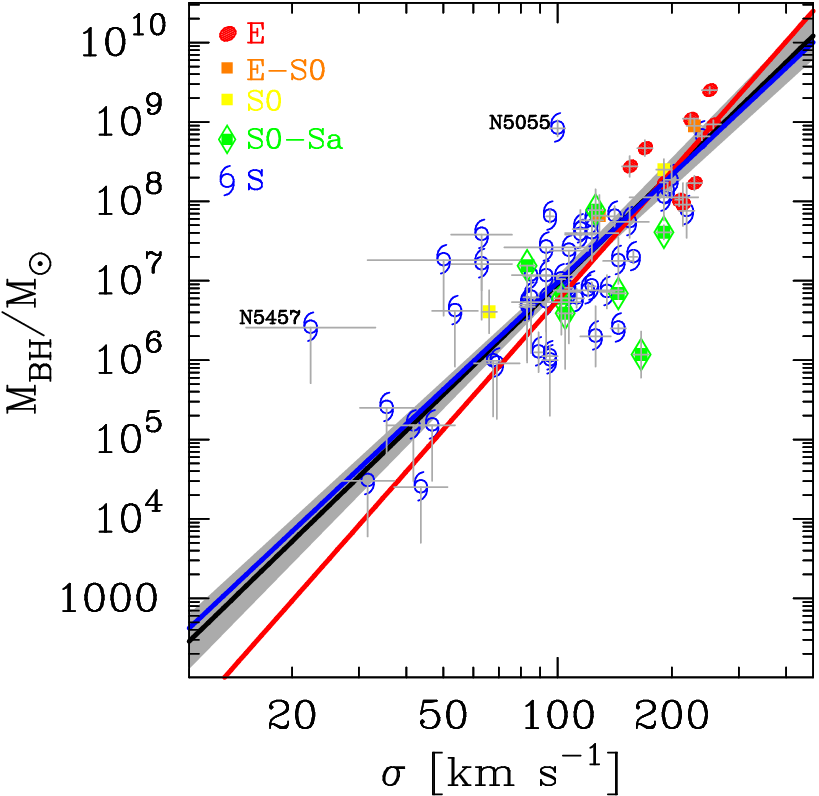}
\includegraphics[angle=270,scale=0.5]{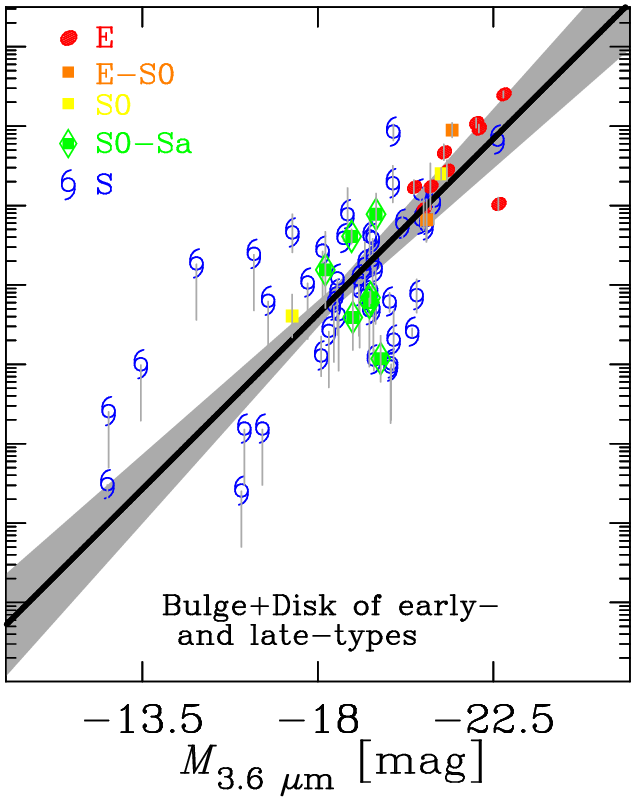}
\label{Fig5} 
\caption{Similar to Fig.~\ref{Fig2}, but shown here are correlations
  between $M_{\rm BH}$ and (left panel) velocity dispersion ($\sigma$,
  \citealt[his Table 2]{2016ApJ...831..134V}) and (right panel) total
  3.6 {\micron} absolute magnitude of our sample galaxies
  ($M_{3.6 ~ {\micron}}$). $M_{3.6 ~ {\micron}}$ are computed using
  the total 3.6 {\micron} apparent magnitudes ($m_{3.6 ~ {\micron}}$,
  Table~\ref{Table6}) and distances for the galaxies from \citet[his
  Table 2]{2016ApJ...831..134V}. We did not fit separate linear
  regressions to our early- and late-type ($M_{\rm BH}$, $M_{3.6 ~
    {\micron}}$) data or to the
  core-S\'ersic and S\'ersic ($M_{\rm BH}$, $M_{3.6 ~
    {\micron}}$) data, see the text for more 
  details. }  
 \end{figure*}

As noted previously, the \mbox{\sc linmix\_err} code and  MCMC
Bayesian analysis---which do account for the 24 galaxies (22 late-type galaxies and 2
early-type galaxies) with $M_{\rm BH}$ upper limits---yield
$M_{\rm BH}- \mathcal{C_{\rm UV,tot}}$ relations consistent with the
{\sc bces} regression analyses (Table~\ref{Table1}). Nonetheless, we
checked for a potential bias for the late-type galaxies due to the
inclusion of $M_{\rm BH}$ upper limits by rerunning the {\sc bces}
bisector regression analysis on the 23 (=45-22) late-type galaxies
with more securely measured $M_{\rm BH}$. We find that the slopes,
intercepts and $\Delta$ of the $M_{\rm BH}- \mathcal{C_{\rm UV,tot}}$
relations are only weakly influenced by the exclusion of $M_{\rm BH}$
upper limits\footnote{The strength of the $M_{\rm BH}- \mathcal{C_{\rm
      UV,tot}}$ correlations  decreases when the SMBH upper
  limits are excluded. The Pearson correlation
  coefficient for the ($\mathcal{C}_{\rm FUV,tot}$, $M_{\rm BH}$) blue
  sequence has reduced from $r \sim 0.60 $ to 0.34 due to the
  exclusion of $M_{\rm BH}$ upper limits, and for the
  ($\mathcal{C}_{\rm NUV,tot}$, $M_{\rm BH}$) blue sequence there is a
  decrease in $r$ from $ \sim 0.65 $ to 0.36.}. Including the upper
limits in the black hole scaling relations is useful
\citep{2009ApJ...698..198G}, given they also follow the $M-\sigma$
relation traced by galaxies with more securely measured $M_{\rm BH}$
(Fig.~\ref{Fig5}).  

 We can compare our work to that of  \citet[][their Figs.~1 and 2]{2017ApJ...844..170T} 
 who used star formation rates (SFRs) determined based on {\it IRAS} far-infrared
imaging and reported 
an inverse correlation between specific star formation rate (sSFR) and
SMBH mass for 91 galaxies with  measured black hole masses. 
Although they did not separate the galaxies into late- and early-types, 
their full sample seems to follow a single $M_{\rm BH}-\mbox{sSFR}$ relation 
with no break, contrary to our results. To explain this discrepancy, we  split the 
galaxies  in \citet{2017ApJ...844..170T} by morphology and find that their late-type
 galaxies (which constitute a third of the full sample) reside at 
the low mass end of their $M_{\rm BH}-\mbox{sSFR}$ relation and they span very small ranges in SMBH mass ($4\times 10^{6} \la M_{\rm BH}/M_{\sun} \la 10^{8}$) 
and in sSFR ($10^{-11} \la  \mbox{sSFR}/\mbox{yr}^{-1} \la 8\times10^{-9}$),  inadequate to establish
the blue $M_{\rm BH}- \mathcal{C_{\rm UV,tot}}$ sequence (Fig.~\ref{Fig2}). 
Furthermore, we note that  the FIR flux may underestimate  the actual SFR  for low-mass late-type galaxies as most of these galaxies' UV photons are unobscured by dust and thus not reprocessed to FIR wavelengths \citep{2015A&A...584A..87C}. We also find that  the \citet{2017ApJ...844..170T}  early-type $M_{\rm BH}-\mbox{sSFR}$ relation  has more scatter than that of ours.  We suspect this may be due to a variable  contamination of  the  {\it IRAS} FIR emissions in massive galaxies by  the AGN\footnote{While the IR emissions in massive galaxies have contributions from the dusty AGN, the NUV emissions arise from main-sequence turn-off stars and are less prone to contamination from the AGN \citep{2005ApJ...629L..29B}.}  which  heats the surrounding dust. While the dusty AGN in some massive galaxies might  led to an increase in the SFR values based on FIR luminosities    \citep[e.g.,][]{2015A&A...584A..87C,2017ApJ...840...21T},   star formation activities are likely  the dominant contributor to the SFR values reported by \citet{2017ApJ...844..170T}, AGN contamination being mainly responsible for the larger scatter observed  their $M_{\rm BH}-\mbox{sSFR}$ relation.

\begin{figure}
\includegraphics[angle=270,scale=0.60305]{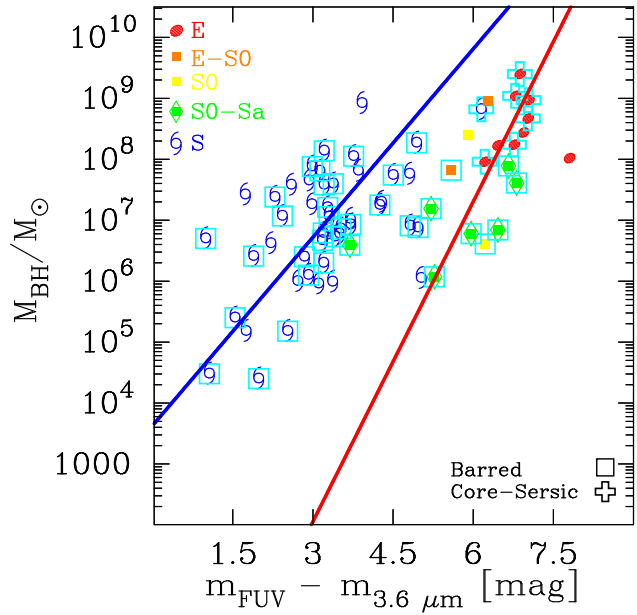}
\caption{Similar to Fig.~\ref{Fig2}(a), but here we also show host galaxy properties. Barred
  galaxies are enclosed in boxes. Seven core-S\'ersic galaxies (6 Es +
  1 S) with partially depleted cores published in the literature are 
  enclosed in crosses (see Section~\ref{Sec4.3.4}). 
}
\label{Fig6} 
 \end{figure}

\begin{figure}
\hspace{.03207088186340cm}
\vspace{-.049983088186340cm}
\includegraphics[angle=270,scale=0.63805]{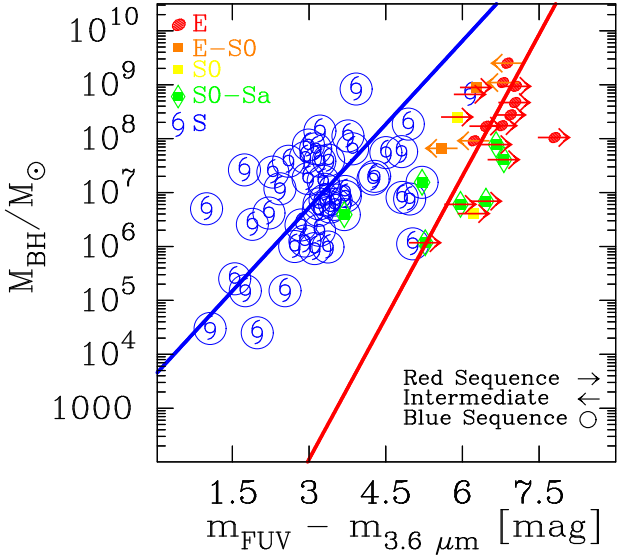}
\caption{Similar to Fig.~\ref{Fig2}, but here comparing the \mbox{$M_{\rm BH}- \mathcal{C}$} red and
  blue sequences and the color-color red/intermediate and blue
  sequences. The color-color relation red and intermediate sequence
  galaxies (\citealt[their section 4.4]{2018ApJS..234...18B}) are
  marked by rightward- and leftward-pointing arrows, respectively,
  blue sequence galaxies are enclosed in circles. }
\label{FigC1} 
 \end{figure}

\begin{figure}
\includegraphics[angle=270,scale=0.69]{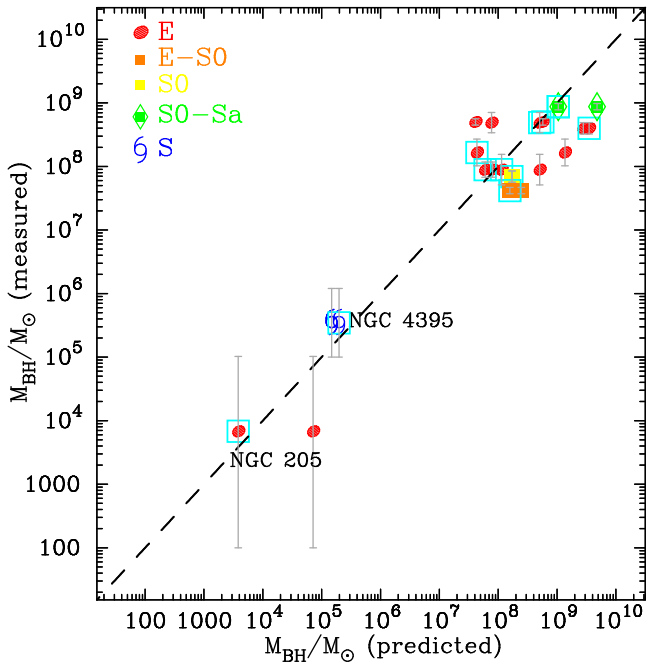}
\caption{Comparison between SMBH masses predicted using our
  $M_{\rm BH}$ $-$ $\mathcal{C_{\rm FUV,tot}}$ and $M_{\rm BH}$ $-$
  $\mathcal{C_{\rm NUV,tot}}$ relations (Table~\ref{Table1}) and those
  determined dynamically \citep{2016ApJ...831..134V,2019ApJ...872..104N} for a selected
  sample of 11 galaxies that are not in our sample (see Table
  \ref{Table6}). SMBH masses predicted based on the \mbox{$M_{\rm BH}$ $-$
  $\mathcal{C_{\rm NUV,tot}}$} relations are enclosed in boxes.}
\label{Fig7} 
 \end{figure}

\subsection{ $M_{\rm BH}-\sigma$ and $M_{\rm
    BH}-L$ relations}\label{Sec3.2}

In Fig.~\ref{Fig2} we have shown, for the first time to our knowledge,
a correlation between SMBH mass and total  (i.e., bulge+disk) colors for early- and late-type
galaxies. For such a correlation to be evident, it is important that the  color is determined using a wide 
wavelength baseline. In this section, we present correlations between
$M_{\rm BH} $ and velocity dispersion ($\sigma$) and 3.6 $\micron$
total luminosity ($L_{3.6,\rm tot}$) for the same galaxy sample to
allow a direct statistical comparison with the 
\mbox{$M_{\rm BH}- \mathcal{C}$} relations (Fig.~\ref{Fig5} and 
Table~\ref{Table4}). 
Assuming a $10\%$ uncertainty\footnote {After comparing the HyperLeda
  individual velocity dispersion measurements and mean homogenized
  values for 100 sample galaxies, we adopt a conservative upper limit
  uncertainty of 10\% on $\sigma$.} on $\sigma$, the {\sc bces}
bisector regression yields $M_{\rm BH}-\sigma$ relations with slopes
5.42 $\pm$ 0.90, 4.49 $\pm$ 0.48 and 4.65 $\pm$ 0.35 for the 
early-type galaxies, late-type galaxies and for the full
ensemble. These relations are in good agreement with each other within
their $1 \sigma$ uncertainties. The late-type galaxies \mbox{NGC 5055
  and NGC 5457}, being the most deviant outliers in the
$M_{\rm BH}-\sigma$ diagram, were excluded.  The unification of early-
and late-type galaxies in the $M_{\rm BH}-\sigma$ diagram is nothing
new (e.g., \citealt{2012MNRAS.419.2497B};
\citealt{2013ApJ...764..151G}; \citealt{2016ApJ...831..134V}; Dullo
et al.\ 2020, submitted). The
$M_{\rm BH}-\sigma$ relations (Table~\ref{Table4}) are consistent with
those from \citet{2009ApJ...698..198G}, \citet{2012MNRAS.419.2497B,
  2013ApJ...764..184M, 2013ARA&A..51..511K,
  2013ApJ...764..151G}\footnote{Our slopes for the full sample of 66
  galaxies are slightly (i.e., $\sim 1.4 \sigma$ $\approx$ $15\%$)
  shallower than that of \citet[slope $\sim$ 5.35 $\pm$
  0.23]{2016ApJ...831..134V}. This is because the \citet[his
  Fig.~1]{2016ApJ...831..134V} sample contains extremely bright
  galaxies with $\sigma \ga 270$ which tend to steepen the
  $M_{\rm BH}-\sigma$ relation (see Fig.~\ref{Fig1}). }.

To determine the \mbox{$M_{\rm BH}-L_{3.6,\rm tot}$} relation, we
converted the inclination and dust corrected 3.6 {\micron} total
apparent magnitudes into absolute magnitudes ($M_{3.6,\rm tot}$,
Appendix~\ref{AppD}) using distances given in
\citet{2016ApJ...831..134V}. The {\sc bces} bisector
\mbox{$M_{\rm BH}-L_{3.6,\rm tot}$} relation for the full sample of 67
galaxies was performed without accounting for the error on
$M_{3.6,\rm tot}$, yielding
$M_{\rm BH} \propto L_{\rm 3.6,tot}^{1.23 \pm 0.15}$ (Fig.~\ref{Fig5}
and Table~\ref{Table5}). \citet{2013ApJ...764..151G} reported two
distinct \mbox{$M_{\rm BH}-L$} relations for the bulges of S\'ersic
galaxies ($M_{\rm BH} \propto L_{\rm K_{\rm s},\rm bulge}^{2.73}$) and
core-S\'ersic galaxies
($M_{\rm BH} \propto L_{\rm K_{\rm s},\rm bulge}^{1.10}$). Since there
are only 7 galaxies we were able to identify as core-S\'ersic galaxies
(see Fig.~\ref{Fig6}), we refrain from separating the galaxies into
S\'ersic and core-S\'ersic galaxies in the
\mbox{$M_{\rm BH}-L_{3.6,\rm tot}$} diagram.

\section{Discussion}\label{Sec4}

\subsection{Comparison between the \mbox{$M_{\rm BH}- \mathcal{C}$},
  $M_{\rm BH}-\sigma$ and $M_{\rm BH}-L$ relations}

Due to the different number of galaxies used to define the
\mbox{$M_{\rm BH}- \mathcal{C}$}, $M_{\rm BH}-\sigma$ and
$M_{\rm BH}-L$ relations, a direct comparison of the strength and
scatter of the relations is difficult. Nonetheless, the correlation
between the color $ \mathcal{C_{\rm UV,tot}}$ and BH mass $M_{\rm BH}$
($r \sim 0.60 - 0.70$, see Table~\ref{Table1}) is slightly weaker than
that between the stellar velocity dispersion $\sigma$ and $M_{\rm BH}$
($r \sim 0.72 - 0.78$, Table~\ref{Table4}).  These two relations have
comparable intrinsic scatter (See Tables~\ref{Table1} and
\ref{Table4}).  In terms of scatter in the log $M_{\rm BH}$ direction,
the $M_{\rm BH}- \mathcal{C_{\rm UV,tot}}$ relations have typically
\mbox{$5\%-27\%$} more scatter (i.e., $\Delta \sim 0.72-0.87$ dex)
than the $M_{\rm BH}-\sigma$ relations ($\Delta \sim 0.68-0.70$
dex). The $M_{\rm BH}-\sigma$ relation appears to be the most
fundamental SMBH scaling relation. However, the $M_{\rm BH}-\sigma$
relations for late- and early-type galaxies are not notably offset from
each other. This contrasts with the formation
models of galaxies which predict the SMBH growth in the two Hubble types to be
completely different (see Section~\ref{Sec4.5}). Furthermore, in Dullo
et al. (2020, submitted) we showed that the $M_{\rm BH}-\sigma$
relation tends to underpredict the actual BH masses for the most
massive galaxies with $M_{*} \ga 10^{12} M_{\sun}$.  In contrast, the
$M_{\rm BH}- \mathcal{C_{\rm UV,tot}}$ relations are in accordance
with models of galaxy formation (Section~\ref{Sec4.5}).

As for the comparison between the \mbox{$M_{\rm BH}- \mathcal{C}$} and
$M_{\rm BH}-L$ relations, these two relations display similar level of
strength and vertical scatter (Table~\ref{Table5},
$r \sim 0.60 - 0.70$ and $\Delta \sim 0.72-0.87$ dex for
\mbox{$ M_{\rm BH}-\mathcal{C_{\rm UV,tot}}$} and $r \sim -0.66$, and
$\Delta \sim 0.84$ dex for \mbox{$M_{\rm BH}-L_{3.6,\rm tot}$}). In passing, we note that  the existence of 
the $M_{\rm BH}-L$  relation coupled with the red sequence and  blue cloud traced by early-and late-type 
galaxies in the color-magnitude 
diagram does not necessitate  a correlation  between the black 
hole mass and galaxy color.

For comparison, our $M_{\rm BH}- \mathcal{C_{\rm UV,tot}}$ relation
for the 47 late-type galaxies is stronger ($r \sim 0.60 -0.65$, and
$\Delta \sim 0.87$ dex) than the $ M_{\rm BH}-M_{*,\rm tot}$ relation
based on 3.6 $\micron$ data by \citet{2018ApJ...869..113D} for their
sample of 40 late-type galaxies ($r \sim 0.47$ and $\Delta \sim 0.79$
dex).

\subsection{ Core-S\'ersic versus S\'ersic}\label{Sec4.3.4}

As noted in the Introduction, SMBH scaling relations may differ depending
on the galaxy core structure (i.e., Core-S\'ersic versus S\'ersic
type).  We identify seven core-S\'ersic galaxies (6 Es + 1 S) in our
sample with partially depleted cores published in the literature:  
NGC~1052 \citep{2007ApJ...662..808L}, NGC~3608, NGC~4278 and NGC~4472
\citep{2012ApJ...755..163D,2014MNRAS.444.2700D}, and NGC~4374,
NGC~4594 and NGC~5846 (\citealt{2013ApJ...764..151G}). They are among
the reddest ($\mathcal{C_{\rm FUV,tot}} \ga 6$) galaxies in our sample
with massive SMBHs ($M_{\rm BH} \ga 10^{8} M_{\sun}$),
Fig.~\ref{Fig6}. While structural analysis of high-resolution {\it
  HST} images are needed to identify a partially depleted core (or
lack thereof) in the remaining sample galaxies (e.g.,
\citealt{2013ApJ...768...36D,2014MNRAS.444.2700D,2016MNRAS.462.3800D,
  2017MNRAS.471.2321D,2018MNRAS.475.4670D}, the majority ($\sim$ 80\%)
of our spiral galaxies have $\sigma \la 140$ km s$^{-1}$ and they are
likely S\'ersic galaxies with no partially depleted cores
(e.g.,
\citealt{2012ApJ...755..163D,2013ApJ...768...36D,2014MNRAS.444.2700D,2017MNRAS.471.2321D}). For
early-type galaxies, we did not find bends or offsets from the
\mbox{$M_{\rm BH}-\mathcal{C_{\rm UV,tot}}$} relations because of
core-S\'ersic or S\'ersic galaxies (Fig.~\ref{Fig6}).


\subsection{Red, intermediate and blue sequences}\label{BpB}

To locate our galaxies in color-color diagrams, we used the
classification by \citet[their section 4.4]{2018ApJS..234...18B} who
compared the (FUV $-$NUV) and (NUV $−$ [3.6]) colors to separate their
galaxies into red, intermediate and blue sequences. Fig.~\ref{FigC1}
shows excellent coincidence between the red and blue
\mbox{$M_{\rm BH}- \mathcal{C}$} (early- and late-type) morphological
sequences (Sections~\ref{Sec3.1}) and the canonical color-color
relation (red/intermediate) and blue sequences, respectively, only
with three exceptions (\mbox{NGC 2685}, \mbox{NGC 4245} and \mbox{NGC
  4594}). The case of \mbox{NGC 2685} was discussed in
Section~\ref{Sec4.2.1}. NGC~4245 is a barred S0-Sa galaxy with a
prominent ring (\citealt{2007AJ....134.1195T}).  The spiral Sa
NGC~4594 (also referred to as the Sombrero Galaxy) exhibits properties
similar to massive early-type galaxies. \cite{2008MNRAS.389.1150S}
found that the number of blue globular cluster in NGC~4594 is
comparable to massive early-type
galaxies.  Also, \cite{2011ApJ...739...21J} noted
 that the galaxy's dark matter density and core radius resemble those
 expected for early-type galaxies with massive bulges. It is the only
red-sequence spiral in our sample (\citealt{2018ApJS..234...18B}),
which is also unique in being the only core-S\'ersic late-type galaxy
in the sample. Interestingly, Fig.~\ref{FigC1} reveals that all the
intermediate sequence galaxies \citep{2018ApJS..234...18B} reside
toward the left of the \mbox{$M_{\rm BH}- \mathcal{C_{\rm UV,tot}}$}
relation defined by early-type galaxies.

\subsection{Predicting SMBH masses using  $M_{\rm BH}- \mathcal{C_{\rm
  UV}}$
  relations} 

It is of interest to assess the robustness of BH masses estimated
using the \mbox{$M_{\rm BH}- \mathcal{C_{\rm FUV,tot}}$} and
\mbox{$M_{\rm BH}- \mathcal{C_{\rm NUV,tot}}$} relations found in this
work. We do so using literature FUV, NUV and 3.6 {\micron} magnitudes
for a selected sample of 11 galaxies with direct SMBH masses that are
not in our sample (see Table~\ref{TableBH}).  These 11 galaxies were not 
included in the main sample as we endeavor to establish the  
\mbox{$M_{\rm BH}- \mathcal{C_{\rm UV,tot}}$} relations using 
homogeneously determined  UV and 3.6 {\micron} magnitudes. Doing this,
 the observed trends in the  \mbox{$M_{\rm BH}- \mathcal{C_{\rm UV,tot}}$} 
 diagrams (Fig.~\ref{Fig2}) cannot be attributed to differences in methodologies 
 and/or data sources. We use the BH mass
measurement of NGC 205 by \citet{2019ApJ...872..104N} and for the
remaining 10 galaxies the SMBH masses are from
\citet{2016ApJ...831..134V}. Off the 11 galaxies, 9 are in common
between \cite{2009MNRAS.398.2028J} and
\cite{2016ApJS..222...10S}. \citet [their
table~1]{2009MNRAS.398.2028J} published total apparent FUV and NUV
magnitudes derived from growth curves for the galaxies, while
\citet[their table~2]{2016ApJS..222...10S} presented their 3.6
{\micron} galaxy apparent magnitudes\footnote{We have converted the
  \mbox{3.6 {\micron}} VEGA magnitudes from \cite{2016ApJS..222...10S}
  into AB magnitudes.} which were computed using their best-fitting
structural parameters. We also included the low-mass elliptical galaxy
NGC~205 and the Seyfert SAm bulge-less galaxy NGC~4395
\citep{2003ApJ...588L..13F, 2005ApJ...632..799P}. NGC~205 potentially
harbors the lowest central BH mass measured for any galaxy to date
\citep{2019ApJ...872..104N} and NGC~4395 is known for being an outlier
from the $M_{\rm BH}- \sigma$ diagrams (e.g.,
\citealt{2017MNRAS.471.2187D}). For NGC~205, we use the UV and 3.6
{\micron} magnitudes from \citet{2007ApJS..173..185G} and
\citet{2006ApJ...646..929M}, respectively.  For NGC~4395, the total UV
and 3.6 {\micron} magnitudes are from \cite{2009ApJ...703..517D} and
\cite{2011ApJS..192....6L}; a caveat here is that these magnitudes are
not corrected for internal dust attenuation. 

Before applying our $M_{\rm BH}- \mathcal{C_{\rm UV,tot}}$ relations
to estimate $M_{\rm BH}$, we homogenize the FUV, NUV and 3.6 {\micron}
data from the literature by comparing our magnitudes with those from
\citet{2009MNRAS.398.2028J} and \cite{2016ApJS..222...10S} for
galaxies in common with them. We find that, compared to our
magnitudes, the \citet{2009MNRAS.398.2028J} total FUV and NUV
magnitudes are fainter typically by 0.52 mag, while the galaxy
magnitudes from \cite{2016ApJS..222...10S} are brighter typically by
0.18 mag. Having applied these corrections (i.e.,
$m_{\rm UV}=m_{\rm UV,Jeo}-0.52$ and
$m_{\rm 3.6}=m_{\rm 3.6,Sav}+0.18$) for the 9 early-type galaxies, we
computed the $\mathcal{C_{\rm FUV,tot}}$ and
$\mathcal{C_{\rm FUV,tot}}$ colors listed in Table~\ref{TableBH}.

Fig.~\ref{Fig7} reveals good agreement between the directly measured
$M_{\rm BH}$ and predicted $M_{\rm BH}$ determined using
\mbox{$M_{\rm BH}- \mathcal{C_{\rm FUV,tot}}$} and
\mbox{$M_{\rm BH}- \mathcal{C_{\rm NUV,tot}}$} relations for the 10
galaxies in Table~\ref{TableBH}. On average,
$|{\rm log} (M_{\rm BH,predicted}/M_{\rm BH,measured})| \sim 0.67 {\rm
  ~dex}\pm 0.29 ~{\rm dex}~ ({\rm FUV})$ and
$ \sim 0.32 ~{\rm dex} \pm 0.29 ~{\rm dex}~ ({\rm NUV})$.  In
Fig.~\ref{Fig7}, the direct BH mass appear to correlate better with that
predicted using the NUV color than using the FUV color, and for the
massive early-type galaxies, this may be due to contributions to the
FUV flux from the extreme horizontal branch stars (see
Section~\ref{Sec3.1}). The approach of using homogenized galaxy colors
obtained through different methods may introduce some systematic
errors in the determination $M_{\rm BH}$. We caution that when using
the \mbox{$M_{\rm BH}- \mathcal{C}$} relations to predict BH masses,
one should use FUV, NUV and \mbox{3.6 {\micron}} magnitudes obtained
in a homogeneous way.  Furthermore, the \mbox{$M_{\rm BH}- \mathcal{C}$} 
relations {\it should not} be used to predict BH masses in galaxies that are  
highly inclined (e.g., edge-on) and obscured with dust.

As noted in the introduction, a clear benefit of the
\mbox{$M_{\rm BH}- \mathcal{C_{\rm UV,tot}}$} relation is its
applicability to early- and late-type galaxies including those with low 
central velocity dispersions ($\sigma \la100$ km s$^{-1}$) and with 
small or no bulges. Moreover, photometry has the advantage of being 
cheaper than spectroscopy. Using our relations (Table~\ref{Table1}) together with  galaxy 
colors derived from the \citet[][their Table~1]{2018ApJS..234...18B} asymptotic FUV, NUV 
and 3.6 $\micron$ magnitudes, we tentatively predict BH masses in a sample of 1382 
\textit{GALEX}/S$^{4}$G galaxies (Table~\ref{Table7}) with no measured  BH masses, see Appendix~\ref{AppE}.  
We show that late-type galaxies with
$\mathcal{C_{\rm FUV,tot}} \la 1.33 $ AB mag or 
$\mathcal{C_{\rm NUV,tot} }\la 1.28$ AB mag  may harbor  
intermediate-mass black holes ($M_{\rm BH} \sim 100 - 10^{5} M_{\sun}$). Similarly,  early-type galaxies with
$\mathcal{C_{\rm FUV,tot}} \la 4. 68$ AB mag or 
$\mathcal{C_{\rm NUV,tot} }\la 3.4$ AB mag are potential IMBH hosts.
 While Sloan Digital Sky Survey (SDSS) velocity dispersion measurements are available for hundreds of thousands of galaxies and one can use them together with the \mbox{$M_{\rm BH}- \sigma$} relation to estimate BH masses, as cautioned by SDSS Data Release\footnote{\url{https://www.sdss.org/dr12/algorithms/redshifts/}}  12 \citep{2015ApJS..219...12A}, velocity dispersion values less than $100$ km s$^{-1}$ reported by SDSS  are below the resolution limit of the SDSS spectrograph and are regarded as unreliable. Note that galaxies with $\sigma \la100$ km s$^{-1}$ are expected to have low stellar masses ($M_{*} \la 2 \times 10^{10} M_{\sun}$), and such galaxies make up 
 a significant  fraction of the SDSS galaxy sample \citep[][see their Fig.~9]{2015ApJS..219....8C}. In addition, the SDSS spectra measure the light within a fixed aperture of radius 1$\farcs$5, thus the SDSS velocity dispersion values of more distant galaxies can be systematically lower than those of similar, nearby galaxies.

\begin{center}
\begin{table} 
\setlength{\tabcolsep}{0.0206in}
\caption{Supermassive black hole masses}
\label{TableBH}
\begin{tabularx}{0.478\textwidth}{@{}lllccccc@{}}\hline
\hline
\hline
Galaxy &Type & $\mathcal{C_{\rm UV,tot}}$ 
                                                                 &${\rm log} (M_{\rm BH}/M_{\sun})$
                            &  ${\rm log} (M_{\rm BH}/M_{\sun})$ \\
 &&[AB mag]&(directly measured)&(predicted)&
  \\
&&(FUV/NUV)&&(FUV/NUV)&\\
(1)&(2)&(3)&(4)&(5)&\\
\multicolumn{1}{c}{} \\              
\hline   
\hline    
NGC 0205                &E5 pec&    4.61/2.67&   $3.83^{+1.18}_{-1.83}$& 4.85/3.58\\     
NGC 0524                &SA0&    7.30/5.48&$8.94^{+0.05}_{-0.05}$ &9.68/9.02\\     
NGC 0821               & E6&     7.02/4.69& $8.22^{+0.21 }_{-0.21 }$   & 9.13/7.64\\    
NGC 1023             & SB0&     6.64/5.01&  $7.62^{+0.05 }_{-0.05 }$  & 8.39/8.20\\   
NGC 4395             &SAm &       1.49/1.37&  $5.54^{+0.54 }_{-0.54 } $&  5.17/5.29\\   
NGC 4459              &SA0&     6.56/5.03&   $7.84^{+0.09  }_{-0.09 }$   &  8.24/8.23\\   
NGC 4473             &E5&     6.80/4.93&  $7.95 ^{+0.24 }_{-0.24 }$    &   8.71/8.06\\   
NGC 4552              &E&       6.24/5.33&  $8.70^{+0.05 }_{-0.05}$     &   7.61/8.76\\   
NGC 4564              &E6&      6.34/4.77&   $7.95^{+0.12}_{-0.12}$ &  7.91/7.78\\   
NGC 4621              &E5&     7.18/5.78&    $8.60^{+0.09}_{-0.09}$   & 9.41/9.55\\   
NGC 5845             &E&       6.38/5.30&    $8.69^{+0.16}_{-0.16}$  &  7.89/8.71\\                                              
\hline
\end{tabularx} 
Notes.---  Col. (1) galaxy name. Col. (2) morphological type from NED.   Col. (3) total
FUV- [3.6] and NUV- [3.6] colors ($\mathcal{C_{\rm FUV,tot}}$ and
$\mathcal{C_{\rm NUV,tot}}$).   Cols.  (4) directly measured
SMBH masses.  Cols.  (5)  BH masses predicted using the  $M_{\rm
  BH}- \mathcal{C_{\rm FUV,tot}}$ and  $M_{\rm BH}- \mathcal{C_{\rm
    NUV,tot}}$ relations (Table~\ref{Table1}) and the appropriate
colors given in Col. (3). We assign a typical uncertainty of 0.85 dex
on ${\rm log} (M_{\rm BH}$) for these predicted BH masses.  \\\\
\end{table}
\end{center}

\subsection{The $M_{\rm BH}- \mathcal{C_{\rm
  UV}}$ correlation as evidence for the co-evolution of SMBHs and galaxies}\label{Sec4.5}

Morphologically splitting the sample galaxies, we have demonstrated
that late-type hosts do not correlate with SMBHs in the same manner as
early-type hosts (Table~\ref{Table1}). This reconciles very well with
the prediction that early- and late-type galaxies have fundamentally
different formation histories (e.g.,
\citealt{1978MNRAS.183..341W,2001ApJ...561..517K,2002NewA....7..155S,2003MNRAS.341...54K,2004ARA&A..42..603K,2010ApJ...711..284S,2014MNRAS.444.2700D,2014MNRAS.440..889S,2016MNRAS.459.4109T,2018ApJ...869..113D,2019ApJ...871....9D}).
For example, the red and blue $M_{\rm BH}- \mathcal{C_{\rm FUV,tot}}$
relations are such that
$M_{\rm BH,early-type} \propto (L_{\rm FUV,tot}/L_{\rm
  3.6,tot})^{-4.38}$ and
$M_{\rm BH,late-type} \propto (L_{\rm FUV,tot}/L_{\rm
  3.6,tot})^{-2.58}$.  Given that $L_{\rm UV}/L_{3.6}$ is a good proxy 
for the sSFR, (\citealt[][their
Appendix B]{2018ApJS..234...18B}), it therefore implies that both
early- and late-type galaxies exhibit log-linear inverse correlations
between $M_{\rm BH}$ and sSFR, the latter having a steeper dependence
on sSFR than the former (see Table~\ref{Table1} and Fig.~\ref{Fig2}).

A correlation between $M_{\rm BH}$ and sSFR is not
unexpected. Observations have shown that bright quasars and local
Seyferts tend to reside in strong starburst galaxies or in galaxies
with an ongoing
star formation (e.g.,
\citealt{1988ApJ...325...74S,2003MNRAS.346.1055K,
  2005Natur.434..738A,2008ApJ...684..853L, 2009MNRAS.399.1907N,
  2010MNRAS.405..933W,2012A&A...545A..45R,
  2017ApJ...842...72Y,2017ApJ...850...27B}). Other findings lending further support to the
link between star formation and SMBH growth are the correlation
between black hole accretion rate and host galaxy star formation rate
\citep[e.g.,][]{2004MNRAS.354L..37M,2004ApJ...613..109H,2010MNRAS.407.1529H,2013ApJ...773....3C,
  2014ARA&A..52..415M,2015MNRAS.452..575S,2017ApJ...842...72Y}, the
inverse correlation between sSFR and specific supermassive black hole
mass \citep{2017ApJ...844..170T} and the trend between SMBH mass and
host galaxy star formation histories over cosmic time
\citep[e.g.,][]{2018Natur.553..307M,2019MNRAS.tmp..406V}.

Within the self-regulated SMBH growth model, the correlation between
SMBH masses and the host galaxy properties (e.g., stellar luminosity,
\citealt{2002MNRAS.331..795M,2003ApJ...589L..21M}) is interpreted as
reflecting a link between the growth of SMBHs and star
formation events in the host
\citep{1998A&A...331L...1S,1999MNRAS.308L..39F,2003ApJ...596L..27K,
  2005Natur.435..629S,2005Natur.433..604D,2005ApJ...618..569M,2006MNRAS.365...11C,
  2006ApJS..163....1H,2006Natur.442..888S,2009Natur.460..213C,2017MNRAS.465.3291W}. In
this scenario, the same cold gas reservoir that fuels the AGN/quasar,
feeds starburst events. The energy or momentum released by the
AGN/quasar can heat the interstellar medium and cause the expulsion of
gas from the host galaxy, shutting off star formation and halting
accretion onto the SMBH. However, whether AGN accretion and star
formation are precisely coincidental is unclear (e.g.,
\citealt{2005ApJ...629..680H}).
 
We (see also Dullo et al.\ 2020, submitted) argue that the significantly different
\mbox{$M_{\rm BH}- \mathcal{C_{\rm FUV}}$} relations for early- and
late-type galaxies (i.e., the
\mbox{$M_{\rm BH}- \mathcal{C_{\rm FUV}}$} red and blue sequences)
suggest that the two Hubble types follow two distinct  channels of SMBH growth,
the former is major merger driven while the latter involves
(major merger)-free processes \cite[see ][]{2009ApJ...694..599H}, in
broad accordance with \citet{2010ApJ...711..284S,2014MNRAS.440..889S}.

The standard cosmological  formation paradigm is that SMBHs in
early-type galaxies are built up during the period of rapid galaxy
growth at high redshift (z $\sim$ 2$-$5) when the major, gas-rich
mergers of disk galaxies
\citep[e.g.,][]{1972ApJ...178..623T,1978MNRAS.183..341W} drives gas
infall into the nuclear regions of the newly formed merger remnant,
leading to starburst events and AGN accretion processes
\citep[e.g.,][]{1991ApJ...370L..65B,1996ApJ...471..115B,
  2006MNRAS.372..839N,2010MNRAS.407.1529H,2015MNRAS.454.1742K}. 
Accretion onto more massive SMBHs trigger stronger AGN feedback,
efficient at quenching of star formation rapidly. As such, the position of an
early-type galaxy on the $M_{\rm BH}- \mathcal{C_{\rm UV}}$ red
sequence is dictated by the complex interplay between the details of
its SMBH growth, efficiency of AGN feedback, regulated star formation
histories and major merger histories, rather than this being set by
simple hierarchical merging
\citep{2007ApJ...671.1098P,2011ApJ...734...92J}.

Massive early-type galaxies (i.e., total stellar mass
$M_{*,k} \sim 8 \times 10^{10} M_{\sun} - 10^{12} M_{\sun}$) with
$M_{\rm BH} \ga 10^{8} M_{\sun}$ and $\mathcal{C_{\rm FUV}} \ga 6.3$
mag, at the high-mass end of the $M_{\rm BH}- \mathcal{C_{\rm UV}}$
red sequence (Fig.~\ref{Fig2}, Section~\ref{BpB}), are consistent
with the scenario where (gas-rich) major merger at high redshift
drives intense bursts of star formation, efficient SMBH growth and
ensuing quenching of star formation by strong AGN feedback in short
timescales
\citep[e.g.,][]{2005ApJ...621..673T,2010MNRAS.404.1775T,2011MNRAS.418L..74D,2015MNRAS.448.3484M,2016MNRAS.461L.102S}. This
is accompanied by a few (0.5$-$2) successive gas-poor (dry) major mergers
since $z \sim 1.5-2$
\citep[e.g.,][]{2004ApJ...608..752B,2006ApJ...640..241B,2009MNRAS.397..506K,2012ApJ...744...85M,2012ApJ...755..163D,2013ApJ...768...36D,2014MNRAS.444.2700D,2015MNRAS.449...49R},
involving low level star formation\footnote{Massive early-type
  galaxies and some BCGs can acquire cold gas through the cooling of
  hot gas and/or via cannibalism of a gas-rich satellite and they may
  undergo episodes of low level star formation at low redshift
  \citep{2003A&A...412..657S,2008ApJ...681.1035O,2009ApJ...694..599H,2010A&A...523A..75S,2011MNRAS.414..940Y,2012MNRAS.426.2751Z,2014ApJ...784...78R,2017MNRAS.471L..66S,2019arXiv190209227R,2020A&A...635A.129K}.}
(i.e., `red but not strictly dead', see \citealt
{2011MNRAS.418L..74D,2019MNRAS.484.4413H, 2019arXiv190308884D})
detected by the {\it GALEX} FUV and NUV detectors
\citep[e.g.,][]{2007ApJS..173..185G,2018ApJS..234...18B}.  The bulk of
these objects are core-S\'ersic elliptical galaxies a fraction of which may
gradually grow stellar disk structures and transform into massive
lenticular galaxies
\citep{2013pss6.book...91G,2013ApJ...768...36D,2014ASPC..480...75D,2015ApJ...804...32G,2016MNRAS.457.1916D}.

As for the less massive (S\'ersic) early-type galaxies
($M_{*,k} \sim 10^{10} M_{\sun} - 2\times10^{11} M_{\sun}$) with
smaller SMBH masses
($10^{6} M_{\sun} \la M_{\rm BH} \la 10^{8} M_{\sun}$) and
$\mathcal{C_{\rm FUV}} \la 6.3$ mag (Fig.~\ref{Fig2},
Section~\ref{BpB}) likely grow primarily via gas-rich (wet) major mergers
and form their stellar populations over an extended period of time
\citep{2005ApJ...621..673T,
  2010MNRAS.404.1775T,2011MNRAS.418L..74D,2015MNRAS.448.3484M}. Our
findings disfavor a scenario where S\'ersic early-type galaxies with
intermediate colors are late-type galaxies quenching star formation
and moving away from the \mbox{$M_{\rm BH}- \mathcal{C_{\rm FUV}}$}
blue sequence. Collectively, core-S\'ersic and S\'ersic early-type
galaxies define a red sequence in the
$M_{\rm BH}- \mathcal{C_{\rm FUV}}$ diagram. Lacking the most luminous
and massive BCGs with $M_{*,k} \ga 10^{12} M_{\sun}$ in our sample, we
note that our $M_{\rm BH}- \mathcal{C_{\rm UV}}$ relation is not
constrained at the highest-mass end. Since such galaxies are generally
expected to have negligible star formation \citep{2019ApJ...886...80D}, the 
$M_{\rm BH}- \mathcal{C_{\rm UV}}$ relations, the $M_{\rm BH}- \mathcal{C_{\rm FUV}}$ in particular, may not apply to them.
Our interpretation of the assembly of the red sequence for early-type
galaxies is in accordance with the `downsizing' scenario, where more
massive galaxies form stars earlier and over a shorter time scale than
less massive galaxies (e.g.,
\citealt{1996AJ....112..839C,2000ApJ...536L..77B,2008MNRAS.389..567C,2008ApJ...675..234P,2009ApJ...698L.116P})

The (major merger)-free scenario---secular processes and minor
mergers---may be naturally consistent with the observed late-type
\mbox{$M_{\rm BH}- \mathcal{C_{\rm FUV}}$} blue sequence
(Fig.~\ref{Fig2}, Section~\ref{BpB}) and dynamically cold stellar disks of late-type
galaxies. Recently, \citet{2018MNRAS.tmp..321M} reported that massive
SMBHs in disk galaxies can grow primarily via secular processes with
small contributions (i.e., only 35 \% of the SMBH mass) from
mergers. We tentatively hypothesize secular-driven processes involving
non-axisymmetric stellar structures, such as bars and spiral arms can
trigger a large inflow of gas from the large scale disk into nuclear
regions of late-type galaxies, slowly feeding SMBHs and fueling star
formation \citep[e.g.,][]{1993IAUS..153..209K,2004ARA&A..42..603K,
  2008AJ....136..773F,2012ApJ...745..149L,2013seg..book....1K,2013ApJ...776...50C,2016MNRAS.459.4109T,2019ApJ...871....9D}.
 Barred galaxies make up the bulk ($\sim$62\%) of the late-type galaxies
 in our sample (Fig.~\ref{Fig6}). 
In addition, gas-rich minor mergers have been suggested in the
literature to trigger enhanced star formation and SMBH growth in
late-type galaxies without destroying the disks
\citep{2013MNRAS.429.2199S,2014MNRAS.437L..41K,2014MNRAS.440.2944K,2017MNRAS.470.1559S,2018MNRAS.476.2801M,2018MNRAS.tmp..321M}. A
question remains however whether pure (major merger)-free processes
could be the main mechanism for the formation of late-type galaxies
with massive bulges and high-velocity dispersion, as in the case of NGC
4594.

As noted above, we find that core-S\'ersic\footnote{All the seven
  core-S\'ersic galaxies in this paper are ``normal-core''  galaxies
  \citep{2019ApJ...886...80D}. } and S\'ersic galaxies collectively
define a single early- or late-type morphology sequence, in agreement
with the conclusions by \citet{2019arXiv190304738S,
  2019ApJ...886...80D}; Dullo et al. (2020, submitted). Moreover,
although \citet{2016ApJ...817...21S} reported a blue, spiral galaxy
$M_{\rm BH}-M_{*,\rm bulge}$ sequence, their sample of 17 spiral
galaxies trace the red end of the blue
\mbox{$M_{\rm BH}-M_{*,\rm bulge}$} sequence \citep[their Section
2.2]{2018ApJ...869..113D}.

We remark that bulgeless spirals and spiral galaxies with classical
bulges or pseudo-bulges all follow the blue, late-type
\mbox{$M_{\rm BH}- \mathcal{C_{\rm UV}}$} relations (Fig.~\ref{Fig2}).
Pseudo-bulges are hosted typically by late-type galaxies and a few
early-type galaxies, while classical bulges are generally associated
with early-type galaxies and massive late-type galaxies.  A key point
to note here is that the relatively high sSFR for pseudo-bulges
coupled with the steeper dependence of SMBH masses on sSFRs
 for late-type galaxies likely explain
why pseudo-bulges seem to obey a different \mbox{$M_{\rm BH}- \sigma$}
relation than classical bulges \citep[e.g.,
][]{2013ARA&A..51..511K}. This also explains as to why bulgeless
spirals and low-mass spirals offset from the
\mbox{$M_{\rm BH}- L_{\rm Bulge}$},
\mbox{$M_{\rm BH}- M_{*,\rm Bulge}$} and \mbox{$M_{\rm BH}- \sigma$}
relations defined by the massive early- and late-type galaxies as
reported in the literature
\citep[e.g.,][]{2008ApJ...688..159G,2010ApJ...721...26G,2016ApJ...826L..32G,2017ApJ...850..196B}.

\section{Conclusions}\label{Sec5} 

Using a sample of 67 \textit{GALEX}/S$^{4}$G galaxies with directly
measured supermassive black hole masses ($M_{\rm BH}$), comprised of
20 early-type galaxies and 47 late-type galaxies, for the first time
we establish a correlation between ($M_{\rm BH}$) and the host galaxy total  (i.e., bulge+disk)  UV$-$ [3.6]
color ($\mathcal{C_{\rm UV}}$). More massive SMBHs are hosted by
galaxies with redder colors. The \textit{GALEX} FUV/NUV and S$^{4}$G
\textit{Spitzer} \mbox{3.6 $\micron$} asymptotic magnitudes of the
sample galaxies determined in a homogeneous manner
\citep{2018ApJS..234...18B} along with their \mbox{3.6
  $\micron$} multi-component decomposition by
\citet{2015ApJS..219....4S} were used to derive dust-corrected total
(bulge+disk) magnitudes in FUV, NUV and \mbox{3.6 $\micron$} bands. We
provide these magnitudes in Table~\ref{Table6}.

We fit our ($M_{\rm BH}, \mathcal{C_{\rm UV,tot}}$) dataset using
several statistical techniques, focusing on the symmetric BCES
bisector regressions. Our key
findings are as
follows.\\

(1) Investigating the nature of the
$M_{\rm BH}- \mathcal{C_{\rm UV,tot}}$ relations, our results show
that early-type galaxies define a red-sequence in the
$M_{\rm BH}- \mathcal{C_{\rm UV,tot}}$ diagrams different from the
late-type blue-sequence. We found a strong tendency  for the
galaxies which lie on the red/blue \mbox{$M_{\rm BH}- \mathcal{C}$}
morphological sequences to also be on the (red plus intermediate)/blue
sequences in the canonical color-color relation  (See Section~\ref{BpB} and Fig.~\ref{FigC1}).

(2)  The
\mbox{$M_{\rm BH}-\mathcal{C_{\rm FUV,tot}}$} and
\mbox{$M_{\rm BH}-\mathcal{C_{\rm NUV,tot}}$} relations for early-type
galaxies  have slopes of 1.75 $\pm $ 0.41 and 1.95 $\pm $ 0.28,
respectively, whereas for 
late-type galaxies the slopes are  substantially shallower, i.e., 1.03 $\pm $
0.13 and 1.38 $\pm $ 0.23. The early- and late-type
\mbox{$M_{\rm BH}-\mathcal{C_{\rm UV,tot}}$} relations have
root-mean-square (rms) scatters ($\Delta $) in the \mbox {log
  $M_{\rm BH}$} direction of $\Delta_{\rm UV,early} \sim 0.72- 0.86$
dex and $\Delta_{\rm UV,late} \sim 0.86$ dex and Pearson correlation
coefficients ($r$) of $r_{\rm early} \sim 0.61 - 0.70$ and
$r_{\rm late} \sim 0.60 -0.65$. 

(3) Given $L_{\rm UV,tot}/L_{\rm 3.6,tot}$ is a good proxy for
specific star formation rate (sSFR), it follows that both early- and
late-type galaxies exhibit log-linear inverse correlations between
$M_{\rm BH}$ and sSFR, the latter having a steeper dependence on sSFR
(i.e., \mbox{$M_{\rm BH} \propto \rm sSFR_{\rm FUV}^{-2.58}$}) than the former
(\mbox{$M_{\rm BH} \propto \rm sSFR_{\rm FUV}^{-4.38}$}). This suggests
different channels for  SMBH growth in early- and late-type
galaxies.

(4) We have compared the \mbox{$M_{\rm BH}-\mathcal{C_{\rm UV,tot}}$}
relations with the \mbox{$M_{\rm BH}-\sigma$} and
\mbox{$M_{\rm BH}-L_{\rm 3.6,tot}$} relations for our sample
galaxies. While the \mbox{$M_{\rm BH}-\mathcal{C_{\rm UV}}$} relations are marginally
weaker ($r \sim 0.60-0.70$) and have typically $5\%-27\%$ more scatter than the
$M_{\rm BH}-\sigma$ relations ($r \sim 0.72 -0.78$), the  former potentially constrains  SMBH-galaxy co-evolution models that predict different SMBH growth for different morphologies.
In contrast,  the $M-\sigma$ relations for late- and early-type galaxies are  similar. The \mbox{$M_{\rm BH}-\mathcal{C_{\rm UV,tot}}$} and
\mbox{$M_{\rm BH}-L_{\rm 3.6,tot}$} relations display similar level of
strength and vertical scatter.

(5) We did not detect departures (bends
or offsets) from  the \mbox{$M_{\rm BH}-\mathcal{C_{\rm UV,tot}}$}
relations because of core-S\'ersic or S\'ersic galaxies. However,  we 
cannot firmly rule out the presence of such substructures, as our sample does not
consist of a large number of core-S\'ersic galaxies.

(6) We argue that the different
\mbox{$M_{\rm BH}- \mathcal{C_{\rm UV}}$} relations for early- and
late-type galaxies reflect that the two Hubble types have two distinct
SMBH feeding mechanisms. {\it
  Massive early-type galaxies}
($M_{*,k} \sim 8 \times 10^{10} M_{\sun} - 10^{12} M_{\sun}$)
 at the high-mass end of the $M_{\rm BH}- \mathcal{C_{\rm UV}}$
red sequence, are core-S\'ersic galaxies. Their formation is
consistent with the scenario in which gas-rich (wet) major merger at high
redshift drives intense bursts of star formation, efficient SMBH
growth and the ensuing rapid quenching of star formation by strong AGN
feedback. This is accompanied by gas-poor (dry) 
major mergers since $z \sim 1.5-2$, involving low level star
formation. In contrast, the {\it less massive (S\'ersic) early-type
  galaxies} ($M_{*,k} \sim 10^{10} M_{\sun} - 2\times10^{11} M_{\sun}$)  
at the low-mass part of the
$M_{\rm BH}- \mathcal{C_{\rm UV}}$ red sequence are likely built-up
primarily via gas-rich major mergers and form their stellar
populations over an extended period of time.  We tentatively hypothesize
that {\it late-type galaxies} ($M_{*,k}  \sim 10^{9} M_{\sun} - 2\times10^{11} M_{\sun}$)  which define the
$M_{\rm BH}- \mathcal{C_{\rm UV}}$ blue sequence form via
secular-driven processes involving non-axisymmetric stellar
structures, such as bars and spiral arms. Gas-rich minor mergers could
also account for the build-up of late-type galaxies. 

(7) Having demonstrated the potential of our
\mbox{$M_{\rm BH}-\mathcal{C_{\rm UV}}$} relations to predict the SMBH
masses in 10 galaxies, we employ these new relations to estimate the central BH masses in 1382 \textit{GALEX}/S$^{4}$G galaxies with no measured BH masses, after excluding highly inclined and dust obscured \textit{GALEX}/S$^{4}$G  galaxies \citep{2018ApJS..234...18B}. We suggest the \mbox{$M_{\rm BH}-\mathcal{C_{\rm UV}}$} relations can be used to estimate BH masses, without the need 
for high-resolution spectroscopy. However, we warn that to do so one should use galaxy colors determined based on UV and 3.6
{\micron} magnitudes obtained in a homogeneous way. Furthermore,
the \mbox{$M_{\rm BH}-\mathcal{C_{\rm UV}}$} relations can be used to
identify low-mass and bulgeless galaxies that potentially harbor
intermediate-mass black holes.

\section{ACKNOWLEDGMENTS}

We thank the referee for their useful comments. B.T.D acknowledges support from a Spanish postdoctoral fellowship
`Ayudas 1265 para la atracci\'on del talento investigador. Modalidad
2: j\'ovenes investigadores.' funded by Comunidad de Madrid under
grant number 2016-T2/TIC-2039. B.T.D acknowledges support from 
grant   `Ayudas para la realizaci\'on de proyectos de I+D para jóvenes doctores 2019.'
 funded by Comunidad de Madrid and Universidad Complutense 
 de Madrid under grant number PR65/19-22417.
We acknowledge financial support from
the Spanish Ministry of Economy and Competitiveness (MINECO) under
grant numbers AYA2016-75808-R and RTI, which is partly funded by the European
Regional Development Fund, and from the Excellence Network
MaegNet (AYA2017-90589-REDT). A.Y.K.B. acknowledges financial
support from the Spanish Ministry of Economy and Competitiveness
(MINECO), project Estallidos AYA2016-79724-C4-2-P. J.H.K. acknowledges
financial support from the European Union's Horizon 2020 research and
 innovation programme under Marie Sk\l odowska-Curie 
grant agreement No 721463 to the SUNDIAL ITN network, from the State 
Research Agency (AEI) of the Spanish Ministry of Science, Innovation and 
Universities (MCIU) and the European Regional Development Fund (FEDER) 
under the grant with reference AYA2016-76219-P, from IAC project P/300724, 
financed by the Ministry of Science, Innovation and Universities, through the 
State Budget and by the Canary Islands Department of Economy, Knowledge 
and Employment, through the Regional Budget of the Autonomous Community, 
and from the Fundaci\'on BBVA under its 2017 programme of assistance to scientific 
research groups, for the project ``Using machine-learning techniques to drag galaxies 
from the noise in deep imaging". J.G.
acknowledges financial support from the Spanish Ministry of Economy
and Competitiveness under grant number AYA2016-77237-C3-2P.
\bibliographystyle{apj} \bibliography{Bil_Paps_biblo.bib}

\section{Appendix }\label{A}

\begin{appendices}

\section{Appendix A}\label{AppA}
\subsection{Bayesian approach}

We have performed linear regression fits to the
($M_{\rm BH}, \mathcal{C_{\rm UV,tot}}$) dataset applying a Bayesian
statistical inference with Markov Chain Monte Carlo (MCMC)
technique\footnote{Interested readers are referred to \citet[page
  134]{2015bmps.book.....A} and \citet[page 278]{2017bmad.book.....H}
  for a description of the MCMC method.}. The regression fits take
into account errors in $M_{\rm BH}$ and $\mathcal{C_{\rm UV,tot}}$ and
an additional spread not explained by the error bars. The  MCMC runs
were implemented in our work through the Stan programming language\footnote{https://mc-stan.org} (e.g.,
\citealt{2017JSS....76....1C}). In addition to the slope, intercept,
and dispersion, the model parameters include `true' color and SMBH
mass values. In our implementation, the observed SMBH mass upper 
limits were modeled by drawing samples from an asymmetric
normal distribution centered around the true values with standard
deviation having an upper wing which equals to the error bar of the
measurements and  a lower wing with a much larger size.
As such, the regression fits probe a large range in SMBH masses for
galaxies with SMBH upper limits. For the Bayesian analysis, we used
non-informative prior distributions (e.g., \citealt{2013AndrewGelman})
and checked that the results did not depend on the choice of the prior
details.
 
\begin{figure*}
\hspace{-1.0628088186340cm}
\includegraphics[angle=0,scale=0.575305,trim=-1cm 0cm 0 -2cm ]{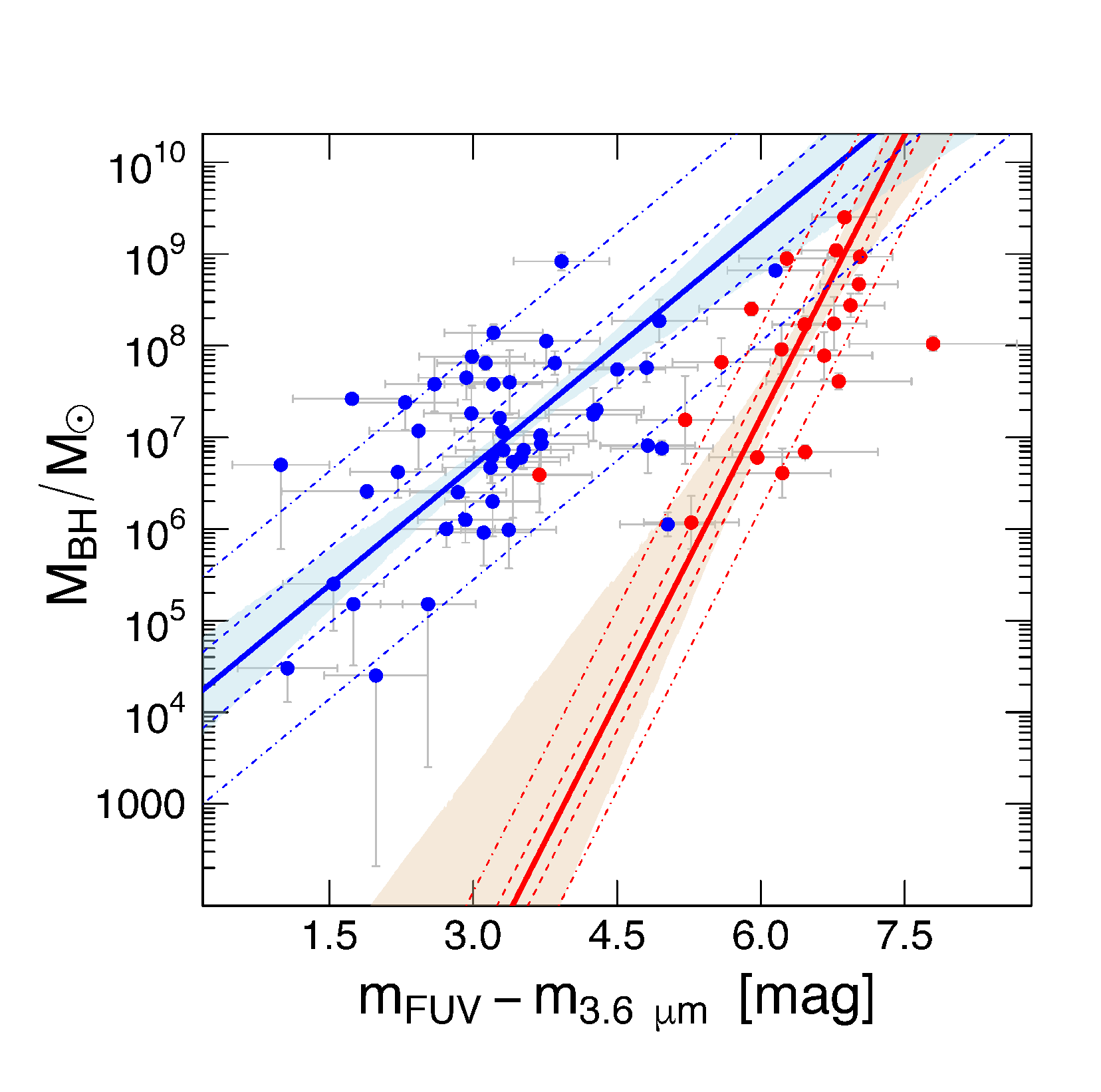}
\hspace{7.37088186340cm}
\includegraphics[angle=0,scale=0.575305,trim=14cm 0 0cm -2cm]{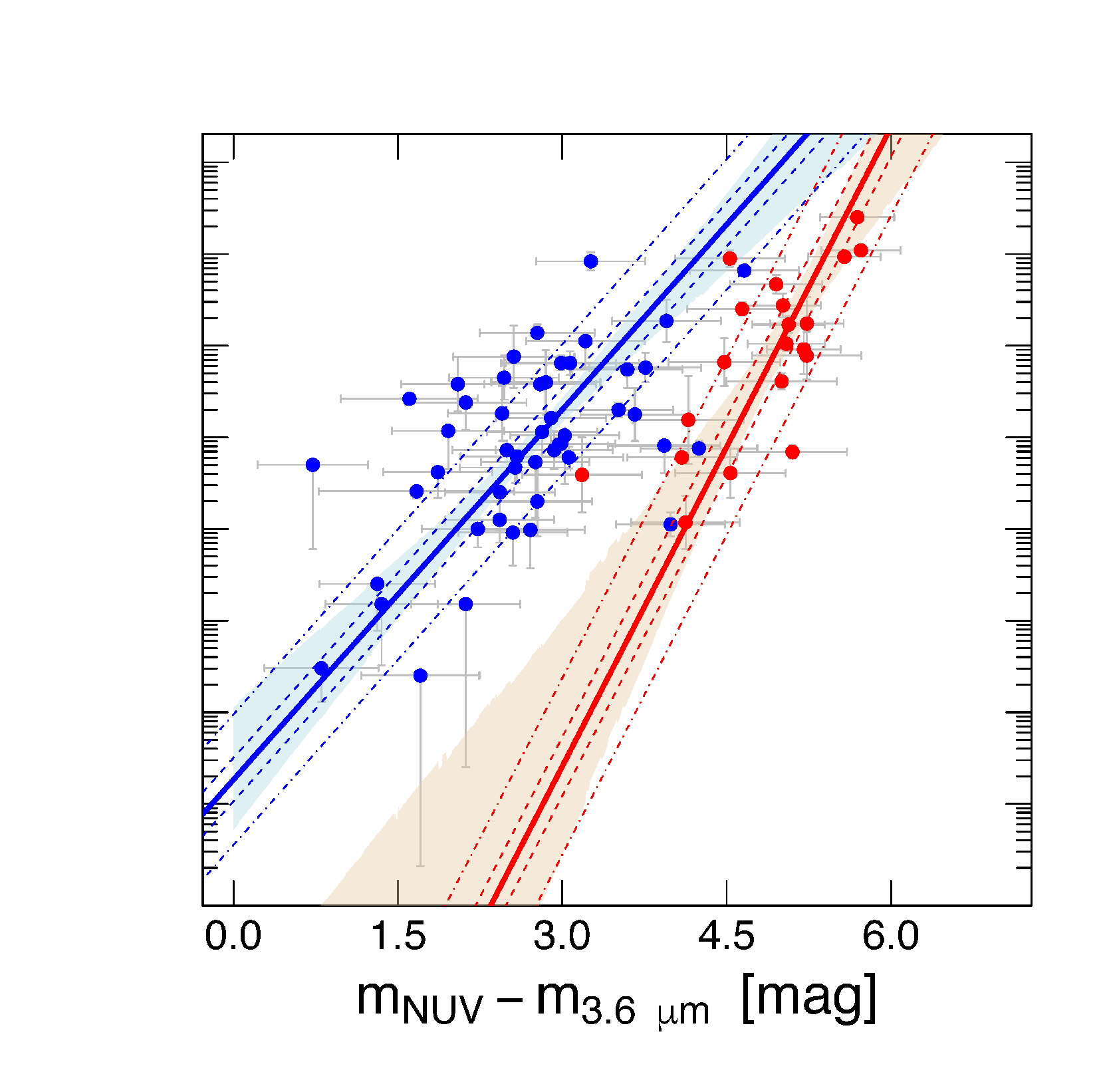}
\vspace{-.558088186340cm}
\caption{Similar to Fig.~\ref{Fig2}, but here plotting 
the results of our symmetric linear Bayesian regression analysis. Early-type  and late-type galaxies are shown in
red and blue, respectively. The shaded regions indicate the 95\% Highest
Density Interval for the derived fits. The dotted lines mark the 1$\sigma$ and 3$\sigma$
intervals for the additional real dispersion not explained by the
error bars,  see the text for further detail.  
 } 
\label{FigAA1} 
 \end{figure*}

Fig.~\ref{FigAA1} shows the results of the Bayesian linear
regressions. The shaded regions indicate the 95\% Highest
Density Interval (HDI) for the derived fits. Note that the HDIs are
not symmetric, since the derived parameters exhibit skewed posterior
distributions. The dotted lines mark the 1$\sigma$ and 3$\sigma$
intervals for the additional real dispersion that is not explained by the
error bars (Fig.~\ref{FigAA1}). The probability distribution functions for these
additional dispersions are not symmetrical, and the derived values
exhibit 95\% HDIs of $(0.00, 0.74)$ and $(0.14, 0.68)$ for the early-
and late-type galaxies. For the former, the additional
dispersion is not significantly different from zero, therefore, the
observed dispersion could be fully explained by the error bars. In
contrast, an additional real dispersion is needed for the latter to
explain the residuals from the fitted relation. 
 
\begin{figure*}
\vspace{7.59088186340cm}
\hspace{1.51088186340cm}
\includegraphics[angle=0,scale=0.447 ,bb=0 15 100
10]{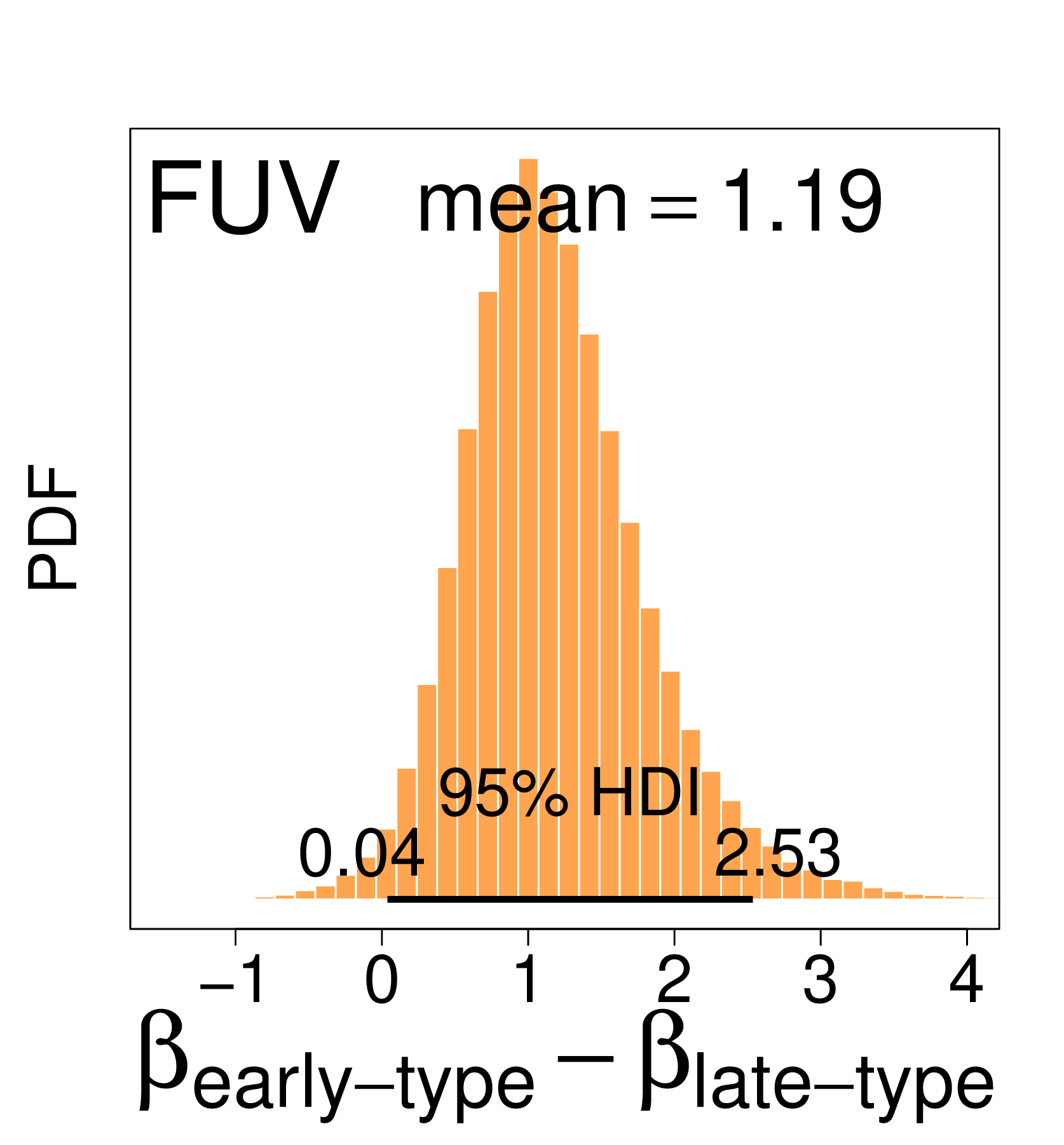}
\hspace{6.0288186340cm}
\includegraphics[angle=0,scale=0.447,bb=15 15 100 10]{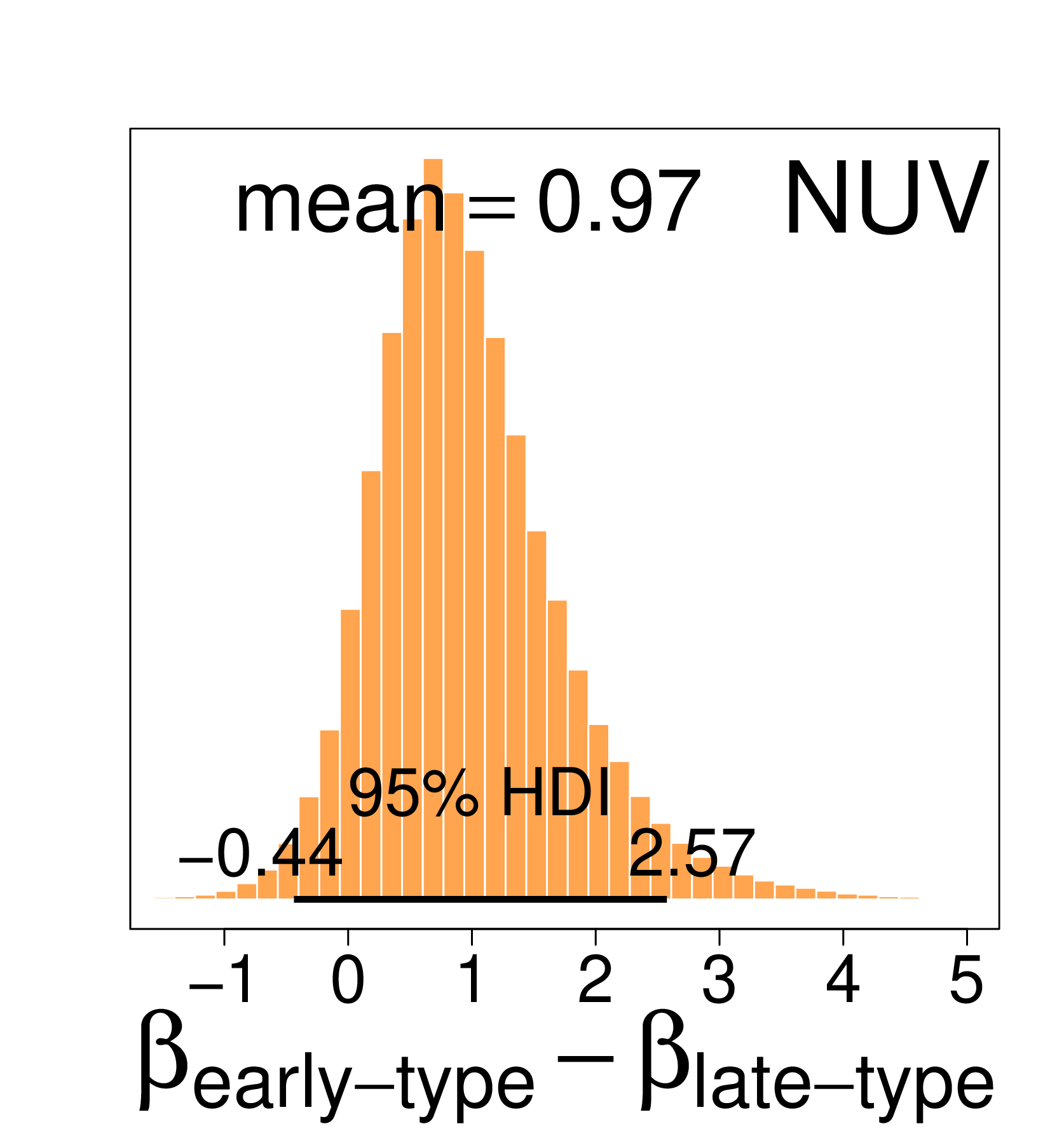}
\caption{probability distribution function (PDF) for the difference in
  slopes ($\beta$)
between the  linear Bayesian regression fits to the  early- and late-type ($M_{\rm BH}, \mathcal{C_{\rm
    UV,tot}}$) data (see Fig.~\ref{FigAA1}).}
\label{FigAA2} 
 \end{figure*}

In Fig.~\ref{FigAA2}, we show the probability distribution functions
(PDF) for the  disparity  in slope between early- and late-type
$M_{\rm BH}- \mathcal{C_{\rm UV,tot}}$ relations obtained by applying the
linear Bayesian regression fits. The PDFs are determined using all the
individual steps of the computed chains, thus avoiding any assumption
about the probability distribution of the derived parameters.  We find that the
significance levels for rejecting the null hypothesis of both
morphological types having the same slope are 1.7\% and 6.7\%  for the FUV
and NUV relations, respectively.

\section{Appendix B}\label{ApB}

\subsection{Outliers in the $M_{\rm BH}- \mathcal{C_{\rm UV,tot}}$ diagrams 
  }\label{outlier}

  Five galaxies in our sample, which offset notably from the
  \mbox{$M_{\rm BH}- \mathcal{C}$} relations (Fig.~\ref{Fig2}).

\subsubsection{ NGC 2685} 

The Helix Galaxy NGC~2685 is a well-studied polar ring lenticular
galaxy (\citealt{1978AJ.....83.1360S}, \citealt{1980A&A....82..314S,
  1994ApJ...420L..21W, 1997ApJ...486..259E},
\citealt{2004AJ....127..789K, 2002ApJ...577L.103S,
  2007MNRAS.381..245R, 2009A&A...494..489J}).
\citet{1961hag..book.....S} referred to it as to the most unusual of
all the galaxies in his atlas, and subsequent studies 
confirmed its rare nature (e.g., \citealt{1997ApJ...486..259E};
\citealt{2002ApJ...577L.103S}; \citealt{2009A&A...494..489J}). The
galaxy contains a large concentration of molecular, neutral and atomic
hydrogen gas in its polar ring.  Fig.~\ref{Fig2} shows it resides in the
$M_{\rm BH}- \mathcal{C}$ blue sequence defined by late-type galaxies
and its $\mathcal{C_{\rm UV,tot}}$ color is abnormally
($\sim$$1.90/1.00$ FUV/NUV mag) bluer than that predicted for
early-type galaxies given the galaxy's SMBH mass. This result agrees 
with the two formation scenarios considered in the literature for NGC
2685: formation via accretion of a small gas-rich companion (e.g.,
\citealt{1980A&A....82..314S, 1994ApJ...420L..21W,
  2002ApJ...577L.103S}) and a merger of two disk galaxies
(e.g., \citealt{2009A&A...494..489J}). \\

\subsubsection{ NGC 3310}

The starburst galaxy NGC 3310 is a well-studied peculiar
spiral galaxy (SAB) known for its very blue color, circumnuclear ring
of star formation, tidal features and very bright infrared luminosity
(e.g., \citealt{1981A&A....96..271B, 1995A&A...300..687M,
  1996ApJ...473L..21S,2000AJ....119...79C, 2002AJ....123.1381E,
  2005ApJ...618L..21W, 2006MNRAS.371.1047W, 2010MNRAS.402.1005H}). The
favored scenario for the formation of the galaxy is through accretion of
a small gas-rich dwarf galaxy $\sim$10 Myr ago
(\citealt{1981A&A....96..271B, 1995A&A...300..687M,
  1996ApJ...473L..21S, 2000AJ....119...79C, 2005ApJ...618L..21W,
  2006MNRAS.371.1047W}). Consistent with this picture, Fig.~\ref{Fig2}
shows that the galaxy's $\mathcal{C_{\rm UV,tot}}$ color is ($\sim 1.99/1.79$
FUV/NUV mag) bluer than expected for late-type galaxies. \\

\subsubsection{ NGC 3368} 

The case of the double-barred spiral (SAB) galaxy NGC~3368 is less
clear. It is the brightest galaxy in the nearby NGC 3368 group
\citep{2003ApJ...591..185S,2014ApJ...791...38W}, containing a dominant
pseudo-bulge, a small classical bulge and box/peanut component
\citep{2010MNRAS.403..646N,2015MNRAS.446.4039E}.  Not only its
$\mathcal{C_{\rm UV,tot}}$ color is
($\sim$$1.81/1.61$ FUV/NUV mag) redder than that expected for
late-type galaxies given its $M_{\rm
  BH}$, the galaxy also falls on the $M_{\rm BH}- \mathcal{C_{\rm
    NUV,tot}}$ red sequence defined by early-type galaxies
(Fig.~\ref{Fig2}).  Fig.~\ref{Fig2} shows that NGC 3368 lies 0.45 dex
below the best-fitting $M_{\rm BH}-\sigma$ line in the log $M_{\rm
  BH}$ direction, this might suggest that the galaxy has abnormally
low SMBH mass, rather than a discrepant redder color. We cannot rule
out the possibility where both the black hole mass and color conspire,
causing the offset in the $M_{\rm BH}-
\mathcal{C}$ diagrams (Fig. \ref{Fig2}).

\subsubsection{ NGC 4826}\label{Sec4.2.1}

The SABa galaxy NGC 4826 is also referred to as the Evil- or Black-Eye
Galaxy, attributed to its dusty disk with a radius $R \sim
50\arcsec$. It has a large-scale counter-rotating gas disk
\citep{1993PASJ...45L..47V, 1994ApJ...420..558B,
  2003A&A...407..485G}. The galaxy is the most significant outlier in
the $M_{\rm BH}-\mathcal{C_{\rm UV,tot}}$ diagrams, having
$\mathcal{C_{\rm UV,tot}}$ color ($\sim 2.68/2.04$ FUV/NUV mag) redder
than what is expected for late-type galaxies (Fig.~\ref{Fig2}). The
dust disk likely explains the deviant (redder)
UV$-[3.6]$ color for the galaxy.  Fig.~\ref{Fig2} shows that the
galaxy falls on the red, early-type sequence, despite being a
late-type galaxy.\\

\subsubsection{ NGC 5018}

The giant elliptical (E3) NGC~5018 is the most intriguing case. It is
the brightest member of the poor NGC 5018 group containing 5 galaxies
\citep{1992A&A...255...69G} known for its morphological peculiarity
including shells, ripples and dust lane \citep{1987IAUS..127..109S,
  1988ApJ...330..684K, 1997PASA...14...52M, 2000ApJ...534..650L,
  2014A&A...569A..91K}.  There is a tidal bridge between NGC~5018 and a
gas-rich spiral companion NGC~5022, detected in optical
\citep{1987IAUS..127..109S,1997PASA...14...52M} and H{\sc i}
observations \citep{1988ApJ...330..684K, 2005ApJ...623..815G}, which
was built up via accretion of gaseous material from NGC~5022 onto
NGC~5018. NGC~5018 is currently forming stars
\citep{2005ApJ...623..815G} and about 12 \% (by mass) of the stars in
the galaxy are younger than 3.4 Gyr (\citealt[see also
\citealt{2000ApJ...534..650L}]{2007MNRAS.375..381N}). The galaxy has
low metallicity and far- and mid-UV flux deficits unusual for its
luminosity and velocity dispersion \citep{1993ApJ...403..573B,
  2004A&A...423..965B}.  While \citet{1994MNRAS.270..743C} attributed
the observed unusual metallicity and UV fluxes to nuclear dust
extinction, the galaxy is an outlier from the
$\mathcal{C_{\rm FUV,tot}}-M_{\rm BH}$ diagram but not
$\mathcal{C_{\rm NUV,tot}}-M_{\rm BH}$ diagram (Fig.~\ref{Fig2}),
revealing that the offset nature of the galaxy is not because of dust
extinction but rather owing to the mixing of young and old populations
of stars at the galaxy center washing out pre-existing metallicity
\citep{1993ApJ...403..573B, 1996A&A...314..357H, 1988ApJ...330..684K,
  2004A&A...423..965B}. We also note that the galaxy offsets slightly
from the \mbox{$M_{\rm BH}- L_{\rm 3.6,tot}$} relation towards a brighter
magnitude, this may partly explain the deviant red color of the
galaxy.

\section{Appendix C}\label{AppD}
Table~\ref{Table6} presents total magnitudes, bulge-to-total
($B/T$) and disk-to-total ($D/T$) ratios, dust corrections  
and SMBH
masses for our sample galaxies. 
\newpage
\setcounter{table}{4}

\begin{table} 
\begin{sideways}
\setlength{\tabcolsep}{.08638880in}
\begin {minipage}{233mm}
\caption{Basic data for our sample of 67  \textit{GALEX}/S$^{4}$G  galaxies with directly measured supermassive black hole masses}
\label{Table6}
\begin{tabular}{@{}lclclclclclccccccccccccccc@{}}
  \hline
  \hline
Galaxy&Type & $m_{\rm FUV}$&
                                               $m_{\rm
                                                                   NUV}$&$m_{\rm
                                                                          3.6
                                                                          }$&$B/T$&$B/T$&$D/T$&$D/T$&Dust$_{\rm
                                                                                                                          corr}$&Dust$_{\rm
                                                                                                                                  corr}$&Dust$_{\rm
                                                                                                                                          corr}$&log
                                                                                                                                                  $M_{\rm
                                                                                                                                                  BH}$
  \\
&&B+D&B+D&B+D&FUV/NUV&$3.6$ &FUV/NUV&$3.6$&(B and D)$_{3.6}$ &
                                                                                                                                        (B
                                                                          and
                                                                          D)$_{\rm
                                                                          FUV}$
                                                                                                                                        &
                                                                      
                                                                                                                                                (B
                                                                                                                                          and
                                                                                                                                          D)$_{\rm
                                                                                                                                          NUV}$
                                                                                                                                                & [$M_{\sun}$]\\ 
(1)&(2)&(3)&(4)&(5)&(6)&(7)&(8)&(9)&(10)&(11)&
                                               (12)&(13)&\\
  \multicolumn{1}{c}{} \\              
  \hline      

NGC~0289&SBbc	& 	12.84$^{+0.39}_{-0.39}$ &           12.67$^{+0.40}_{-0.40}$& 	10.55$^{+0.38}_{-0.38}$ &---&0.02&0.92&0.92&0.14/0.05& ---/0.66&---/0.66&7.38$^{+0.30}_{-7.38}$\\
NGC~0428 &SABm  	& 	12.91$^{+0.37}_{-0.37}$ &  12.65$^{+0.37}_{-0.37}$   &   11.85$^{+0.35}_{-0.35}$ &---&0.00&0.93&0.93&0.13/0.05&---/0.70&---/0.70&4.48$^{-0.37}_{-4.48}$\\
NGC~0613 & SBbc&		12.99$^{+0.36}_{-0.36}$& 	12.46$^{+0.36}_{-0.36}$	& 9.61$^{+0.34}_{-0.34}$ 	&0.04&0.13&0.92&0.75&0.13/0.05&1.29/0.62&1.29/0.62&7.60$^{+0.35}_{-0.35}$\\
NGC~1042&SABc &		13.08$^{+0.39}_{-0.39}$ & 12.80$^{+0.39}_{-0.39}$& 	11.10$^{+0.37}_{-0.37}$ &---&0.02&0.97&0.97&0.13/0.05&---/0.70&---/0.70&4.40$^{+2.08}_{-4.40}$\\
NGC~1052&E4&	16.81$^{+0.24}_{-0.24}$&		15.28$^{+0.24}_{-0.24}$&	10.05$^{+0.23}_{-0.23}$&0.67&0.72&---&---&0.01/---&0.23/---&0.23/---&8.24$^{+0.29}_{-0.29}$\\
NGC~1097&SBb &	12.10$^{+0.37}_{-0.37}$& 	11.66$^{+0.37}_{-0.37}$	& 8.89$^{+0.35}_{-0.35}$ 	&0.11&0.22&0.79&0.56&0.15/0.05&1.41/0.73&1.41/0.73&8.14$^{+0.09}_{-0.09}$\\
NGC~1300&SBc & 		13.21$^{+0.40}_{-0.40}$&12.78$^{+0.40}_{-0.40}$& 		10.23$^{+0.38}_{-0.38}$ &---&0.07&0.87&0.87&0.16/0.05&---/0.76&---/0.76&7.88$^{+0.34}_{-0.34}$\\
NGC~1386&SB0-a&	15.79$^{+0.35}_{-0.35}$&		14.64$^{+0.36}_{-0.36}$&		10.51$^{+0.32}_{-0.32}$&0.26&0.36&0.70&0.58&0.32/0.10&1.54/1.00&1.54/1.00&6.07$^{+0.29}_{-0.29}$\\
NGC~1493&SBc & 12.99$^{+0.40}_{-0.40}$& 	12.76$^{+0.38}_{-0.38}$& 		11.45$^{+0.36}_{-0.36}$ &---&0.00&0.97&0.97&0.11/0.04&---/0.53&---/0.53&5.40$^{+0.51}_{-5.40}$\\
NGC~2685&SB0-a&		14.66$^{+0.39}_{-0.39}$&		14.15$^{+0.39}_{-0.39}$&	10.97$^{+0.36}_{-0.36}$	&0.36&0.46&0.54&0.41&0.23/0.08&1.82/1.17&1.83/1.18&6.59$^{+0.41}_{-6.59}$\\
NGC~2748&Sbc &		13.98$^{+0.36}_{-0.36}$& 	13.52$^{+0.36}_{-0.36}$	 & 11.05$^{+0.34}_{-0.34}$ 	&---&0.03&0.97&0.97&0.34/0.11&---/1.16&---/1.17&7.65$^{+0.24}_{-0.24}$\\
NGC~2787&SB0-a &		16.66$^{+0.54}_{-0.54}$&		14.85$^{+0.36}_{-0.36}$&		9.85$^{+0.35}_{-0.35}$&0.39&0.51&0.59&0.45&0.19/0.07&2.32/1.65&2.34/1.66&7.61$^{+0.09}_{-0.09}$\\
NGC~2903&SBbc  &		11.66$^{+0.33}_{-0.33}$& 		11.17$^{+0.36}_{-0.36}$& 		8.35$^{+0.34}_{-0.34}$ &---&0.07&0.90&0.90&0.25/0.08&---/0.96&---/0.96&7.06$^{+0.28}_{-7.06}$\\
NGC~2964&SBbc &		14.13$^{+0.35}_{-0.35}$&		13.47$^{+0.35}_{-0.35}$& 	10.72$^{+0.34}_{-0.34}$ 	&---&0.10&0.90&090&0.20/0.06&---/0.75&---/0.75&6.73$^{+0.61}_{-6.73}$\\
NGC~2974&E4&	     16.98$^{+0.23}_{-0.23}$&          15.59$^{+0.24}_{-0.24}$&                   10.52$^{+0.22}_{-0.22}$	&0.49&0.54&---&---&0.01/---&0.44/---&0.45/0.45&8.23$^{+0.09}_{-0.09}$\\
NGC~3021&Sbc&		14.60$^{+0.35}_{-0.35}$&		14.07$^{+0.35}_{-0.35}$&	11.62$^{+0.34}_{-0.34}$ &---&0.01&0.98&0.98	&0.19/0.06&---/0.71&---/0.71&7.26$^{+0.30}_{-7.26}$\\
NGC~3031&Sab&		10.07$^{+0.36}_{-0.36}$&		9.29$^{+0.45}_{-0.45}$&		6.22$^{+0.39}_{-0.39}$ &0.14&0.46&0.86&0.54&0.23/0.08&1.96/1.30&1.97/1.31&7.81$^{+0.13}_{-0.13}$\\
NGC~3079&SBc&		12.10$^{+0.36}_{-0.36}$&		11.69$^{+0.36}_{-0.36}$&		9.26$^{+0.34}_{-0.34}$ &0.07&0.24&0.93&0.77&0.57/0.24&1.81/1.73&1.82/1.73&6.40$^{+0.05}_{-0.05}$\\
NGC~3115&E-S0& 		14.45$^{+0.36}_{-0.36}$&		12.71$^{+0.35}_{-0.35}$&		8.18$^{+0.33}_{-0.33}$&0.63&0.74&0.37&0.26&0.36/0.13&1.86/1.37&1.87/1.38&8.95$^{+0.09}_{-0.09}$\\
NGC~3310&SBbc&		11.68$^{+0.36}_{-0.36}$&	11.41$^{+0.36}_{-0.36}$&	10.69$^{+0.34}_{-0.34}$ &0.16&0.33&0.83&0.63&0.13/0.05&1.33/0.65&1.33/0.66&6.70$^{+0.92}_{-6.70}$\\
NGC~3368&SBab&			13.76$^{+0.38}_{-0.38}$&	13.04$^{+0.38}_{-0.38}$& 	8.79$^{+0.36}_{-0.36}$ &---&0.07&0.92&0.92&0.15/0.05&---/0.71&---/0.71&6.88$^{+0.08}_{-0.08}$\\
NGC~3414&SB0  & 		16.39$^{+0.39}_{-0.39}$&		15.13$^{+0.36}_{-0.36}$&		10.49$^{+0.34}_{-0.34}$&0.57&0.69&0.41&0.29&0.13/0.05&1.36/0.69&1.37/0.69&8.40$^{+0.07}_{-0.07}$\\
NGC~3423&Sc & 		12.92$^{+0.37}_{-0.37}$&		12.52$^{+0.37}_{-0.37}$	 &	11.17$^{+0.35}_{-0.35}$&---&0.06&0.95&0.95&0.12/0.05&---/0.71&---/0.71&5.18$^{+0.67}_{-5.18}$\\
NGC~3489&SB0-a &	16.01$^{+0.36}_{-0.36}$&		14.14$^{+0.36}_{-0.36}$&	10.05$^{+0.33}_{-0.33}$&0.23&0.32&0.71&0.58&0.19/0.06&1.38/0.71&1.39/0.71&6.78$^{+0.05}_{-0.05}$\\
NGC~3608&E2 &	17.58$^{+0.29}_{-0.29}$&	                      15.51$^{+0.29}_{-0.29}$&		10.56$^{+0.27}_{-0.27}$	&1.00&1.00&---&---&0.01/---&0.17/---&0.18/---&8.67$^{+0.10}_{-0.10}$\\
NGC~3627&SBb&		12.07$^{+0.35}_{-0.35}$&	11.35$^{+0.35}_{-0.35}$&		8.36$^{+0.34}_{-0.34}$ &0.03&0.12&0.93&0.78&0.26/0.08&1.62/1.00&1.62/1.00&6.93$^{+0.05}_{-0.05}$\\
NGC~3642&Sbc&		13.25$^{+0.44}_{-0.44}$&		13.12$^{+0.39}_{-0.39}$&		11.51$^{+0.42}_{-0.42}$ &0.03&0.14&0.95&0.80&0.12/0.042&1.21/0.55&1.21/0.55&7.42$^{+0.04}_{-7.42}$\\
NGC~4041&Sbc&	13.45$^{+0.36}_{-0.36}$&		12.96$^{+0.36}_{-0.36}$&	10.73$^{+0.35}_{-0.35}$ &0.03&0.14&0.95&0.80&0.12/0.04&1.28/0.61&1.28/0.61&6.00$^{+0.20}_{-6.00}$\\
NGC~4051&SABb&		12.89$^{+0.36}_{-0.36}$&		12.40$^{+0.36}_{-0.36}$&		9.97$^{+0.34}_{-0.34}$ &0.03&0.15&0.96&0.80&0.13/0.04&1.26/0.59&1.27/0.59&6.10$^{+0.25}_{-0.25}$\\
NGC~4088&SABc &		12.92$^{+0.37}_{-0.37}$&			12.31$^{+0.37}_{-0.37}$&	9.72$^{+0.35}_{-0.35}$&---&0.02&0.98&0.98&0.32/0.11&---/1.08&---/1.08&6.79$^{+0.29}_{-6.79}$\\
NGC~4151&SBab&		13.16$^{+0.36}_{-0.36}$&	13.02$^{+0.36}_{-0.36}$&		10.03$^{+0.35}_{-0.35}$ &0.22&0.44&0.78&0.56&0.14/0.05&1.39/0.71&1.39/0.71&7.81$^{+0.08}_{-0.08}$\\
NGC~4203&E-S0&		15.55$^{+0.37}_{-0.37}$&		14.44$^{+0.36}_{-0.36}$&		9.96$^{+0.33}_{-0.33}$&1.0&1.0&---&---&0.11/0.04&1.20/---&1.20/---&7.82$^{+0.26}_{-0.26}$\\
NGC~4212&Sc&			14.08$^{+0.36}_{-0.36}$&		13.42$^{+0.36}_{-0.36}$&		10.71$^{+0.34}_{-0.34}$ &---&0.04&0.96&0.96&0.18/0.06&---/0.82&---/0.82&5.99$^{+0.42}_{-5.99}$\\
NGC~4245&SB0-a&		16.20$^{+0.36}_{-0.36}$&		15.14$^{+0.37}_{-0.37}$&	10.99$^{+0.34}_{-0.34}$&0.11&0.18&0.82&0.70&0.13/0.05&1.32/0.64&1.32/0.64&7.19$^{+0.48}_{-7.19}$\\
\hline      
\\
  \hline

\end{tabular} 
Notes.---Col. (1) galaxy name. Col. (2) morphological type from
HyperLeda, in good agreement with the classification in
NED. Cols. (3)-(5) dust-corrected, FUV, NUV and 3.6 {\micron}  total
($B+D$) apparent magnitudes ($m_{\rm FUV}$, $m_{\rm
  NUV}$ and $m_{\rm 3.6}$). Cols. (6)-(9) FUV, NUV and 3.6 {\micron} bulge-to-total
($B/T$) and disk-to-total ($D/T$) ratios.  Cols. (10)-(12) bulge and
disk dust corrections in $m_{\rm 3.6}$, FUV- and NUV-bands, see
Section~\ref{Sec2.3}. Col. (13) supermassive black hole mass from
\cite{2016ApJ...831..134V}.
\end {minipage}
\end{sideways}
\end{table}

\begin{table} 
\begin{sideways}
\setcounter{table}{4} 
\setlength{\tabcolsep}{.08638880in}
\begin {minipage}{233mm}
\caption{({\it Continued})}
\label{Ta0000000}
\begin{tabular}{@{}lclclclclclccccccccccccccc@{}}
  \hline
  \hline
Galaxy&Type & $m_{\rm FUV}$& $m_{\rm NUV}$& 
                                                                                  $m_{\rm
                                                                                  3.6
                                                                                  }$&$B/T$&$B/T$&$D/T$&$D/T$&Dust$_{\rm
                                                                                                                          corr}$&Dust$_{\rm
                                                                                                                                  corr}$&Dust$_{\rm
                                                                                                                                          corr}$&log
                                                                                                                                                  $M_{\rm
                                                                                                                                                  BH}$
  \\
&&B+D&B+D&B+D&FUV/NUV&$3.6$ &FUV/NUV&$3.6$& (B and D)$_{3.6}$ &
                                                                                                                                        (B
                                                                           and
                                                                           D)$_{\rm
                                                                           FUV}$
                                                                                                                                        &
                                                                      
                                                                                                                                                (B
                                                                                                                                          and
                                                                                                                                          D)$_{\rm
                                                                           NUV}$
                                                                                                                                                & [$M_{\sun}$]&\\ 
(1)&(2)&(3)&(4)&(5)&(6)&(7)&(8)&(9)&(10)&(11)& (12)&(13)&\\         
  \multicolumn{1}{c}{} \\              
  \hline 
NGC~4258&SBbc&			11.10$^{+0.36}_{-0.36}$&	  10.69$^{+0.36}_{-0.36}$&		7.89$^{+0.35}_{-0.35}$ &---&0.07&0.96&0.96	&0.32/0.10&---/1.02&---/1.02&7.58$^{+0.03}_{-0.03}$\\
NGC~4278&E1-2&		16.00$^{+0.24}_{-0.24}$&	14.99$^{+0.24}_{-0.24}$&		9.79$^{+0.22}_{-0.22}$&0.65&0.71&---&---&0.01/---&0.25/---&0.26/---&7.96$^{+0.27}_{-0.27}$\\
NGC~4314&SBa&	15.16$^{+0.35}_{-0.35}$&		14.27$^{+0.35}_{-0.35}$&		10.34$^{+0.34}_{-0.34}$ 	&0.08&0.17&0.79&0.57&0.12/0.05&1.32/0.67&1.33/0.67&6.91$^{+0.30}_{-6.91}$\\
NGC~4321&SABb&		12.17$^{+0.36}_{-0.36}$&		11.56$^{+0.36}_{-0.36}$&		8.99$^{+0.34}_{-0.34}$ &---&0.10&0.90&0.90&0.12/0.05&---/0.67&---/0.67&6.67$^{+0.17}_{-6.67}$\\
NGC~4371&SB0-a&		16.90$^{+0.54}_{-0.54}$&		15.54$^{+0.36}_{-0.36}$&		10.44$^{+0.32}_{-0.32}$&0.16&0.23&0.74&0.62&0.12/0.05&1.41/0.76&1.41/0.76&6.84$^{+0.07}_{-0.07}$\\
NGC~4374&E1&      	15.77$^{+0.24}_{-0.24}$&		14.31$^{+0.23}_{-0.23}$&		8.74$^{+0.22}_{-0.22}$&0.60&0.65&---&---&0.01/---&0.35/---&0.35/---&8.97$^{+0.05}_{-0.05}$\\
NGC~4388&SBb&		13.42$^{+0.35}_{-0.35}$&		12.60$^{+0.36}_{-0.36}$&	 	10.11$^{+0.34}_{-0.34}$ &0.06&0.16&0.94&0.84	&0.56/0.24&1.97/1.84&1.97/1.85&6.86$^{+0.04}_{-0.04}$\\
NGC~4472&E2&		14.78$^{+0.21}_{-0.21}$&		13.60$^{+0.22}_{-0.22}$&    	7.91$^{+0.20}_{-0.20}$&0.58&0.63&---&---&0.01/---&0.19/---&0.19/---&9.40$^{+0.04}_{-0.04}$\\
NGC~4501&Sb&		13.04$^{+0.35}_{-0.35}$&		12.27$^{+0.35}_{-0.35}$&		8.76$^{+0.39}_{-0.39}$ &---&0.06&0.94&0.94	&0.21/0.07&---/0.92&---/0.93&7.30$^{+0.08}_{-0.08}$\\
NGC~4548&SBb&		14.07$^{+0.37}_{-0.37}$&		13.48$^{+0.36}_{-0.36}$    &	9.82$^{+0.35}_{-0.35}$ &0.03&0.12&0.93&0.78&0.13/0.05&1.45/0.78&1.46/0.79&7.25$^{+0.29}_{-0.29}$\\
NGC~4593&SBb  &		14.26$^{+0.37}_{-0.37}$&		13.66$^{+0.38}_{-0.38}$&		10.74$^{+0.36}_{-0.36}$ &0.10&0.16&0.81&0.61&0.13/0.05&1.36/0.69&1.37/0.69&6.86$^{+0.21}_{-0.21}$\\
NGC~4594&Sa&		13.35$^{+0.36}_{-0.36}$&		11.86$^{+0.37}_{-0.37}$&	7.20$^{+0.35}_{-0.35}$ &0.49&0.83&0.51&0.17&0.31/0.10&1.84/1.27&1.84/1.28&8.82$^{+0.05}_{-0.05}$\\
NGC~4596&SB0-a&		16.72$^{+0.36}_{-0.36}$&		15.29$^{+0.35}_{-0.35}$&		10.06$^{+0.32}_{-0.32}$&0.13&0.19&0.77&0.65&0.13/0.05&1.32/0.64&1.32/0.65&7.89$^{+0.26}_{-0.26}$\\
NGC~4698&Sab&		14.89$^{+0.38}_{-0.38}$&	13.84$^{+0.36}_{-0.36}$&		10.08$^{+0.35}_{-0.35}$ &0.22&0.44&0.78&0.56	&0.17/0.05&1.43/0.74&1.43/0.75&7.76$^{+0.16}_{-0.16}$\\
NGC~4736&Sab&		11.51$^{+0.36}_{-0.36}$&	11.07$^{+0.36}_{-0.36}$&		8.01$^{+0.34}_{-0.34}$ &0.24&0.42&0.54&0.21&0.12/0.04&1.27/0.61&1.28/0.61&6.78$^{+0.12}_{-0.12}$\\
NGC~4800&Sb& 	14.64$^{+0.34}_{-0.34}$&	13.96$^{+0.35}_{-0.35}$&		10.94$^{+0.34}_{-0.34}$ &0.04&0.17&0.95&0.80&0.13/0.05&1.28/0.60&1.28/0.61&7.02$^{+0.53}_{-7.02}$\\
NGC~4826&SABa&		12.94$^{+0.36}_{-0.36}$&		11.90$^{+0.36}_{-0.36}$&7.91$^{+0.34}_{-0.34}$	&0.12&0.28&0.87&0.68&0.21/0.07&1.62/0.96&1.63/0.96&6.05$^{+0.13}_{-0.13}$\\
NGC~5005&SABb&		14.11$^{+0.35}_{-0.35}$&		13.12$^{+0.35}_{-0.35}$&	9.17$^{+0.33}_{-0.33}$&0.06&0.15&0.88&0.69&0.24/0.07&1.44/0.81&1.45/0.81&8.27$^{+0.23}_{-0.23}$\\
NGC~5018&E3&		18.07$^{+0.63}_{-0.63}$&		15.32$^{+0.25}_{-0.25}$&		10.28$^{+0.27}_{-0.27}$&0.83&0.88&---&---&0.02/---&0.79/---&0.80/---&8.02$^{+0.08}_{-0.08}$\\
NGC~5055&Sbc&		11.89$^{+0.36}_{-0.36}$&	11.23$^{+0.36}_{-0.36}$&		7.97$^{+0.34}_{-0.34}$&0.03&0.17&0.97&0.81&0.19/0.06&1.40/0.73&1.40/0.73&8.92$^{+0.10}_{-0.10}$\\
NGC~5194&SABb&		10.60$^{+0.35}_{-0.35}$&			10.04$^{+0.34}_{-0.34}$&		7.49$^{+0.32}_{-0.32}$&---&0.10&0.87&0.88&0.17/0.06&---/0.83&---/0.83&5.96$^{+0.36}_{-5.96}$\\
NGC~5248&SBb&	12.86$^{+0.36}_{-0.36}$&		12.43$^{+0.36}_{-0.36}$&		9.65$^{+0.33}_{-0.33}$&0.05&0.20&0.95&0.80&0.14/0.05&1.37/0.69&1.37/0.69&6.30$^{+0.38}_{-0.38}$\\
NGC~5273&S0 	&	17.46$^{+0.36}_{-0.36}$&		15.77$^{+0.36}_{-0.36}$&		11.24$^{+0.34}_{-0.34}$&---&0.09&0.99&0.91&0.11/0.04&---/0.53&---/0.53&6.61$^{+0.27}_{-0.27}$\\
NGC~5347&SBab& 		15.27$^{+0.37}_{-0.37}$&		14.89$^{+0.38}_{-0.38}$&		11.99$^{+0.34}_{-0.34}$&0.17&0.33&0.80&0.59&0.13/0.05&1.31/0.64&1.32/0.64&7.21$^{+0.42}_{-7.21}$\\
NGC~5427&SABc&		13.59$^{+0.37}_{-0.37}$&		13.04$^{+0.37}_{-0.37}$&		10.99$^{+0.35}_{-0.35}$&---&0.10&0.90&0.90&0.12/0.05&---/0.68&---/0.69&7.58$^{+0.30}_{-7.58}$\\
NGC~5457&SABc &			9.65$^{+0.64}_{-0.64}$&		9.43$^{+0.63}_{-0.63}$&		7.76$^{+0.61}_{-0.61}$&---&0.05&0.95& 0.95&0.11/0.04&---/0.53&---/0.53&6.41$^{+0.08}_{-6.41}$\\
NGC~5576&E3 &  	17.35$^{+0.26}_{-0.26}$&		15.43$^{+0.24}_{-0.24}$&		10.42$^{+0.23}_{-0.23}$&0.69&0.74&---&---&0.01/---&0.25/---&0.26/---&8.44$^{+0.13}_{-0.13}$\\
NGC~5728&SBa&		14.75$^{+0.40}_{-0.40}$&		14.20$^{+0.38}_{-0.38}$&		10.99$^{+0.38}_{-0.38}$&0.12&0.26&0.72&0.39&0.20/0.07&2.09/1.41&2.10/1.42&8.05$^{+0.29}_{-0.29}$\\
NGC~5846&E0-1&		16.24$^{+0.26}_{-0.26}$&		15.18$^{+0.25}_{-0.25}$&		9.46$^{+0.21}_{-0.21}$&0.60&0.65&---&---&0.01/---&0.46/---&0.46/---&9.04$^{+0.06}_{-0.06}$\\
NGC~5879&SBbc&		13.14$^{+0.35}_{-0.35}$&		12.79$^{+0.35}_{-0.35}$&		10.93$^{+0.34}_{-0.34}$&0.16&0.46&0.83&0.51&0.44/0.16&1.67/1.33&1.67/1.33&6.62$^{+0.28}_{-6.62}$\\
NGC~5921&SBbc&		13.23$^{+0.37}_{-0.37}$&		12.76$^{+0.37}_{-0.37}$&		10.80$^{+0.35}_{-0.35}$	&0.01&0.11&0.99&0.83&0.13/0.05&1.45/0.78&1.45/0.79&7.07$^{+0.42}_{-7.07}$\\
NGC~7418&SBc&		13.35$^{+0.36}_{-0.36}$&	12.94$^{+0.35}_{-0.35}$&		10.82$^{+0.33}_{-0.33}$&---&0.02&0.97&0.97&0.13/0.04&---/0.61&---/0.61&5.18$^{+1.78}_{-5.18}$\\
NGC~7582&SBab &		14.48$^{+0.36}_{-0.36}$&		13.57$^{+0.36}_{-0.36}$&		9.97$^{+0.34}_{-0.34}$	&0.10&0.25&0.78&0.45&0.29/0.09&1.51/0.93&1.51/0.93&7.74$^{+0.20}_{-0.20}$\\

\hline      
\\
  \hline 

\end{tabular} 
                                     Table~\ref{Table6} {\it continued.}
\end {minipage}
\end{sideways}
\end{table}

\section{Appendix D}\label{AppE}

We use our $M_{\rm
  BH}- \mathcal{C_{\rm FUV,tot}}$ and  $M_{\rm BH}- \mathcal{C_{\rm
    NUV,tot}}$ relations (Table~\ref{Table1}) together with  the  appropriate asymptotic  galaxy 
colors  derived based on the asymptotic FUV, NUV 
and 3.6 $\micron$  magnitudes from \citet[][their Table~1]{2018ApJS..234...18B} to predict tentative BH masses in a sample of 1382 
\textit{GALEX}/S$^{4}$G galaxies with no measured BH masses
(Table~\ref{Table7}).  From the \citet{2018ApJS..234...18B} sample, we
excluded galaxies that are:  highly inclined, dust obscured and with
prominent large-scale bars and rings. In contrast to the total (B+D)
magnitudes in this paper (Table~\ref{Table6}), the
\citet{2018ApJS..234...18B}  asymptotic magnitudes, which were not
corrected for internal dust attenuation, contain additional fluxes
from bars, rings and nuclear components as such we caution about
overinterpreting these predicted BH masses. Furthermore, {\it Chandra}
X-ray data or/and high-resolution radio data are important to confirm the presence of a central black
hole in the low-mass ($M_{*}\la 10^{10} M_{\sun}$) \textit{GALEX}/S$^{4}$G galaxies.

\begin{table} 
\begin{sideways}
\setlength{\tabcolsep}{.038880in}
\begin {minipage}{233mm}
\caption{Predicted black hole masses for 1382 \textit{GALEX}/S$^{4}$G galaxies with no measured  BH masses}
\label{Table7}
\begin{tabular}{@{}lclclclclclccccccccccccccc@{}}
  \hline
  \hline
Galaxy&Type & $M_{\rm BH}$&
                                               $M_{\rm BH}$&Galaxy&Type & $M_{\rm BH}$&
                                               $M_{\rm BH}$&Galaxy&Type & $M_{\rm BH}$&
                                               $M_{\rm BH}$&Galaxy&Type & $M_{\rm BH}$&
                                               $M_{\rm BH}$&Galaxy&Type & $M_{\rm BH}$&
                                               $M_{\rm BH}$\\
& &(FUV)& (NUV)&& &(FUV)& (NUV)&& &(FUV)&(NUV)&& &(FUV)& (NUV)&& &(FUV)&(NUV)\\                                               
(1)& (2)&(3)& (4)&(1)& (2)&(3)& (4)&(1)& (2)&(3)& (4)&(1)& (2)&(3)& (4)&(1)& (2)&(3)& (4)\\
  \hline                                                                                                                                                        
ESO013-016     &   7.5 &  5.35 &  5.42 & ESO347-008     &   9.0 &  4.50 &  4.16 & ESO440-046     &   8.7 &  5.91 &  5.50 & ESO548-032     &   9.0 &  4.31 &  3.94 & IC1125         &   7.3 &  5.46 &  5.50 \\
ESO026-001     &   5.9 &  5.03 &  4.96 & ESO347-017     &   9.0 &  4.83 &  4.52 & ESO440-049     &   6.9 &  5.52 &  5.39 & ESO548-082     &   4.2 &  5.33 &  4.66 & IC1151         &   5.1 &  6.29 &  6.13 \\
ESO027-001     &   5.0 &  6.42 &  6.45 & ESO347-029     &   7.9 &  5.02 &  4.78 & ESO441-014     &   4.2 &  5.80 &  5.48 & ESO549-002     &   9.6 &  5.28 &  4.97 & IC1158         &   5.1 &  6.38 &  6.34 \\
ESO027-008     &   5.1 &  7.98 &  8.15 & ESO355-026     &   4.2 &  6.13 &  6.16 & ESO441-017     &   4.4 &  5.91 &  5.61 & ESO549-018     &   4.8 &  7.41 &  7.80 & IC1210         &   2.1 &  7.07 &  7.17 \\
ESO048-017     &   6.9 &  5.10 &  4.86 & ESO356-018     &   5.0 &  5.36 &  5.24 & ESO442-013     &   5.9 &  4.50 &  3.91 & ESO549-035     &   6.0 &  4.93 &  4.59 & IC1251         &   6.0 &  5.20 &  4.81 \\
ESO079-005     &   7.0 &  5.08 &  4.75 & ESO357-007     &   9.0 &  5.18 &  4.93 & ESO443-069     &   6.0 &  5.65 &  5.35 & ESO550-024     &   6.7 &  5.28 &  5.06 & IC1265         &   2.0 &  6.49 &  6.64 \\
ESO079-007     &   4.0 &  5.32 &  5.12 & ESO357-012     &   7.0 &  5.03 &  4.83 & ESO443-079     &   9.8 &  4.84 &  4.24 & ESO551-016     &   4.1 &  5.15 &  4.72 & IC1438         &   1.2 &  7.56 &  7.95 \\
ESO085-014     &   9.0 &  4.75 &  4.08 & ESO358-005     &   9.0 &  5.56 &  5.40 & ESO443-085     &   8.0 &  6.26 &  5.87 & ESO551-031     &   6.7 &  5.92 &  5.55 & IC1447         &   3.2 &  6.75 &  6.85 \\
ESO085-047     &   9.0 &  3.81 &  3.19 & ESO358-015     &   8.9 &  5.86 &  5.70 & ESO444-033     &   8.4 &  5.65 &  5.79 & ESO553-017     &   5.0 &  4.77 &  4.50 & IC1532         &   4.0 &  5.85 &  5.86 \\
ESO119-016     &   9.7 &  5.12 &  4.82 & ESO358-020     &   9.3 &  6.69 &  6.60 & ESO444-037     &   7.5 &  4.22 &  3.61 & ESO572-012     &   4.8 &  5.39 &  5.26 & IC1553         &   7.0 &  6.93 &  7.24 \\
ESO120-012     &   7.1 &  4.43 &  4.03 & ESO358-025     &  -2.6 &  5.05 &  5.35 & ESO445-089     &   6.7 &  5.37 &  5.06 & ESO572-018     &   4.7 &  5.49 &  5.33 & IC1555         &   7.0 &  5.18 &  4.72 \\
ESO145-025     &   9.0 &  4.08 &  3.69 & ESO358-051     &   0.0 &  6.33 &  6.23 & ESO476-010     &   8.8 &  4.83 &  4.70 & ESO572-030     &   9.1 &  5.24 &  5.06 & IC1558         &   9.0 &  5.06 &  4.74 \\
ESO149-001     &   8.0 &  4.95 &  4.39 & ESO358-054     &   8.0 &  5.02 &  4.79 & ESO479-004     &   8.5 &  5.26 &  5.04 & ESO576-003     &   6.7 &  5.41 &  4.88 & IC1574         &   9.9 &  5.51 &  5.08 \\
ESO149-003     &   9.7 &  3.56 &  3.07 & ESO358-060     &   9.9 &  3.74 &  3.30 & ESO480-025     &   8.8 &  5.71 &  5.03 & ESO576-005     &   8.0 &  5.96 &  7.13 & IC1596         &   2.0 &  6.04 &  5.91 \\
ESO150-005     &   7.8 &  5.07 &  4.86 & ESO358-063     &   7.7 &  8.37 &  8.55 & ESO481-014     &   8.9 &  4.42 &  3.97 & ESO576-008     &  -1.6 &  4.56 &  5.43 & IC1727         &   8.9 &  4.91 &  4.56 \\
ESO154-023     &   8.9 &  4.65 &  4.75 & ESO359-003     &   1.4 &  6.28 &  5.88 & ESO482-035     &   2.2 &  6.81 &  6.81 & ESO576-017     &   7.0 &  5.75 &  5.57 & IC1892         &   7.7 &  5.01 &  4.71 \\
ESO159-025     &  10.0 &  4.84 &  4.37 & ESO359-022     &   8.6 &  4.14 &  3.70 & ESO483-008     &   9.0 &  4.70 &  4.73 & ESO576-032     &   5.8 &  6.56 &  6.53 & IC1914         &   6.9 &  5.10 &  4.93 \\
ESO187-035     &   9.0 &  4.78 &  4.57 & ESO361-009     &   9.9 &  4.23 &  3.65 & ESO485-021     &   8.2 &  4.27 &  3.88 & ESO576-059     &   9.8 &  5.28 &  5.47 & IC1933         &   6.2 &  5.00 &  4.74 \\
ESO187-051     &   9.0 &  4.79 &  4.52 & ESO361-019     &   1.1 &  5.10 &  4.77 & ESO486-003     &   8.9 &  4.81 &  4.71 & ESO580-022     &   7.7 &  5.73 &  5.61 & IC1952         &   4.0 &  7.72 &  7.84 \\
ESO202-041     &   8.9 &  3.86 &  3.34 & ESO362-009     &   9.1 &  4.57 &  4.16 & ESO486-021     &   4.2 &  4.43 &  4.01 & ESO580-030     &   5.5 &  5.56 &  5.50 & IC1954         &   3.2 &  6.58 &  6.62 \\
ESO234-043     &   8.9 &  5.19 &  4.75 & ESO362-019     &   8.9 &  4.17 &  3.66 & ESO501-080     &   4.9 &  5.01 &  4.46 & ESO580-041     &   4.1 &  5.94 &  5.80 & IC1959         &   8.5 &  4.71 &  4.38 \\
ESO234-049     &   4.1 &  5.28 &  5.00 & ESO399-025     &   0.4 &  7.73 &  7.75 & ESO502-023     &  10.0 &  4.40 &  3.79 & ESO581-025     &   6.9 &  7.34 &  7.44 & IC1986         &   8.8 &  4.65 &  4.37 \\
ESO237-052     &   7.5 &  5.27 &  5.06 & ESO400-025     &   7.8 &  5.64 &  5.26 & ESO503-022     &   9.8 &  5.81 &  5.85 & ESO582-004     &   5.0 &  5.82 &  5.25 & IC1993         &   3.0 &  7.91 &  8.13 \\
ESO238-018     &   5.5 &  4.62 &  4.35 & ESO402-025     &   4.8 &  5.67 &  5.51 & ESO504-010     &   8.7 &  5.29 &  4.79 & ESO601-007     &   8.2 &  6.30 &  6.27 & IC2007         &   3.7 &  6.54 &  6.64 \\
ESO245-005     &   9.9 &  3.64 &  3.04 & ESO402-026     &   1.9 &  8.42 &  8.89 & ESO504-024     &   9.0 &  5.93 &  5.76 & ESO601-031     &   9.9 &  4.41 &  3.97 & IC2032         &   9.9 &  4.36 &  4.01 \\
ESO245-007     &   9.5 &  6.22 &  5.17 & ESO402-030     &  -0.8 &  2.77 &  3.69 & ESO504-028     &   7.0 &  6.26 &  6.16 & ESO602-003     &   9.9 &  4.42 &  3.92 & IC2056         &   3.8 &  6.48 &  6.37 \\
ESO248-002     &   6.9 &  6.34 &  6.26 & ESO403-024     &   6.0 &  6.28 &  6.13 & ESO505-002     &   9.7 &  5.08 &  4.67 & ESO602-015     &   6.7 &  4.20 &  3.68 & IC2135         &   5.9 &  6.83 &  6.93 \\
ESO249-008     &   1.3 &  5.37 &  5.24 & ESO404-012     &   5.1 &  6.39 &  6.38 & ESO505-008     &   4.0 &  5.57 &  5.24 & ESO602-030     &   6.8 &  5.30 &  5.31 & IC2574         &   8.9 &  4.92 &  4.74 \\
ESO249-026     &   7.2 &  4.07 &  3.44 & ESO404-017     &   7.6 &  5.00 &  4.90 & ESO505-009     &   5.0 &  5.19 &  4.52 & ESO603-031     &   1.0 &  5.82 &  5.75 & IC2604         &   9.1 &  4.73 &  4.31 \\
ESO249-035     &   5.9 &  4.28 &  3.68 & ESO404-027     &   5.0 &  6.55 &  6.62 & ESO505-013     &   8.5 &  4.74 &  4.24 & IC0167         &   5.0 &  5.16 &  4.90 & IC2627         &   4.6 &  6.42 &  6.31 \\
ESO249-036     &  10.0 &  4.66 &  4.31 & ESO406-042     &   8.8 &  4.49 &  4.13 & ESO506-029     &   6.0 &  5.56 &  5.41 & IC0223         &  10.0 &  4.17 &  3.66 & IC2764         &  -0.1 &  4.21 &  5.53 \\
ESO285-048     &   5.9 &  5.73 &  5.55 & ESO407-009     &   6.7 &  6.31 &  6.35 & ESO507-065     &   8.6 &  5.36 &  5.67 & IC0600         &   7.8 &  4.98 &  4.63 & IC2828         &   3.6 &  5.25 &  4.75 \\
ESO286-044     &  -0.8 &  6.33 &  6.25 & ESO407-014     &   5.1 &  5.72 &  5.62 & ESO508-007     &   7.0 &  8.01 &  3.88 & IC0718         &   9.8 &  6.01 &  5.81 & IC2969         &   3.9 &  5.29 &  5.30 \\
ESO287-037     &   8.5 &  5.61 &  5.64 & ESO407-018     &   9.8 &  4.54 &  4.64 & ESO508-011     &   6.6 &  5.90 &  5.51 & IC0719         &  -2.0 &  4.84 &  6.05 & IC2995         &   5.5 &  6.36 &  6.15 \\
ESO288-013     &   9.0 &  5.46 &  5.26 & ESO408-012     &   6.8 &  5.20 &  4.93 & ESO508-051     &   7.9 &  4.94 &  4.72 & IC0749         &   5.9 &  6.38 &  6.39 & IC2996         &   3.5 &  7.03 &  6.78 \\
ESO289-026     &   7.9 &  4.98 &  4.72 & ESO409-015     &   5.4 &  3.70 &  3.22 & ESO510-058     &   5.9 &  6.68 &  6.62 & IC0755         &   3.5 &  5.10 &  4.78 & IC3021         &   9.0 &  6.49 &  6.35 \\
ESO291-024     &   5.0 &  5.57 &  5.40 & ESO410-012     &   4.6 &  4.35 &  3.84 & ESO510-059     &   5.9 &  4.74 &  4.24 & IC0758         &   6.0 &  5.75 &  5.57 & IC3023         &   9.7 &  4.65 &  4.17 \\
ESO293-034     &   6.2 &  6.89 &  7.06 & ESO410-018     &   8.9 &  4.51 &  4.16 & ESO532-014     &   5.9 &  4.32 &  3.92 & IC0764         &   5.0 &  5.92 &  5.95 & IC3033         &   7.7 &  5.52 &  5.14 \\
ESO293-045     &   7.8 &  4.14 &  3.69 & ESO411-013     &   9.0 &  5.52 &  5.06 & ESO532-022     &   5.5 &  5.18 &  4.99 & IC0769         &   4.0 &  6.36 &  6.16 & IC3044         &   6.0 &  5.67 &  5.33 \\
ESO298-015     &   6.2 &  5.50 &  5.33 & ESO411-026     &   9.0 &  5.32 &  5.57 & ESO533-028     &   4.6 &  6.06 &  5.99 & IC0776         &   7.9 &  4.84 &  4.64 & IC3059         &   9.8 &  6.07 &  5.84 \\
ESO298-023     &   6.0 &  4.68 &  4.35 & ESO418-008     &   7.7 &  5.13 &  4.88 & ESO539-007     &   8.7 &  4.67 &  4.32 & IC0796         &  -0.2 &  6.13 &  5.66 & IC3102         &  -0.9 &  8.09 &  8.26 \\
ESO300-014     &   8.9 &  5.68 &  5.50 & ESO420-009     &   5.0 &  5.96 &  6.02 & ESO541-004     &   4.2 &  6.56 &  6.68 & IC0797         &   6.0 &  6.37 &  6.28 & IC3105         &   9.8 &  4.00 &  3.37 \\
ESO302-021     &   5.0 &  4.43 &  4.08 & ESO421-019     &   9.0 &  4.98 &  4.77 & ESO541-005     &   8.0 &  5.28 &  5.40 & IC0800         &   5.2 &  6.56 &  6.55 & IC3115         &   5.5 &  5.65 &  5.47 \\
ESO305-009     &   8.0 &  4.27 &  3.74 & ESO422-005     &   9.5 &  4.10 &  3.68 & ESO544-030     &   7.9 &  6.00 &  5.80 & IC0851         &   3.8 &  6.13 &  5.99 & IC3258         &   9.7 &  4.76 &  4.39 \\
ESO305-017     &   9.9 &  4.73 &  4.31 & ESO422-033     &   9.7 &  4.66 &  4.42 & ESO545-002     &   8.9 &  5.27 &  5.18 & IC0863         &   0.4 &  7.51 &  7.43 & IC3259         &   7.9 &  7.17 &  7.32 \\
ESO340-017     &   8.0 &  5.50 &  5.72 & ESO423-002     &   6.5 &  6.19 &  6.09 & ESO545-005     &   7.6 &  6.83 &  6.79 & IC1014         &   8.1 &  5.80 &  5.55 & IC3267         &   5.9 &  7.07 &  7.10 \\
ESO340-042     &   7.8 &  5.31 &  5.41 & ESO438-017     &   4.9 &  5.75 &  5.77 & ESO545-016     &   9.0 &  5.10 &  4.90 & IC1024         &  -2.0 &  4.35 &  5.66 & IC3268         &  10.0 &  5.39 &  5.11 \\
ESO341-032     &   9.0 &  5.18 &  4.92 & ESO440-004     &   7.9 &  4.87 &  4.20 & ESO546-034     &   8.8 &  4.31 &  3.91 & IC1055         &   3.1 &  8.03 &  8.24 & IC3322         &   6.3 &  7.82 &  7.97 \\
ESO342-050     &   5.0 &  6.80 &  6.99 & ESO440-011     &   6.9 &  5.82 &  5.55 & ESO547-005     &  10.0 &  5.27 &  5.14 & IC1066         &   3.2 &  6.97 &  7.04 & IC3355         &   9.7 &  4.32 &  3.88 \\
ESO345-046     &   7.0 &  5.16 &  5.11 & ESO440-044     &   8.7 &  5.19 &  4.67 & ESO547-020     &   9.5 &  4.91 &  4.75 & IC1067  &   3.0 &  7.06 &  7.24 & IC3356         &   9.7 &  4.52 &  4.14 \\
 \hline        
 \hline      
\end{tabular}                                                                                                                 
Notes.---Col. (1) galaxy name. Col. (2)  galaxy T-type  from
HyperLeda. Cols. (3 and 4) Tentative BH masses (log $M_{\rm BH}/M_{\sun}$) estimated using our $M_{\rm
  BH}- \mathcal{C_{\rm FUV,tot}}$ and  $M_{\rm BH}- \mathcal{C_{\rm
    NUV,tot}}$ relations  (Table~\ref{Table1}) and the asymptotic FUV, NUV 
and 3.6 $\micron$  magnitudes from \citet [][their Table~1]{2018ApJS..234...18B}.  We adopt a typical uncertainty of 0.85 dex
on ${\rm log} (M_{\rm BH}$) for these predicted BH masses.         
                  
\end {minipage}
\end{sideways}
\end{table}

\begin{table} 
\begin{sideways}
\setcounter{table}{5} 
\setlength{\tabcolsep}{.038880in}
\begin {minipage}{233mm}
\caption{({\it Continued})}
\label{Table7}
\begin{tabular}{@{}lclclclclclccccccccccccccc@{}}
  \hline
  \hline
Galaxy&Type & $M_{\rm BH}$&
                                               $M_{\rm BH}$&Galaxy&Type & $M_{\rm BH}$&
                                               $M_{\rm BH}$&Galaxy&Type & $M_{\rm BH}$&
                                               $M_{\rm BH}$&Galaxy&Type & $M_{\rm BH}$&
                                               $M_{\rm BH}$&Galaxy&Type & $M_{\rm BH}$& 
                                               $M_{\rm BH}$\\
& &(FUV)& (NUV)&& &(FUV)& (NUV)&& &(FUV)&(NUV)&& &(FUV)& (NUV)&& &(FUV)&(NUV)\\                                               
(1)& (2)&(3)& (4)&(1)& (2)&(3)& (4)&(1)& (2)&(3)& (4)&(1)& (2)&(3)& (4)&(1)& (2)&(3)& (4)\\
  \hline                                                                                                                                                        
IC3371         &   6.0 &  4.92 &  4.46 & NGC0470        &   3.1 &  7.22 &  7.44 & NGC1326        &  -0.8 &  4.94 &  6.64 & NGC2591        &   5.9 &  8.41 &  8.84 & NGC3147        &   3.9 &  8.16 &  8.48 \\
IC3391         &   5.9 &  6.14 &  6.10 & NGC0473        &  -0.3 &  3.88 &  4.96 & NGC1326A       &   8.9 &  4.60 &  4.35 & NGC2608        &   3.3 &  7.53 &  7.71 & NGC3155        &   3.5 &  6.46 &  6.62 \\
IC3476         &   9.6 &  5.55 &  5.33 & NGC0474        &  -2.0 &  8.64 &  7.36 & NGC1326B       &   8.9 &  4.70 &  4.34 & NGC2633        &   3.0 &  7.51 &  7.88 & NGC3162        &   4.6 &  5.95 &  5.90 \\
IC3517         &   8.5 &  5.85 &  5.72 & NGC0485        &   6.0 &  6.95 &  6.82 & NGC1337        &   6.0 &  6.09 &  6.04 & NGC2701        &   5.2 &  6.08 &  6.02 & NGC3165        &   8.5 &  5.95 &  5.76 \\
IC3576         &   8.6 &  4.83 &  4.50 & NGC0493        &   5.9 &  6.11 &  5.98 & NGC1338        &   3.1 &  7.05 &  7.04 & NGC2710        &   3.1 &  6.46 &  6.60 & NGC3169        &   1.2 &  8.47 &  8.85 \\
IC3583         &   9.6 &  4.94 &  4.50 & NGC0578        &   5.0 &  6.02 &  5.97 & NGC1339        &  -4.3 &  8.46 &  8.39 & NGC2712        &   3.1 &  7.15 &  7.44 & NGC3182        &   0.5 &  8.73 &  9.03 \\
IC3687         &   9.9 &  4.01 &  3.51 & NGC0584        &  -4.7 &  9.77 &  9.41 & NGC1341        &   1.3 &  6.56 &  6.42 & NGC2726        &   1.0 &  8.25 &  8.42 & NGC3187        &   5.0 &  5.66 &  5.58 \\
IC3718         &  -5.0 &  4.58 &  4.18 & NGC0600        &   7.0 &  5.75 &  5.62 & NGC1345        &   5.2 &  5.34 &  5.18 & NGC2731        &   4.2 &  7.06 &  6.92 & NGC3193        &  -4.8 &  9.14 &  8.47 \\
IC3881         &   5.8 &  5.27 &  5.11 & NGC0672        &   6.0 &  5.90 &  5.69 & NGC1357        &   1.9 &  8.15 &  8.60 & NGC2735        &   3.0 &  7.63 &  8.27 & NGC3206        &   6.0 &  5.09 &  4.75 \\
IC4182         &   8.8 &  5.04 &  4.72 & NGC0691        &   4.0 &  7.76 &  7.97 & NGC1359        &   8.5 &  4.17 &  3.79 & NGC2742        &   5.3 &  7.24 &  7.37 & NGC3220        &   3.0 &  5.35 &  5.18 \\
IC4213         &   5.8 &  5.92 &  5.75 & NGC0723        &   3.8 &  5.98 &  5.98 & NGC1365        &   3.2 &  7.28 &  7.62 & NGC2743        &   7.9 &  6.92 &  6.89 & NGC3225        &   5.8 &  5.80 &  5.75 \\
IC4216         &   5.8 &  5.90 &  5.98 & NGC0755        &   3.4 &  5.86 &  5.77 & NGC1367        &   1.1 &  7.79 &  8.34 & NGC2750        &   5.3 &  6.46 &  6.45 & NGC3239        &   9.8 &  4.32 &  3.80 \\
IC4221         &   4.7 &  6.43 &  6.26 & NGC0772        &   3.1 &  7.97 &  8.17 & NGC1385        &   5.9 &  6.29 &  6.16 & NGC2764        &  -1.8 &  5.42 &  6.42 & NGC3246        &   7.9 &  5.84 &  5.50 \\
IC4231         &   4.1 &  7.50 &  7.73 & NGC0784        &   7.9 &  5.30 &  4.83 & NGC1390        &   0.4 &  6.79 &  6.71 & NGC2768        &  -4.5 &  9.56 &  8.22 & NGC3252        &   7.4 &  7.05 &  7.51 \\
IC4237         &   3.5 &  7.19 &  7.21 & NGC0803        &   5.3 &  6.49 &  6.63 & NGC1411        &  -3.0 &  9.16 &  8.55 & NGC2776        &   5.2 &  6.26 &  6.21 & NGC3254        &   4.0 &  6.85 &  7.10 \\
IC4247         &   6.9 &  5.10 &  4.48 & NGC0855        &  -4.8 &  3.07 &  3.69 & NGC1421        &   4.1 &  6.83 &  6.84 & NGC2780        &   3.0 &  7.55 &  7.42 & NGC3259        &   3.7 &  6.11 &  6.21 \\
IC4263         &   6.6 &  5.97 &  5.80 & NGC0895        &   6.0 &  6.16 &  6.13 & NGC1422        &   2.3 &  8.46 &  8.51 & NGC2782        &   1.1 &  7.26 &  7.39 & NGC3264        &   7.9 &  4.48 &  4.17 \\
IC4316         &   9.9 &  5.86 &  5.80 & NGC0899        &   9.9 &  5.01 &  4.82 & NGC1425        &   3.2 &  6.93 &  7.25 & NGC2793        &   8.7 &  5.41 &  5.15 & NGC3274        &   6.6 &  4.79 &  4.38 \\
IC4407         &   8.8 &  5.46 &  5.36 & NGC0907        &   7.6 &  6.80 &  6.86 & NGC1427A       &   9.9 &  5.35 &  5.41 & NGC2799        &   8.8 &  6.23 &  6.08 & NGC3277        &   1.9 &  8.19 &  8.41 \\
IC4468         &   4.7 &  7.07 &  7.36 & NGC0908        &   5.1 &  7.41 &  7.57 & NGC1436        &   2.0 &  8.34 &  8.56 & NGC2805        &   6.9 &  5.20 &  5.07 & NGC3287        &   7.6 &  6.68 &  6.68 \\
IC4951         &   7.7 &  4.58 &  4.14 & NGC0918        &   5.2 &  6.11 &  5.83 & NGC1461        &  -2.0 &  9.93 &  9.95 & NGC2814        &   2.8 &  5.91 &  5.95 & NGC3294        &   5.1 &  7.47 &  7.53 \\
IC4986         &   7.6 &  5.55 &  5.12 & NGC0941        &   5.3 &  5.66 &  5.51 & NGC1473        &   9.7 &  4.90 &  4.57 & NGC2841        &   3.0 &  8.77 &  9.42 & NGC3299        &   8.0 &  7.10 &  7.00 \\
IC5007         &   6.7 &  5.66 &  5.50 & NGC0986A       &   9.9 &  5.22 &  4.95 & NGC1476        &   1.4 &  5.00 &  4.64 & NGC2844        &   0.6 &  7.97 &  8.56 & NGC3306        &   7.8 &  6.87 &  7.03 \\
IC5039         &   4.0 &  6.35 &  6.33 & NGC0991        &   5.0 &  5.66 &  5.64 & NGC1483        &   4.0 &  5.54 &  5.44 & NGC2854        &   3.1 &  7.56 &  7.86 & NGC3319        &   6.0 &  5.36 &  5.18 \\
IC5069         &   3.4 &  5.97 &  5.91 & NGC1035        &   5.2 &  8.00 &  8.30 & NGC1494        &   7.0 &  5.50 &  5.59 & NGC2859        &  -1.2 &  8.46 &  8.72 & NGC3320        &   5.8 &  6.35 &  6.31 \\
IC5078         &   5.0 &  6.83 &  6.92 & NGC1047        &  -0.5 &  7.46 &  6.40 & NGC1511        &   1.8 &  7.50 &  7.48 & NGC2882        &   5.0 &  7.69 &  7.71 & NGC3321        &   5.1 &  6.02 &  5.79 \\
IC5152         &   9.7 &  5.60 &  5.43 & NGC1051        &   8.8 &  5.68 &  5.60 & NGC1512        &   1.1 &  7.13 &  7.83 & NGC2893        &   0.3 &  6.91 &  6.84 & NGC3338        &   5.1 &  6.45 &  6.44 \\
IC5269A        &   8.9 &  5.10 &  4.85 & NGC1073        &   5.3 &  5.52 &  5.63 & NGC1518        &   8.2 &  5.29 &  4.99 & NGC2894        &   1.0 &  8.85 &  8.92 & NGC3346        &   5.9 &  6.58 &  6.60 \\
IC5269C        &   7.0 &  5.75 &  5.57 & NGC1079        &   0.6 &  8.24 &  8.83 & NGC1519        &   3.0 &  6.76 &  6.82 & NGC2906        &   5.9 &  7.99 &  8.24 & NGC3353        &   3.4 &  5.51 &  5.39 \\
IC5273         &   5.7 &  6.10 &  6.13 & NGC1084        &   4.9 &  7.05 &  7.08 & NGC1533        &  -2.5 &  7.48 &  8.51 & NGC2919        &   3.7 &  7.16 &  7.28 & NGC3361        &   5.0 &  6.59 &  6.64 \\
IC5321         &   1.8 &  5.10 &  4.82 & NGC1087        &   5.2 &  6.40 &  6.27 & NGC1546        &  -0.4 &  8.32 &  7.93 & NGC2938        &   5.8 &  5.19 &  4.92 & NGC3364        &   5.0 &  6.68 &  6.84 \\
IC5325         &   4.2 &  6.81 &  6.77 & NGC1090        &   3.8 &  7.08 &  7.26 & NGC1553        &  -2.3 &  9.90 &  9.52 & NGC2962        &  -1.1 &  7.13 &  8.47 & NGC3370        &   5.1 &  6.37 &  6.45 \\
IC5334         &   3.7 &  7.98 &  8.37 & NGC1110        &   8.9 &  5.01 &  4.66 & NGC1556        &   2.6 &  5.29 &  5.03 & NGC2966        &   4.2 &  7.73 &  7.93 & NGC3377A       &   8.9 &  6.00 &  5.86 \\
NGC0007        &   4.8 &  5.13 &  4.83 & NGC1140        &   9.2 &  4.82 &  4.60 & NGC1640        &   3.0 &  6.92 &  7.28 & NGC2967        &   5.2 &  6.74 &  6.44 & NGC3380        &   1.0 &  7.50 &  7.69 \\
NGC0024        &   5.1 &  6.33 &  6.41 & NGC1179        &   5.9 &  5.95 &  5.94 & NGC1679        &   9.4 &  5.10 &  4.78 & NGC2976        &   5.2 &  6.92 &  7.10 & NGC3381        &   3.2 &  6.02 &  5.87 \\
NGC0059        &  -2.9 &  2.67 &  3.44 & NGC1187        &   5.0 &  6.50 &  6.63 & NGC1703        &   3.2 &  6.22 &  5.97 & NGC3003        &   4.3 &  5.85 &  5.76 & NGC3389        &   5.3 &  6.11 &  5.80 \\
NGC0063        &  -3.4 &  5.50 &  5.72 & NGC1232        &   5.0 &  6.25 &  6.39 & NGC1744        &   6.7 &  5.00 &  4.67 & NGC3018        &   3.1 &  5.32 &  4.81 & NGC3395        &   5.8 &  5.15 &  4.78 \\
NGC0115        &   3.9 &  5.08 &  4.83 & NGC1249        &   6.0 &  5.42 &  5.17 & NGC1800        &   8.0 &  5.60 &  5.47 & NGC3020        &   5.9 &  4.60 &  4.23 & NGC3396        &   9.4 &  5.55 &  5.30 \\
NGC0131        &   3.0 &  6.76 &  6.84 & NGC1253        &   5.9 &  5.45 &  5.32 & NGC1824        &   8.9 &  5.44 &  5.01 & NGC3023        &   5.5 &  5.07 &  4.77 & NGC3403        &   4.0 &  6.63 &  6.93 \\
NGC0150        &   3.4 &  7.00 &  7.29 & NGC1255        &   4.0 &  6.16 &  6.21 & NGC1827        &   5.9 &  6.23 &  6.02 & NGC3024        &   5.0 &  5.50 &  5.69 & NGC3437        &   5.3 &  7.45 &  7.62 \\
NGC0157        &   4.0 &  7.26 &  7.22 & NGC1258        &   5.7 &  6.13 &  6.46 & NGC1879        &   8.6 &  5.05 &  4.90 & NGC3026        &   9.7 &  6.12 &  5.99 & NGC3440        &   3.0 &  5.27 &  4.95 \\
NGC0178        &   8.7 &  4.63 &  4.34 & NGC1292        &   5.0 &  6.20 &  6.21 & NGC2101        &   9.9 &  4.51 &  3.97 & NGC3032        &  -1.9 &  3.75 &  4.41 & NGC3443        &   6.6 &  5.20 &  4.88 \\
NGC0210        &   3.1 &  6.93 &  7.33 & NGC1299        &   3.0 &  6.58 &  6.57 & NGC2104        &   8.7 &  5.39 &  5.03 & NGC3049        &   2.5 &  6.47 &  6.33 & NGC3445        &   8.9 &  5.03 &  4.83 \\
NGC0254        &  -1.2 &  7.87 &  8.12 & NGC1309        &   3.9 &  5.87 &  5.70 & NGC2460        &   1.9 &  7.80 &  8.40 & NGC3055        &   5.3 &  6.50 &  6.45 & NGC3447        &   8.8 &  4.50 &  3.95 \\
NGC0255        &   4.1 &  5.41 &  5.30 & NGC1310        &   5.0 &  6.44 &  6.50 & NGC2500        &   7.0 &  5.43 &  5.33 & NGC3057        &   7.9 &  5.07 &  4.78 & NGC3448        &   1.8 &  6.22 &  6.13 \\
NGC0275        &   6.0 &  5.92 &  5.81 & NGC1311        &   8.8 &  5.47 &  5.21 & NGC2537        &   8.7 &  5.95 &  5.94 & NGC3061        &   5.3 &  6.34 &  6.50 & NGC3455        &   3.1 &  5.74 &  5.46 \\
NGC0298        &   5.9 &  5.56 &  5.15 & NGC1313        &   7.0 &  4.93 &  4.70 & NGC2541        &   6.0 &  4.88 &  4.71 & NGC3073        &  -2.8 &  4.49 &  5.04 & NGC3485        &   3.1 &  6.20 &  6.20 \\
NGC0337        &   6.7 &  6.07 &  5.52 & NGC1316C       &  -1.9 &  5.40 &  5.99 & NGC2543        &   3.0 &  6.96 &  7.18 & NGC3094        &   1.1 &  8.58 &  8.89 & NGC3486        &   5.2 &  5.76 &  5.68 \\
NGC0337A       &   8.0 &  4.87 &  4.34 & NGC1325        &   4.0 &  7.22 &  7.53 & NGC2551        &   0.5 &  7.69 &  8.20 & NGC3104        &   9.9 &  4.77 &  4.43 & NGC3488        &   5.2 &  6.24 &  6.27 \\
NGC0450        &   5.8 &  5.22 &  4.86 & NGC1325A       &   6.9 &  6.58 &  6.70 & NGC2552        &   9.0 &  5.12 &  4.99 & NGC3118        &   4.1 &  5.69 &  5.28 & NGC3495        &  6.4 &  7.23 &  7.26 \\
 \hline        
 \hline      
\end{tabular}                                                                                                                 
Table~\ref{Table7} {\it continued.}
\end {minipage}
\end{sideways}
\end{table}

\begin{table} 
\begin{sideways}
\setcounter{table}{5} 
\setlength{\tabcolsep}{.038880in}
\begin {minipage}{233mm}
\caption{({\it Continued})}
\label{Table7}
\begin{tabular}{@{}lclclclclclccccccccccccccc@{}}
  \hline
  \hline
Galaxy&Type & $M_{\rm BH}$&
                                               $M_{\rm BH}$&Galaxy&Type & $M_{\rm BH}$&
                                               $M_{\rm BH}$&Galaxy&Type & $M_{\rm BH}$&
                                               $M_{\rm BH}$&Galaxy&Type & $M_{\rm BH}$&
                                               $M_{\rm BH}$&Galaxy&Type & $M_{\rm BH}$&
                                               $M_{\rm BH}$\\
& &(FUV)& (NUV)&& &(FUV)& (NUV)&& &(FUV)&(NUV)&& &(FUV)& (NUV)&& &(FUV)&(NUV)\\                                               
(1)& (2)&(3)& (4)&(1)& (2)&(3)& (4)&(1)& (2)&(3)& (4)&(1)& (2)&(3)& (4)&(1)& (2)&(3)& (4)\\
  \hline                                                                                                                                                        
NGC3504        &   2.1 &  7.46 &  7.53 & NGC3898        &   1.7 &  8.40 &  9.09 & NGC4189        &   6.0 &  7.16 &  7.14 & NGC4434        &  -4.8 &  9.00 &  8.74 & NGC4658        &   4.0 &  6.59 &  6.34 \\
NGC3510        &   8.6 &  5.05 &  4.63 & NGC3900        &  -0.2 &  4.47 &  5.84 & NGC4190        &   9.9 &  5.03 &  4.56 & NGC4442        &  -1.9 &  9.46 &  9.72 & NGC4668        &   7.4 &  5.52 &  5.19 \\
NGC3512        &   5.1 &  6.45 &  6.57 & NGC3901        &   6.0 &  5.55 &  5.36 & NGC4193        &   4.1 &  7.80 &  7.94 & NGC4451        &   2.4 &  7.59 &  7.53 & NGC4680        &   0.0 &  7.09 &  7.08 \\
NGC3513        &   5.1 &  5.70 &  5.44 & NGC3906        &   6.7 &  6.02 &  5.93 & NGC4194        &   9.5 &  7.71 &  7.61 & NGC4455        &   7.0 &  5.03 &  4.63 & NGC4682        &   5.8 &  7.08 &  7.25 \\
NGC3522        &  -4.9 &  8.67 &  6.50 & NGC3912        &   3.1 &  7.34 &  7.36 & NGC4204        &   8.0 &  5.30 &  5.00 & NGC4460        &  -0.9 &  3.63 &  4.53 & NGC4684        &  -1.1 &  5.57 &  6.48 \\
NGC3547        &   3.1 &  6.07 &  5.87 & NGC3917        &   5.9 &  7.55 &  7.83 & NGC4220        &  -0.3 &  7.87 &  8.41 & NGC4461        &  -0.8 &  9.56 &  9.17 & NGC4688        &   6.0 &  5.03 &  4.75 \\
NGC3549        &   5.1 &  7.53 &  7.82 & NGC3922        &  -0.1 &  7.36 &  7.26 & NGC4234        &   8.8 &  6.22 &  6.24 & NGC4480        &   5.1 &  6.69 &  6.78 & NGC4691        &   0.3 &  6.59 &  6.52 \\
NGC3583        &   3.1 &  7.82 &  7.99 & NGC3930        &   5.2 &  5.60 &  5.39 & NGC4238        &   6.5 &  5.78 &  5.46 & NGC4485        &   9.7 &  4.97 &  4.61 & NGC4694        &  -2.0 &  5.35 &  4.92 \\
NGC3589        &   7.0 &  5.01 &  4.81 & NGC3949        &   4.0 &  6.28 &  6.19 & NGC4242        &   7.9 &  6.03 &  5.76 & NGC4487        &   6.0 &  6.02 &  5.94 & NGC4700        &   4.9 &  5.34 &  4.92 \\
NGC3592        &   5.3 &  8.17 &  8.41 & NGC3952        &   9.3 &  5.11 &  4.66 & NGC4248        &   8.0 &  7.87 &  7.73 & NGC4496A       &   7.6 &  5.53 &  5.37 & NGC4707        &   8.8 &  4.10 &  3.63 \\
NGC3599        &  -2.0 &  7.83 &  7.69 & NGC3955        &   0.3 &  8.22 &  8.23 & NGC4252        &   3.1 &  5.67 &  5.51 & NGC4498        &   6.4 &  6.28 &  6.16 & NGC4713        &   6.8 &  5.38 &  5.18 \\
NGC3611        &   1.0 &  7.83 &  7.75 & NGC3956        &   5.1 &  6.01 &  5.83 & NGC4254        &   5.2 &  6.99 &  6.89 & NGC4502        &   5.8 &  5.74 &  5.55 & NGC4722        &   0.0 &  8.30 &  8.60 \\
NGC3614        &   5.2 &  6.14 &  6.17 & NGC3976        &   3.2 &  7.46 &  7.84 & NGC4262        &  -2.7 &  7.15 &  7.89 & NGC4504        &   6.0 &  5.92 &  5.75 & NGC4723        &   8.8 &  4.60 &  4.08 \\
NGC3619        &  -0.9 &  6.45 &  7.20 & NGC3985        &   8.8 &  6.33 &  6.21 & NGC4267        &  -2.7 &  7.87 &  9.04 & NGC4517A       &   7.8 &  5.20 &  4.90 & NGC4725        &   2.2 &  7.97 &  8.66 \\
NGC3622        &   6.0 &  5.86 &  5.72 & NGC4010        &   6.8 &  7.84 &  8.23 & NGC4273        &   5.2 &  6.90 &  6.89 & NGC4519        &   6.9 &  5.57 &  5.44 & NGC4731        &   5.9 &  5.34 &  5.11 \\
NGC3625        &   3.1 &  6.48 &  6.49 & NGC4020        &   6.9 &  6.24 &  6.08 & NGC4276        &   7.6 &  6.56 &  6.49 & NGC4522        &   6.0 &  6.90 &  6.99 & NGC4746        &   3.1 &  7.67 &  7.97 \\
NGC3629        &   5.9 &  5.17 &  4.82 & NGC4030        &   4.0 &  7.61 &  7.58 & NGC4286        &   1.0 &  7.69 &  7.76 & NGC4523        &   9.1 &  4.50 &  4.02 & NGC4747        &   7.1 &  7.16 &  7.32 \\
NGC3631        &   5.2 &  6.37 &  6.44 & NGC4032        &   9.8 &  5.60 &  5.29 & NGC4288        &   7.0 &  5.20 &  4.96 & NGC4525        &   5.9 &  6.67 &  6.60 & NGC4750        &   2.4 &  8.04 &  8.30 \\
NGC3654        &   4.0 &  6.52 &  6.52 & NGC4034        &   6.0 &  6.36 &  6.53 & NGC4294        &   5.8 &  5.62 &  5.33 & NGC4528        &  -2.0 &  8.72 &  8.82 & NGC4758        &   9.1 &  7.86 &  8.05 \\
NGC3655        &   5.0 &  7.79 &  7.84 & NGC4037        &   2.9 &  6.88 &  6.92 & NGC4298        &   5.2 &  7.97 &  7.98 & NGC4531        &  -0.5 &  6.13 &  6.68 & NGC4765        &   0.0 &  5.29 &  4.92 \\
NGC3664        &   9.0 &  4.61 &  4.20 & NGC4038        &   8.9 &  6.83 &  6.90 & NGC4299        &   8.5 &  4.89 &  4.60 & NGC4532        &   9.7 &  5.45 &  5.10 & NGC4775        &   6.9 &  5.24 &  5.03 \\
NGC3669        &   6.8 &  5.92 &  5.70 & NGC4039        &   8.9 &  6.71 &  6.63 & NGC4303A       &   6.4 &  4.61 &  4.27 & NGC4534        &   7.8 &  4.62 &  4.31 & NGC4779        &   4.6 &  6.55 &  6.59 \\
NGC3672        &   5.0 &  7.36 &  7.46 & NGC4049        &   7.5 &  5.62 &  5.44 & NGC4309        &  -0.9 &  8.13 &  8.08 & NGC4535        &   5.0 &  6.92 &  7.11 & NGC4781        &   7.0 &  6.20 &  6.02 \\
NGC3681        &   4.0 &  6.98 &  7.29 & NGC4050        &   2.2 &  7.80 &  8.22 & NGC4310        &  -0.9 &  6.33 &  6.99 & NGC4540        &   6.2 &  7.58 &  7.46 & NGC4790        &   4.8 &  5.85 &  5.91 \\
NGC3682        &  -0.1 &  4.70 &  5.70 & NGC4062        &   5.3 &  7.82 &  8.04 & NGC4324        &  -0.9 &  5.89 &  7.38 & NGC4545        &   5.6 &  6.26 &  6.24 & NGC4791        &   1.0 &  7.62 &  8.04 \\
NGC3683A       &   5.1 &  6.74 &  6.88 & NGC4067        &   3.1 &  7.13 &  7.35 & NGC4331        &   9.9 &  5.46 &  5.39 & NGC4546        &  -2.7 &  9.62 &  8.78 & NGC4793        &   5.1 &  7.42 &  7.46 \\
NGC3687        &   3.8 &  6.67 &  6.79 & NGC4068        &   9.9 &  4.37 &  3.98 & NGC4336        &  -0.1 &  6.82 &  5.96 & NGC4550        &  -2.1 &  7.95 &  7.77 & NGC4806        &   4.9 &  5.64 &  5.32 \\
NGC3689        &   5.3 &  7.90 &  7.98 & NGC4080        &   9.5 &  6.57 &  6.49 & NGC4344        &  -2.1 &  3.44 &  4.06 & NGC4561        &   7.2 &  4.91 &  4.60 & NGC4808        &   5.9 &  6.65 &  6.60 \\
NGC3691        &   3.0 &  5.99 &  5.83 & NGC4094        &   5.4 &  6.56 &  6.63 & NGC4351        &   2.5 &  6.64 &  6.37 & NGC4562        &   7.1 &  6.68 &  6.62 & NGC4809        &   9.9 &  4.15 &  3.66 \\
NGC3701        &   4.0 &  5.87 &  5.86 & NGC4100        &   4.1 &  7.88 &  8.15 & NGC4353        &   9.9 &  7.23 &  7.32 & NGC4567        &   4.0 &  7.48 &  7.26 & NGC4814        &   3.1 &  7.01 &  7.17 \\
NGC3712        &   6.0 &  5.30 &  4.93 & NGC4108        &   5.2 &  6.30 &  6.30 & NGC4359        &   5.0 &  6.67 &  6.50 & NGC4571        &   6.5 &  7.56 &  7.65 & NGC4866        &  -0.1 &  8.02 &  8.61 \\
NGC3715        &   3.8 &  7.57 &  7.57 & NGC4108B       &   7.0 &  4.81 &  4.50 & NGC4376        &   9.9 &  5.54 &  5.39 & NGC4572        &   5.0 &  7.92 &  8.13 & NGC4897        &   4.0 &  6.05 &  5.98 \\
NGC3733        &   5.6 &  5.57 &  5.48 & NGC4116        &   7.4 &  5.56 &  5.36 & NGC4378        &   1.0 &  8.53 &  9.11 & NGC4591        &   3.3 &  7.24 &  7.47 & NGC4899        &   5.0 &  6.40 &  6.33 \\
NGC3738        &   9.8 &  5.38 &  5.47 & NGC4117        &  -2.1 &  5.66 &  7.20 & NGC4383        &   1.0 &  6.20 &  6.31 & NGC4592        &   8.0 &  5.31 &  5.04 & NGC4900        &   5.2 &  5.99 &  5.87 \\
NGC3755        &   5.2 &  5.21 &  4.93 & NGC4123        &   5.0 &  6.34 &  6.45 & NGC4384        &   1.0 &  6.38 &  6.26 & NGC4595        &   3.8 &  6.23 &  6.10 & NGC4902        &   3.0 &  7.02 &  7.22 \\
NGC3769        &   3.4 &  6.83 &  6.92 & NGC4129        &   2.3 &  7.05 &  6.99 & NGC4389        &   4.1 &  7.21 &  7.18 & NGC4597        &   8.7 &  5.22 &  4.92 & NGC4904        &   5.8 &  6.74 &  6.67 \\
NGC3779        &   6.7 &  4.72 &  4.37 & NGC4136        &   5.2 &  5.64 &  5.51 & NGC4390        &   5.0 &  5.96 &  5.86 & NGC4604        &  10.0 &  6.11 &  5.90 & NGC4920        &  10.0 &  4.35 &  3.95 \\
NGC3780        &   5.2 &  6.83 &  7.06 & NGC4138        &  -0.8 &  4.94 &  6.56 & NGC4396        &   6.9 &  6.15 &  6.16 & NGC4605        &   5.1 &  6.61 &  6.55 & NGC4928        &   4.0 &  5.36 &  5.41 \\
NGC3782        &   6.5 &  5.01 &  4.68 & NGC4141        &   6.0 &  4.64 &  4.74 & NGC4405        &  -0.1 &  5.17 &  5.66 & NGC4618        &   8.7 &  5.64 &  5.48 & NGC4942        &   6.9 &  5.78 &  5.55 \\
NGC3788        &   2.3 &  7.88 &  7.98 & NGC4142        &   6.5 &  5.43 &  5.15 & NGC4409        &   4.7 &  6.31 &  6.26 & NGC4625        &   8.8 &  5.92 &  5.86 & NGC4948        &   7.3 &  7.40 &  7.25 \\
NGC3794        &   6.3 &  5.16 &  4.93 & NGC4144        &   5.9 &  5.88 &  5.61 & NGC4411A       &   5.4 &  5.69 &  5.47 & NGC4630        &   9.8 &  6.75 &  6.77 & NGC4948A       &   7.7 &  5.41 &  5.21 \\
NGC3810        &   5.2 &  6.91 &  6.82 & NGC4145        &   6.9 &  6.32 &  6.31 & NGC4411B       &   6.3 &  5.47 &  5.29 & NGC4632        &   5.1 &  6.53 &  6.44 & NGC4951        &   6.0 &  7.08 &  7.11 \\
NGC3813        &   3.3 &  7.13 &  7.18 & NGC4152        &   5.0 &  6.05 &  5.99 & NGC4412        &   3.1 &  6.78 &  6.86 & NGC4633        &   7.9 &  6.10 &  5.99 & NGC4961        &   5.6 &  5.79 &  5.66 \\
NGC3846A       &   9.7 &  5.57 &  5.37 & NGC4158        &   3.1 &  6.80 &  6.88 & NGC4413        &   2.0 &  7.29 &  7.36 & NGC4634        &   5.9 &  8.37 &  8.57 & NGC4965        &   6.7 &  5.63 &  5.28 \\
NGC3850        &   5.3 &  5.81 &  5.72 & NGC4159        &   6.5 &  6.59 &  6.64 & NGC4414        &   5.2 &  8.41 &  8.69 & NGC4635        &   6.6 &  6.09 &  6.01 & NGC4980        &   1.1 &  5.33 &  4.89 \\
NGC3879        &   8.0 &  5.16 &  4.74 & NGC4162        &   4.0 &  6.87 &  6.93 & NGC4416        &   5.9 &  6.58 &  6.55 & NGC4639        &   3.5 &  6.89 &  7.13 & NGC4981        &   4.0 &  7.03 &  7.00 \\
NGC3887        &   3.9 &  6.81 &  6.88 & NGC4163        &   9.9 &  5.60 &  5.26 & NGC4423        &   7.8 &  5.98 &  5.80 & NGC4641        &  -2.0 &  4.12 &  4.26 & NGC4984        &  -0.8 &  7.03 &  7.61 \\
NGC3888        &   5.3 &  6.96 &  7.06 & NGC4165        &   1.9 &  7.46 &  7.55 & NGC4424        &   1.0 &  8.45 &  8.22 & NGC4642        &   3.9 &  6.65 &  6.62 & NGC4995        &   3.1 &  7.67 &  7.72 \\
NGC3893        &   5.2 &  6.74 &  6.71 & NGC4173        &   7.2 &  5.12 &  4.64 & NGC4428        &   5.0 &  7.56 &  7.73 & NGC4651        &   5.2 &  7.12 &  7.28 & NGC5002        &   9.0 &  5.16 &  4.86 \\
NGC3896        &   0.2 &  5.79 &  5.48 & NGC4183        &   5.8 &  6.37 &  6.20 & NGC4430        &   3.4 &  6.81 &  6.93 & NGC4653        &   6.0 &  5.91 &  5.81 & NGC5012       &   5.1 &  7.17 &  7.42 \\
 \hline        
 \hline      
\end{tabular}                                                                                                                 
Table~\ref{Table7} {\it continued.}
\end {minipage}
\end{sideways}
\end{table}

\begin{table} 
\begin{sideways}
\setcounter{table}{5} 
\setlength{\tabcolsep}{.038880in}
\begin {minipage}{233mm}
\caption{({\it Continued})}
\label{Table7}
\begin{tabular}{@{}lclclclclclccccccccccccccc@{}}
  \hline
  \hline
Galaxy&Type & $M_{\rm BH}$&
                                               $M_{\rm BH}$&Galaxy&Type & $M_{\rm BH}$&
                                               $M_{\rm BH}$&Galaxy&Type & $M_{\rm BH}$&
                                               $M_{\rm BH}$&Galaxy&Type & $M_{\rm BH}$&
                                               $M_{\rm BH}$&Galaxy&Type & $M_{\rm BH}$&
                                               $M_{\rm BH}$\\
& &(FUV)& (NUV)&& &(FUV)& (NUV)&& &(FUV)&(NUV)&& &(FUV)& (NUV)&& &(FUV)&(NUV)\\                                               
(1)& (2)&(3)& (4)&(1)& (2)&(3)& (4)&(1)& (2)&(3)& (4)&(1)& (2)&(3)& (4)&(1)& (2)&(3)& (4)\\
 \hline                                                                                                                                                        
NGC5014        &   1.4 &  7.66 &  7.76 & NGC5448        &   1.4 &  7.66 &  8.02 & NGC5789        &   7.8 &  4.35 &  3.92 & NGC7059        &   5.6 &  6.25 &  6.04 & NGC7750        &   5.5 &  6.61 &  6.74 \\
NGC5016        &   4.4 &  6.52 &  6.60 & NGC5464        &   9.5 &  4.92 &  4.30 & NGC5798        &   9.7 &  5.24 &  5.01 & NGC7064        &   5.1 &  4.66 &  4.26 & NGC7755        &   5.0 &  6.77 &  6.90 \\
NGC5033        &   5.1 &  7.52 &  8.02 & NGC5468        &   6.0 &  5.52 &  5.32 & NGC5806        &   3.2 &  8.25 &  8.49 & NGC7070        &   6.0 &  6.12 &  6.10 & NGC7757        &   5.3 &  5.52 &  5.39 \\
NGC5042        &   5.0 &  5.91 &  5.46 & NGC5472        &   2.1 &  8.45 &  8.60 & NGC5809        &   0.2 &  7.14 &  7.13 & NGC7079        &  -1.8 &  8.55 &  8.30 & NGC7764        &   9.4 &  5.20 &  4.96 \\
NGC5054        &   4.2 &  7.88 &  8.06 & NGC5473        &  -2.7 &  9.74 &  9.31 & NGC5821        &   5.0 &  6.96 &  7.25 & NGC7091        &   7.9 &  4.81 &  4.53 & NGC7793        &   7.4 &  5.67 &  5.64 \\
NGC5085        &   5.0 &  7.29 &  7.48 & NGC5474        &   6.1 &  5.17 &  5.34 & NGC5832        &   3.3 &  5.99 &  5.86 & NGC7107        &   8.6 &  5.60 &  5.46 & NGC7798        &   5.1 &  6.88 &  6.95 \\
NGC5101        &   0.2 &  8.41 &  8.30 & NGC5476        &   7.8 &  6.35 &  6.30 & NGC5850        &   3.1 &  7.84 &  8.31 & NGC7151        &   5.9 &  6.16 &  6.13 & PGC002492      &   2.0 &  5.27 &  5.40 \\
NGC5105        &   5.0 &  5.72 &  5.57 & NGC5477        &   8.8 &  4.24 &  3.95 & NGC5854        &  -1.1 &  8.93 &  8.02 & NGC7154        &   9.5 &  5.63 &  5.40 & PGC002689      &   8.8 &  4.33 &  3.95 \\
NGC5107        &   6.6 &  5.23 &  4.93 & NGC5486        &   8.7 &  5.02 &  4.67 & NGC5861        &   5.0 &  6.95 &  7.06 & NGC7162        &   4.8 &  6.43 &  6.60 & PGC003062      &   6.8 &  5.43 &  5.03 \\
NGC5109        &   5.3 &  5.58 &  5.40 & NGC5496        &   6.5 &  5.36 &  4.90 & NGC5866B       &   7.9 &  5.86 &  5.69 & NGC7162A       &   8.9 &  5.20 &  5.07 & PGC003853      &   7.0 &  6.67 &  6.16 \\
NGC5116        &   4.9 &  7.60 &  7.88 & NGC5507        &  -2.3 &  8.41 &  9.19 & NGC5878        &   3.2 &  7.76 &  7.76 & NGC7167        &   5.1 &  5.89 &  5.81 & PGC003855      &   8.8 &  7.90 &  5.33 \\
NGC5117        &   5.7 &  5.76 &  5.73 & NGC5520        &   3.1 &  6.99 &  6.97 & NGC5892        &   7.0 &  5.69 &  5.46 & NGC7184        &   4.5 &  8.02 &  8.45 & PGC004143      &   9.8 &  5.13 &  4.85 \\
NGC5134        &   2.9 &  7.85 &  8.13 & NGC5523        &   5.8 &  6.30 &  6.28 & NGC5915        &   2.7 &  5.60 &  5.14 & NGC7188        &   3.5 &  6.89 &  7.02 & PGC005329      &   8.0 &  5.38 &  4.99 \\
NGC5147        &   7.9 &  5.57 &  5.43 & NGC5534        &   2.1 &  5.98 &  5.97 & NGC5916A       &   5.0 &  6.04 &  6.05 & NGC7191        &   5.1 &  7.86 &  7.98 & PGC006048      &   3.9 &  5.36 &  4.96 \\
NGC5169        &   4.0 &  5.64 &  5.47 & NGC5569        &   5.8 &  5.71 &  5.58 & NGC5937        &   3.2 &  7.27 &  7.39 & NGC7205        &   4.0 &  7.09 &  7.22 & PGC006190      &   6.8 &  5.53 &  5.69 \\
NGC5173        &  -4.9 &  4.61 &  5.72 & NGC5577        &   3.8 &  7.09 &  7.24 & NGC5949        &   4.0 &  7.15 &  7.24 & NGC7218        &   5.6 &  6.13 &  5.98 & PGC006228      &   8.7 &  5.60 &  5.44 \\
NGC5204        &   8.9 &  4.63 &  4.30 & NGC5584        &   6.0 &  5.77 &  5.61 & NGC5951        &   5.2 &  6.61 &  6.55 & NGC7219        &   0.6 &  7.25 &  7.50 & PGC006244      &  10.0 &  4.79 &  4.46 \\
NGC5205        &   3.5 &  6.68 &  6.81 & NGC5585        &   6.9 &  5.01 &  4.77 & NGC5954        &   6.0 &  6.96 &  6.67 & NGC7241        &   4.0 &  7.16 &  7.58 & PGC006626      &   9.9 &  4.92 &  4.67 \\
NGC5216        &  -4.9 &  7.08 &  7.55 & NGC5587        &  -0.1 &  4.42 &  5.80 & NGC5956        &   5.9 &  6.61 &  6.88 & NGC7247        &   3.1 &  7.40 &  7.50 & PGC006703      &   3.1 &  5.87 &  5.69 \\
NGC5236        &   5.0 &  6.91 &  6.46 & NGC5595        &   4.9 &  6.33 &  6.19 & NGC5957        &   3.0 &  6.63 &  6.85 & NGC7254        &   2.9 &  5.99 &  5.83 & PGC006706      &   9.0 &  5.91 &  5.84 \\
NGC5238        &   8.0 &  5.04 &  4.83 & NGC5597        &   6.0 &  6.44 &  6.34 & NGC5962        &   5.1 &  7.01 &  7.08 & NGC7290        &   4.0 &  5.89 &  5.84 & PGC006864      &   6.9 &  5.26 &  4.82 \\
NGC5240        &   5.8 &  6.96 &  7.10 & NGC5600        &   4.8 &  6.61 &  6.44 & NGC5963        &   4.1 &  5.72 &  5.61 & NGC7314        &   4.0 &  6.98 &  7.21 & PGC007109      &   9.0 &  6.54 &  7.06 \\
NGC5247        &   4.1 &  6.59 &  6.46 & NGC5604        &   1.3 &  6.92 &  7.26 & NGC5964        &   6.9 &  6.08 &  5.93 & NGC7361        &   4.6 &  5.75 &  5.70 & PGC007654      &   9.9 &  4.97 &  4.79 \\
NGC5253        &   8.9 &  5.67 &  5.28 & NGC5608        &   9.8 &  5.15 &  5.01 & NGC5984        &   6.4 &  6.94 &  6.84 & NGC7371        &   0.0 &  7.15 &  7.33 & PGC007682      &   1.0 &  5.94 &  5.72 \\
NGC5254        &   5.0 &  7.26 &  7.64 & NGC5630        &   7.9 &  5.56 &  5.25 & NGC5985        &   3.0 &  7.63 &  7.94 & NGC7378        &   2.2 &  7.63 &  7.86 & PGC007900      &   9.0 &  4.42 &  4.01 \\
NGC5264        &   9.7 &  6.30 &  6.06 & NGC5631        &  -1.9 &  9.35 &  9.00 & NGC6012        &   1.7 &  6.90 &  6.70 & NGC7412        &   3.2 &  6.57 &  6.67 & PGC007998      &   9.0 &  3.92 &  3.48 \\
NGC5289        &   2.0 &  7.96 &  8.53 & NGC5636        &  -0.4 &  4.80 &  5.18 & NGC6014        &  -1.8 &  5.42 &  6.36 & NGC7418A       &   6.5 &  4.05 &  3.66 & PGC008295      &   5.0 &  5.00 &  4.68 \\
NGC5297        &   4.9 &  6.89 &  7.10 & NGC5645        &   6.6 &  5.95 &  5.72 & NGC6015        &   5.9 &  6.41 &  6.55 & NGC7421        &   3.7 &  6.97 &  7.19 & PGC009272      &   8.0 &  5.10 &  4.89 \\
NGC5300        &   5.2 &  6.37 &  6.55 & NGC5660        &   5.2 &  5.96 &  5.81 & NGC6063        &   5.9 &  6.68 &  6.73 & NGC7456        &   6.0 &  5.90 &  5.86 & PGC009354      &   5.1 &  4.50 &  4.14 \\
NGC5301        &   4.7 &  7.79 &  7.93 & NGC5661        &   3.2 &  5.63 &  5.50 & NGC6070        &   6.0 &  6.78 &  6.68 & NGC7462        &   3.6 &  5.77 &  5.54 & PGC009559      &   7.8 &  5.10 &  4.67 \\
NGC5304        &  -3.2 &  7.10 &  7.55 & NGC5665        &   5.0 &  7.08 &  7.04 & NGC6106        &   5.3 &  6.30 &  6.13 & NGC7479        &   4.3 &  7.33 &  7.37 & PGC010813      &   5.0 &  4.07 &  3.55 \\
NGC5311        &  -0.1 &  7.43 &  8.00 & NGC5667        &   6.0 &  5.47 &  5.29 & NGC6140        &   5.6 &  5.40 &  5.22 & NGC7531        &   4.0 &  6.74 &  7.02 & PGC011367      &   6.9 &  6.19 &  6.02 \\
NGC5313        &   3.1 &  8.04 &  8.37 & NGC5668        &   6.9 &  5.26 &  5.01 & NGC6155        &   5.2 &  6.94 &  6.90 & NGC7590        &   4.4 &  7.01 &  7.19 & PGC011677      &   9.1 &  5.47 &  5.29 \\
NGC5320        &   5.2 &  6.46 &  6.59 & NGC5669        &   6.0 &  5.44 &  5.24 & NGC6168        &   8.0 &  7.16 &  7.77 & NGC7599        &   5.2 &  7.01 &  7.17 & PGC011744      &   1.0 &  4.34 &  4.05 \\
NGC5334        &   5.2 &  6.13 &  6.09 & NGC5691        &   1.2 &  6.37 &  6.26 & NGC6181        &   5.2 &  7.32 &  7.44 & NGC7661        &   5.9 &  5.44 &  5.29 & PGC012068      &  10.0 &  4.53 &  4.20 \\
NGC5336        &   5.9 &  5.44 &  5.43 & NGC5693        &   6.9 &  6.03 &  5.90 & NGC6236        &   5.9 &  5.27 &  4.85 & NGC7667        &   8.6 &  3.97 &  3.50 & PGC012439      &   6.2 &  8.22 &  8.22 \\
NGC5337        &   2.0 &  8.10 &  8.44 & NGC5701        &  -0.4 &  4.42 &  6.03 & NGC6237        &   6.4 &  4.94 &  4.49 & NGC7689        &   5.9 &  5.72 &  5.72 & PGC012633      &   2.3 &  6.47 &  6.63 \\
NGC5338        &  -2.0 &  4.89 &  4.96 & NGC5708        &   7.8 &  6.41 &  6.42 & NGC6239        &   3.3 &  5.87 &  5.76 & NGC7690        &   2.9 &  6.79 &  7.02 & PGC012664      &   6.7 &  5.12 &  4.97 \\
NGC5339        &   1.1 &  7.03 &  7.17 & NGC5713        &   4.0 &  7.72 &  7.62 & NGC6255        &   5.9 &  4.54 &  4.17 & NGC7694        &  10.0 &  6.29 &  6.09 & PGC012981      &   8.7 &  4.87 &  4.86 \\
NGC5346        &   5.8 &  6.55 &  6.56 & NGC5714        &   5.8 &  7.71 &  8.15 & NGC6267        &   4.8 &  6.82 &  6.77 & NGC7714        &   3.1 &  5.85 &  5.84 & PGC013716      &   4.0 &  7.22 &  7.35 \\
NGC5350        &   3.6 &  7.16 &  7.24 & NGC5729        &   3.2 &  6.74 &  6.95 & NGC6339        &   6.3 &  6.53 &  6.16 & NGC7715        &   9.6 &  5.12 &  4.38 & PGC013821      &   8.9 &  7.68 &  7.59 \\
NGC5353        &  -2.0 &  9.81 &  7.65 & NGC5730        &   9.5 &  6.61 &  6.81 & NGC6395        &   5.8 &  5.94 &  5.69 & NGC7716        &   3.0 &  6.98 &  7.29 & PGC014487      &   7.9 &  4.92 &  4.78 \\
NGC5360        &   0.1 &  7.71 &  7.57 & NGC5731        &   3.6 &  5.75 &  5.47 & NGC6412        &   5.2 &  6.07 &  5.97 & NGC7721        &   4.9 &  6.95 &  7.15 & PGC016090      &   9.0 &  4.79 &  4.24 \\
NGC5362        &   3.4 &  7.06 &  7.15 & NGC5744        &   0.5 &  5.91 &  5.39 & NGC6861E       &   2.0 &  7.00 &  6.97 & NGC7723        &   3.1 &  7.37 &  7.48 & PGC024469      &   4.2 &  6.50 &  6.28 \\
NGC5371        &   4.0 &  7.43 &  7.62 & NGC5757        &   3.1 &  7.13 &  7.04 & NGC6887        &   3.7 &  7.67 &  7.91 & NGC7724        &   3.1 &  7.50 &  7.66 & PGC027747      &   6.1 &  5.51 &  5.10 \\
NGC5375        &   2.4 &  7.16 &  7.53 & NGC5762        &   2.0 &  5.80 &  5.79 & NGC6889        &   3.7 &  6.24 &  6.09 & NGC7731        &   1.0 &  6.78 &  7.29 & PGC027825      &   7.0 &  5.24 &  4.82 \\
NGC5383        &   3.1 &  7.64 &  7.77 & NGC5768        &   5.3 &  5.99 &  5.91 & NGC6902        &   2.3 &  6.52 &  6.46 & NGC7732        &   6.7 &  5.92 &  5.86 & PGC027833      &   7.4 &  5.53 &  5.35 \\
NGC5425        &   6.5 &  6.24 &  6.19 & NGC5774        &   6.9 &  5.38 &  5.11 & NGC6902B       &   5.5 &  5.16 &  4.78 & NGC7741        &   5.9 &  5.72 &  5.50 & PGC029300      &  -1.9 &  7.71 &  7.28 \\
NGC5426        &   5.0 &  6.65 &  6.64 & NGC5781        &   2.8 &  7.37 &  7.47 & NGC6925        &   4.0 &  7.28 &  7.44 & NGC7742        &   2.8 &  7.15 &  7.43 & PGC029653      &   9.9 &  4.25 &  3.90 \\
NGC5430        &   3.1 &  7.39 &  7.66 & NGC5783        &   5.2 &  6.03 &  5.98 & NGC7051        &   0.9 &  7.17 &  7.77 & NGC7743        &  -0.9 &  8.81 &  7.89 & PGC031979      &   6.7 &  4.40 &  3.97 \\
 \hline        
 \hline      
\end{tabular}                                                                                                                 
Table~\ref{Table7} {\it continued.}
\end {minipage}
\end{sideways}
\end{table}

\begin{table} 
\begin{sideways}
\setcounter{table}{5} 
\setlength{\tabcolsep}{.038880in}
\begin {minipage}{233mm}
\caption{({\it Continued})}
\label{Table7}
\begin{tabular}{@{}lclclclclclccccccccccccccc@{}}
  \hline
  \hline
Galaxy&Type & $M_{\rm BH}$&
                                               $M_{\rm BH}$&Galaxy&Type & $M_{\rm BH}$&
                                               $M_{\rm BH}$&Galaxy&Type & $M_{\rm BH}$&
                                               $M_{\rm BH}$&Galaxy&Type & $M_{\rm BH}$&
                                               $M_{\rm BH}$&Galaxy&Type & $M_{\rm BH}$&
                                               $M_{\rm BH}$\\
& &(FUV)& (NUV)&& &(FUV)& (NUV)&& &(FUV)&(NUV)&& &(FUV)& (NUV)&& &(FUV)&(NUV)\\                                               
(1)& (2)&(3)& (4)&(1)& (2)&(3)& (4)&(1)& (2)&(3)& (4)&(1)& (2)&(3)& (4)&(1)& (2)&(3)& (4)\\
 \hline                                                                                                                                                        
PGC032091      &   7.0 &  5.74 &  5.82 & PGC066559      &   8.0 &  4.77 &  4.17 & UGC04305       &   9.9 &  3.85 &  3.39 & UGC05522       &   6.4 &  4.47 &  4.01 & UGC06399       &   8.8 &  5.77 &  5.73 \\
PGC035271      &   7.0 &  4.47 &  3.83 & PGC067871      &   7.0 &  4.74 &  4.32 & UGC04390       &   6.6 &  5.56 &  5.47 & UGC05540       &   5.9 &  5.40 &  4.90 & UGC06433       &   9.2 &  4.83 &  4.45 \\
PGC035705      &   7.9 &  4.89 &  4.48 & PGC068061      &   0.1 &  6.02 &  5.86 & UGC04426       &   9.8 &  5.01 &  4.53 & UGC05571       &   8.8 &  4.21 &  3.58 & UGC06446       &   6.6 &  4.18 &  3.83 \\
PGC036274      &   9.0 &  5.72 &  5.41 & PGC068771      &   6.6 &  5.42 &  5.24 & UGC04499       &   8.0 &  4.53 &  4.14 & UGC05612       &   8.0 &  5.87 &  5.79 & UGC06512       &   5.3 &  4.63 &  4.30 \\
PGC036551      &   7.6 &  5.44 &  5.14 & PGC069114      &   6.9 &  4.93 &  4.50 & UGC04514       &   5.9 &  5.28 &  5.08 & UGC05642       &   4.1 &  6.75 &  6.67 & UGC06517       &   3.8 &  6.09 &  6.02 \\
PGC036643      &   6.3 &  5.16 &  4.89 & PGC069224      &   9.8 &  4.43 &  4.23 & UGC04543       &   7.9 &  4.23 &  3.77 & UGC05646       &   4.0 &  5.94 &  5.81 & UGC06526       &   7.0 &  6.68 &  6.57 \\
PGC037238      &   8.0 &  4.31 &  3.87 & PGC069293      &   8.9 &  5.65 &  5.57 & UGC04549       &   8.1 &  6.43 &  6.96 & UGC05676       &   8.0 &  5.81 &  5.73 & UGC06534       &   6.4 &  5.53 &  5.24 \\
PGC037373      &   5.8 &  5.00 &  4.74 & PGC069339      &   9.9 &  4.30 &  4.52 & UGC04550       &   3.4 &  7.68 &  7.55 & UGC05677       &   8.0 &  5.01 &  4.43 & UGC06570       &  -0.2 &  5.10 &  5.47 \\
PGC039799      &   9.1 &  4.50 &  3.95 & PGC069404      &   8.0 &  7.10 &  7.42 & UGC04628       &   5.8 &  5.96 &  5.73 & UGC05688       &   8.8 &  5.24 &  4.99 & UGC06575       &   5.8 &  5.36 &  5.40 \\
PGC040408      &   8.9 &  5.60 &  5.40 & PGC069448      &   4.0 &  6.34 &  6.33 & UGC04659       &   8.0 &  5.49 &  5.17 & UGC05707       &   5.9 &  4.97 &  4.78 & UGC06603       &   5.9 &  5.41 &  5.14 \\
PGC040447      &  10.0 &  5.07 &  4.57 & PGC072006      &   9.0 &  4.90 &  4.57 & UGC04701       &   7.0 &  5.82 &  5.65 & UGC05720       &   9.9 &  5.68 &  5.50 & UGC06628       &   8.8 &  4.97 &  4.67 \\
PGC040552      &   5.4 &  7.85 &  7.95 & PGC072252      &   5.0 &  4.60 &  4.27 & UGC04704       &   7.9 &  4.75 &  4.31 & UGC05739       &   9.7 &  7.93 &  8.05 & UGC06713       &   8.6 &  5.16 &  4.86 \\
PGC041743      &   9.0 &  6.30 &  6.17 & PGC091215      &   7.9 &  5.54 &  5.40 & UGC04714       &   3.1 &  6.80 &  6.85 & UGC05740       &   8.8 &  4.83 &  4.55 & UGC06782       &   9.8 &  5.38 &  5.03 \\
PGC041965      &   8.0 &  5.56 &  5.48 & PGC091228      &   7.9 &  5.13 &  4.49 & UGC04722       &   7.9 &  3.90 &  3.28 & UGC05764       &   9.9 &  3.69 &  3.18 & UGC06791       &   6.5 &  7.18 &  7.29 \\
PGC042068      &   3.1 &  6.96 &  6.77 & PGC091413      &   7.9 &  5.43 &  4.79 & UGC04777       &   9.2 &  5.77 &  5.51 & UGC05791       &   3.0 &  4.79 &  4.34 & UGC06816       &   9.9 &  4.60 &  4.23 \\
PGC042160      &   9.5 &  4.83 &  4.55 & PGC1059326     &   3.6 &  4.32 &  3.95 & UGC04787       &   7.9 &  5.28 &  4.96 & UGC05798       &   6.4 &  4.96 &  4.50 & UGC06840       &   8.8 &  5.34 &  5.48 \\
PGC042868      &   8.0 &  5.51 &  5.77 & UGC00017       &   9.1 &  5.67 &  5.59 & UGC04797       &   8.8 &  6.06 &  5.69 & UGC05829       &   9.8 &  3.88 &  3.34 & UGC06849       &   8.8 &  5.39 &  5.18 \\
PGC043236      &   7.6 &  5.89 &  5.65 & UGC00132       &   7.9 &  6.34 &  6.15 & UGC04800       &   5.9 &  6.30 &  6.28 & UGC05832       &   4.2 &  5.36 &  4.99 & UGC06862       &   6.7 &  6.24 &  5.84 \\
PGC043341      &  10.0 &  4.84 &  4.46 & UGC00156       &   9.8 &  6.12 &  5.37 & UGC04834       &   8.0 &  5.38 &  5.28 & UGC05833       &  -2.0 &  2.83 &  4.01 & UGC06879       &   7.1 &  6.60 &  6.66 \\
PGC043345      &   9.0 &  5.06 &  4.72 & UGC00313       &   4.3 &  6.52 &  6.49 & UGC04837       &   8.7 &  5.27 &  4.92 & UGC05844       &   6.6 &  6.10 &  6.02 & UGC06900       &   9.7 &  6.72 &  6.41 \\
PGC043458      &   9.0 &  5.34 &  5.07 & UGC00634       &   8.8 &  4.86 &  4.53 & UGC04841       &   6.9 &  5.51 &  5.40 & UGC05846       &   9.9 &  4.42 &  4.03 & UGC06903       &   5.9 &  5.94 &  6.02 \\
PGC043679      &   6.5 &  6.93 &  6.81 & UGC00711       &   6.5 &  5.64 &  5.32 & UGC04845       &   6.9 &  6.31 &  6.08 & UGC05897       &   5.2 &  6.60 &  6.62 & UGC06917       &   8.8 &  5.67 &  5.46 \\
PGC043851      &   9.8 &  4.24 &  3.68 & UGC00866       &   7.8 &  5.52 &  5.30 & UGC04867       &   7.0 &  4.72 &  4.38 & UGC05918       &   9.8 &  4.39 &  3.87 & UGC06922       &   3.8 &  5.70 &  5.52 \\
PGC044157      &   9.0 &  5.70 &  5.61 & UGC00882       &   9.8 &  5.54 &  5.33 & UGC04871       &   8.8 &  4.76 &  4.28 & UGC05921       &   8.1 &  5.82 &  5.41 & UGC06931       &   9.0 &  4.91 &  4.53 \\
PGC044278      &   7.9 &  4.59 &  4.20 & UGC00891       &   9.0 &  5.05 &  4.81 & UGC04953       &   7.0 &  5.16 &  4.96 & UGC05922       &   6.8 &  5.81 &  5.68 & UGC06955       &   9.8 &  5.27 &  5.00 \\
PGC044735      &   9.7 &  4.98 &  4.94 & UGC00941       &   9.9 &  4.75 &  4.42 & UGC04970       &   5.7 &  7.56 &  7.66 & UGC05934       &   8.0 &  4.91 &  4.53 & UGC06956       &   8.8 &  5.50 &  5.28 \\
PGC044906      &   8.0 &  4.69 &  4.12 & UGC00964       &   3.0 &  6.41 &  6.23 & UGC04982       &   7.9 &  7.05 &  7.11 & UGC05947       &  10.0 &  4.62 &  4.26 & UGC07009       &   9.9 &  4.91 &  4.49 \\
PGC044952      &   5.0 &  6.32 &  6.02 & UGC01014       &   9.3 &  5.19 &  4.81 & UGC04988       &   9.0 &  5.90 &  5.81 & UGC05979       &   9.9 &  4.85 &  4.27 & UGC07019       &   9.9 &  5.41 &  5.06 \\
PGC044954      &   3.0 &  5.44 &  5.11 & UGC01020       &   2.1 &  8.13 &  7.75 & UGC05004       &  10.0 &  6.69 &  6.70 & UGC05989       &   9.9 &  4.99 &  4.61 & UGC07035       &   1.0 &  5.94 &  5.69 \\
PGC044982      &   9.7 &  5.24 &  4.42 & UGC01104       &   9.5 &  5.00 &  4.74 & UGC05015       &   7.9 &  5.99 &  5.93 & UGC05990       &   2.0 &  7.06 &  7.36 & UGC07053       &   9.9 &  4.73 &  4.31 \\
PGC045195      &   7.7 &  4.46 &  3.99 & UGC01110       &   5.9 &  6.38 &  6.55 & UGC05107       &   6.4 &  5.09 &  4.79 & UGC06014       &   8.0 &  5.35 &  4.97 & UGC07089       &   7.9 &  6.10 &  5.91 \\
PGC045257      &   9.0 &  4.95 &  4.59 & UGC01112       &   5.8 &  6.20 &  6.06 & UGC05114       &   9.9 &  4.69 &  4.28 & UGC06023       &   6.6 &  6.16 &  6.08 & UGC07094       &   8.0 &  7.22 &  6.79 \\
PGC045359      &   9.9 &  4.20 &  3.45 & UGC01176       &   9.9 &  4.75 &  4.26 & UGC05179       &   3.7 &  6.89 &  6.31 & UGC06083       &   4.1 &  6.27 &  6.30 & UGC07129       &   2.0 &  8.24 &  8.28 \\
PGC045652      &   6.7 &  6.67 &  6.50 & UGC01195       &   9.8 &  5.12 &  4.72 & UGC05228       &   4.9 &  7.00 &  6.70 & UGC06104       &   4.0 &  5.44 &  5.06 & UGC07133       &   6.6 &  5.82 &  5.61 \\
PGC045824      &   6.7 &  4.55 &  4.82 & UGC01197       &   9.8 &  5.46 &  5.15 & UGC05238       &   7.0 &  6.63 &  5.79 & UGC06112       &   7.4 &  5.17 &  4.89 & UGC07143       &   4.9 &  5.44 &  5.22 \\
PGC045877      &   4.9 &  8.04 &  8.19 & UGC01200       &   9.9 &  5.36 &  5.15 & UGC05249       &   6.7 &  5.19 &  4.78 & UGC06145       &   9.9 &  5.33 &  4.93 & UGC07153       &   5.9 &  6.33 &  6.19 \\
PGC046382      &   9.9 &  5.49 &  5.11 & UGC01240       &   7.9 &  4.78 &  4.34 & UGC05349       &   7.8 &  5.17 &  4.89 & UGC06151       &   8.8 &  4.87 &  4.50 & UGC07175       &   7.8 &  4.67 &  3.92 \\
PGC047721      &   4.0 &  8.31 &  8.62 & UGC01547       &   9.9 &  4.23 &  3.90 & UGC05354       &   4.3 &  4.40 &  3.95 & UGC06157       &   7.8 &  4.91 &  4.60 & UGC07184       &   6.4 &  5.66 &  5.44 \\
PGC047846      &   8.9 &  6.26 &  6.39 & UGC01551       &   6.1 &  5.52 &  5.47 & UGC05358       &   3.1 &  5.45 &  5.25 & UGC06162       &   6.4 &  5.20 &  4.97 & UGC07218       &   9.9 &  5.09 &  4.81 \\
PGC048087      &   8.0 &  6.09 &  5.28 & UGC01670       &   8.8 &  4.86 &  4.57 & UGC05373       &   9.9 &  5.08 &  4.77 & UGC06169       &   3.0 &  6.97 &  7.06 & UGC07239       &   9.8 &  6.04 &  5.97 \\
PGC050229      &   1.1 &  5.52 &  5.33 & UGC01862       &   6.4 &  6.50 &  6.52 & UGC05391       &   8.9 &  4.33 &  3.83 & UGC06171       &   9.9 &  4.88 &  4.61 & UGC07249       &   9.5 &  4.97 &  4.57 \\
PGC051291      &   2.0 &  4.91 &  4.52 & UGC01981       &   9.7 &  6.52 &  6.37 & UGC05393       &   8.0 &  4.65 &  4.84 & UGC06181       &   9.7 &  4.68 &  4.23 & UGC07257       &   8.1 &  4.49 &  4.56 \\
PGC051523      &   9.0 &  4.54 &  4.17 & UGC02275       &   8.8 &  3.92 &  3.29 & UGC05401       &   9.7 &  5.30 &  4.86 & UGC06194       &   4.9 &  5.45 &  5.10 & UGC07267       &   7.8 &  5.32 &  4.88 \\
PGC052940      &   9.9 &  4.16 &  3.44 & UGC02345       &   8.8 &  4.70 &  4.26 & UGC05423       &   9.9 &  5.49 &  5.18 & UGC06249       &   5.9 &  5.12 &  4.93 & UGC07271       &   6.9 &  5.60 &  5.30 \\
PGC053134      &   8.0 &  4.70 &  4.30 & UGC02429       &   9.0 &  5.47 &  5.36 & UGC05427       &   7.7 &  4.89 &  4.46 & UGC06271       &   1.0 &  7.81 &  8.15 & UGC07300       &   9.8 &  4.65 &  4.34 \\
PGC053568      &   8.0 &  4.95 &  4.56 & UGC04024       &   5.8 &  6.18 &  6.05 & UGC05446       &   5.9 &  4.96 &  4.67 & UGC06309       &   4.5 &  7.11 &  7.19 & UGC07332       &   9.9 &  4.29 &  3.79 \\
PGC053764      &   5.5 &  6.54 &  6.38 & UGC04121       &   8.8 &  5.26 &  5.04 & UGC05451       &   9.9 &  6.25 &  6.10 & UGC06320       &   8.0 &  5.64 &  5.30 & UGC07396       &   7.6 &  5.71 &  5.32 \\
PGC054944      &   1.8 &  6.28 &  6.09 & UGC04148       &   7.2 &  4.37 &  3.95 & UGC05456       &   9.0 &  4.68 &  4.13 & UGC06335       &   6.0 &  5.33 &  5.37 & UGC07408       &   9.8 &  5.67 &  5.14 \\
PGC065367      &   9.9 &  5.02 &  3.95 & UGC04151       &   7.9 &  6.01 &  5.91 & UGC05464       &   8.7 &  5.18 &  4.90 & UGC06345       &   9.9 &  4.47 &  3.77 & UGC07557       &   8.8 &  4.30 &  4.40 \\
PGC066242      &   4.1 &  5.56 &  5.29 & UGC04169       &   5.8 &  5.30 &  5.06 & UGC05478       &   9.9 &  5.23 &  5.03 & UGC06378       &   6.5 &  5.44 &  5.03 & UGC07559       &  9.9 &  4.34 &  3.94 \\
 \hline        
 \hline      
\end{tabular}                                                                                                                 
Table~\ref{Table7} {\it continued.}
\end {minipage}
\end{sideways}
\end{table}

\begin{table} 
\setcounter{table}{5} 
\setlength{\tabcolsep}{.02580in}
\begin {minipage}{133mm}
\caption{({\it Continued})}
\label{Table7}
\begin{tabular}{@{}lclclclclclccccccccccccccc@{}}
  \hline
  \hline
Galaxy&Type & $M_{\rm BH}$&
                                               $M_{\rm BH}$&Galaxy&Type & $M_{\rm BH}$&
                                               $M_{\rm BH}$&Galaxy&Type & $M_{\rm BH}$&
                                               $M_{\rm BH}$\\
& &(FUV)& (NUV)&& &(FUV)& (NUV)&& &(FUV)&(NUV)&& \\                                               
(1)& (2)&(3)& (4)&(1)& (2)&(3)& (4)&(1)& (2)&(3)& (4)\\
 \hline                                                                                                                                                        
UGC07577       &   9.8 &  5.51 &  5.15 & UGC08614       &   9.9 &  6.28 &  6.08 & UGC09837       &   5.3 &  5.33 &  5.24 & \\
UGC07590       &   4.1 &  4.86 &  4.50 & UGC08629       &   9.8 &  6.18 &  6.28 & UGC09858       &   4.0 &  6.61 &  6.59 &  \\
UGC07596       &   9.8 &  7.14 &  6.66 & UGC08630       &   5.7 &  5.85 &  5.69 & UGC09875       &   8.9 &  5.58 &  5.46 &  \\
UGC07599       &   8.7 &  4.29 &  3.86 & UGC08639       &   9.7 &  5.47 &  5.19 & UGC09936       &   8.8 &  4.88 &  4.67 &  \\
UGC07605       &   9.9 &  4.33 &  3.88 & UGC08642       &   6.8 &  4.89 &  4.59 & UGC09951       &   6.6 &  4.99 &  4.56 &   \\
UGC07612       &   8.8 &  4.84 &  4.53 & UGC08651       &   9.9 &  4.14 &  3.59 & UGC09977       &   5.3 &  6.89 &  7.39 &   \\
UGC07639       &   9.9 &  5.79 &  5.32 & UGC08658       &   5.0 &  5.94 &  5.93 & UGC09979       &   9.9 &  4.86 &  4.68 &   \\
UGC07673       &   9.9 &  4.93 &  4.68 & UGC08684       &   5.9 &  8.22 &  8.37 & UGC09992       &   9.9 &  4.85 &  4.30 &  \\
UGC07690       &   9.9 &  4.88 &  4.31 & UGC08688       &   7.8 &  4.69 &  4.38 & UGC10020       &   6.6 &  5.46 &  5.35 &  \\
UGC07698       &   9.9 &  4.97 &  4.67 & UGC08693       &   4.2 &  6.95 &  7.18 & UGC10054       &   8.0 &  5.28 &  5.00 &  \\
UGC07699       &   6.1 &  5.23 &  4.86 & UGC08733       &   5.9 &  5.15 &  4.92 & UGC10061       &   9.9 &  4.76 &  4.14 &  \\
UGC07700       &   7.9 &  4.91 &  4.56 & UGC08760       &   9.8 &  4.65 &  4.14 & UGC10290       &   8.8 &  4.67 &  4.30 &  \\   
UGC07719       &   8.0 &  4.76 &  4.31 & UGC08795       &   6.0 &  6.15 &  6.05 & UGC10413       &   5.7 &  6.05 &  6.15 &  \\  
UGC07730       &   8.8 &  5.59 &  5.24 & UGC08839       &   9.9 &  4.86 &  4.28 & UGC10445       &   6.0 &  5.01 &  4.63 &  \\   
UGC07739       &   9.8 &  6.11 &  5.83 & UGC08851       &   7.9 &  4.84 &  4.43 & UGC10608       &   8.0 &  3.58 &  3.15 &  \\   
UGC07774       &   6.3 &  5.82 &  5.41 & UGC08877       &   8.0 &  5.44 &  5.51 & UGC10721       &   5.8 &  6.82 &  6.68 &  \\   
UGC07795       &   9.8 &  3.92 &  3.52 & UGC08892       &   9.9 &  4.89 &  4.59 & UGC10736       &   8.0 &  4.85 &  4.23 &  \\   
UGC07802       &   6.1 &  6.58 &  6.45 & UGC08909       &   6.9 &  5.73 &  5.83 & UGC10791       &   8.8 &  6.44 &  6.39 &  \\   
UGC07824       &   9.0 &  6.13 &  6.05 & UGC08995       &   7.4 &  5.63 &  5.40 & UGC10806       &   8.0 &  4.37 &  4.43 &  \\   
UGC07844       &   6.1 &  7.12 &  7.14 & UGC09057       &   7.0 &  4.53 &  4.24 & UGC10854       &   6.0 &  5.11 &  5.20 &  \\   
UGC07906       &   9.9 &  4.62 &  4.26 & UGC09071       &   5.9 &  6.10 &  5.97 & UGC11782       &   8.8 &  4.98 &  4.61 &  \\   
UGC07911       &   8.8 &  5.32 &  5.21 & UGC09126       &   9.8 &  4.96 &  4.66 & UGC12151       &   9.7 &  5.24 &  5.08 &  \\   
UGC07941       &   7.0 &  4.90 &  4.56 & UGC09128       &   9.9 &  4.71 &  4.14 & UGC12178       &   8.0 &  5.27 &  4.72 &  \\   
UGC08041       &   6.9 &  5.74 &  5.61 & UGC09206       &   3.8 &  5.93 &  5.59 & UGC12613       &   9.8 &  6.06 &  5.19 &  \\   
UGC08048       &   9.5 &  5.09 &  4.64 & UGC09215       &   6.3 &  5.21 &  4.90 & UGC12681       &   4.2 &  4.35 &  3.91 &  \\   
UGC08052       &   4.7 &  6.79 &  6.82 & UGC09240       &   9.9 &  4.87 &  4.41 & UGC12682       &   9.8 &  4.62 &  4.09 &  \\   
UGC08053       &   8.0 &  4.49 &  4.02 & UGC09242       &   6.6 &  5.32 &  4.93 & UGC12709       &   8.7 &  5.36 &  5.25 &  \\   
UGC08056       &   6.4 &  4.74 &  4.42 & UGC09274       &   4.2 &  5.98 &  5.93 & UGC12732       &   8.7 &  4.42 &  4.02 &  \\
UGC08084       &   8.0 &  5.49 &  5.08 & UGC09299       &   6.4 &  4.50 &  4.06 & UGC12791       &   9.9 &  5.15 &  4.79 &  \\
UGC08127       &   9.8 &  4.98 &  4.63 & UGC09310       &   8.0 &  5.87 &  5.55 & UGC12843       &   8.2 &  4.94 &  4.56 &  \\
UGC08146       &   6.4 &  5.05 &  4.64 & UGC09356       &   4.6 &  5.70 &  5.51 & UGC12846       &   8.7 &  4.92 &  4.59 &  \\
UGC08153       &   6.6 &  4.89 &  4.59 & UGC09364       &   7.8 &  5.42 &  5.21 & UGC12856       &   9.6 &  4.48 &  4.59 &  \\
UGC08155       &   2.2 &  6.58 &  6.75 & UGC09380       &   9.9 &  4.46 &  4.16 &     \\
UGC08181       &   8.0 &  5.32 &  5.19 & UGC09389       &   3.2 &  5.47 &  5.12 &     \\
UGC08201       &   9.9 &  4.25 &  3.79 & UGC09392       &  10.0 &  5.29 &  4.95 &    \\
UGC08246       &   5.9 &  4.61 &  4.28 & UGC09394       &   5.9 &  4.85 &  4.53 &     \\
UGC08282       &   5.9 &  5.37 &  4.71 & UGC09448       &   3.3 &  8.17 &  8.34 &     \\
UGC08303       &   9.9 &  4.51 &  4.17 & UGC09469       &   9.7 &  5.43 &  5.18 &     \\
UGC08313       &   5.0 &  5.67 &  5.39 & UGC09470       &   7.9 &  4.91 &  4.55 &     \\
UGC08320       &   9.9 &  4.49 &  3.98 & UGC09482       &   6.6 &  5.60 &  5.07 &     \\
UGC08331       &   9.9 &  4.72 &  4.26 & UGC09569       &   6.6 &  4.64 &  4.19 &     \\
UGC08365       &   6.4 &  5.26 &  5.01 & UGC09601       &   5.9 &  5.41 &  5.24 &     \\
UGC08385       &   9.0 &  4.95 &  4.61 & UGC09661       &   8.0 &  6.05 &  5.97 &     \\
UGC08441       &   9.9 &  4.92 &  4.74 & UGC09663       &   9.8 &  4.92 &  4.63 &     \\
UGC08449       &   7.9 &  4.88 &  4.70 & UGC09682       &   8.6 &  5.30 &  4.79 &     \\
UGC08489       &   8.0 &  4.31 &  3.92 & UGC09730       &   6.6 &  5.07 &  4.68 &     \\
UGC08507       &   9.8 &  5.33 &  5.18 & UGC09746       &   4.0 &  7.09 &  7.07 &     \\
UGC08516       &   5.9 &  5.78 &  5.62 & UGC09760       &   6.6 &  4.82 &  4.23 &     \\
UGC08588       &   8.8 &  5.23 &  5.08 & UGC09815       &   7.9 &  5.34 &  5.29 &     \\
UGC08597       &   6.6 &  4.96 &  4.89 & UGC09816       &   9.8 &  5.65 &  5.50 &     \\
 \hline        
 \hline      
\end{tabular}                                                                                                                 
Table~\ref{Table7} {\it continued.}
\end {minipage}
\end{table}

\end{appendices}

\label{lastpage}
\end{document}